\documentclass{aa}

\usepackage{graphics}
\usepackage{epsfig}

\def\bib{\bibitem{}}

\newcommand{\xia}{\overline{\xi}}
\newcommand{\rhob}{\overline{\rho}}
\newcommand{\rhoa}{\overline{\rho}}
\newcommand{\gam}{\gamma}
\newcommand{\Om}{\Omega_m}
\newcommand{\Ol}{\Omega_{\Lambda}}
\newcommand{\inta}{\int_{-i\infty}^{+i\infty}}

\newcommand{\beq}{\begin{equation}}
\newcommand{\eeq}{\end{equation}}
\newcommand{\De}{{\cal D}}
\newcommand{\dmu}{\delta \mu}
\newcommand{\om}{\omega}
\newcommand{\phimu}{\varphi_{\mu}}
\newcommand{\dmup}{\delta \mu^{\;p}}

\newcommand{\ximu}{\xi_{\mu}}
\newcommand{\mumin}{\mu_{\rm min}}
\newcommand{\phieta}{\varphi_{\eta}}
\newcommand{\xieta}{\xi_{\eta}}
\newcommand{\heta}{h_{\eta}}
\newcommand{\hetat}{\tilde{h}_{\eta}}
\newcommand{\phietat}{\tilde{\varphi}_{\eta}}

\newcommand{\hdmu}{\hat{\delta \mu}}
\newcommand{\hn}{\hat{n}}
\newcommand{\Ib}{I_{\beta}}
\newcommand{\bphimu}{\varphi_{\beta,\mu}}
\newcommand{\bximu}{\xi_{\beta,\mu}}
\newcommand{\bmu}{\mu_{\beta}}
\newcommand{\muminb}{\mu_{\beta, {\rm min}}}
\newcommand{\Deo}{{\cal D}_{obs}}
\newcommand{\tP}{\tilde{P}}
\newcommand{\mumax}{\mu_{\rm max}}
\newcommand{\etamax}{\eta_{\rm max}}
\newcommand{\cP}{{\cal P}}
\newcommand{\RP}{{\cal R_P}}
\newcommand{\cLmin}{{\cal L}_{\rm min}}
\newcommand{\lag}{\langle}
\newcommand{\rag}{\rangle}

\newcommand{\Imu}{I_{\mu}}
\begin{document}
%
%
%
%
\renewcommand{\textfraction}{.01}
\renewcommand{\topfraction}{0.99}
\renewcommand{\bottomfraction}{0.99}
\setlength{\textfloatsep}{2.5ex}
\thesaurus{Sect.02 (12.03.4; 12.07.1; 12.12.1)}
\title{Weak gravitational lensing effects on the determination of $\Omega_m$ and $\Omega_{\Lambda}$ from SNeIa}   
\author{Patrick Valageas}
\institute{Service de Physique Th\'eorique, CEA Saclay, 91191 Gif-sur-Yvette, 
France}
\date{Received / Accepted }
\maketitle
\markboth{P. Valageas: Weak gravitational lensing effects on the determination of $\Omega_m$ and $\Omega_{\Lambda}$ from SNeIa}{P. Valageas: Weak gravitational lensing effects on the determination of $\Omega_m$ and $\Omega_{\Lambda}$ from SNeIa}

\begin{abstract}

In this article we present an analytical calculation of the probability distribution of the magnification of distant sources due to weak gravitational lensing from non-linear scales. We use a realistic description of the non-linear density field, which has already been compared with numerical simulations of structure formation within hierarchical scenarios. Then, we can directly express the probability distribution $P(\mu)$ of the magnification in terms of the probability distribution of the density contrast realized on non-linear scales (typical of galaxies) where the local slope of the initial linear power-spectrum is $n=-2$. We recover the behaviour seen by numerical simulations: $P(\mu)$ peaks at a value slightly smaller than the mean $\lag \mu \rag=1$ and it shows an extended large $\mu$ tail (as described in another article our predictions also show a good quantitative agreement with results from N-body simulations for a finite smoothing angle). 

Then, we study the effects of weak lensing on the derivation of the cosmological parameters from SNeIa. We show that the inaccuracy introduced by weak lensing is not negligible: $\Delta \Omega_m \ga 0.3$ for two observations at $z_s=0.5$ and $z_s=1$. However, observations can unambiguously discriminate between $\Omega_m=0.3$ and $\Omega_m=1$. Moreover, in the case of a low-density universe one can clearly distinguish an open model from a flat cosmology (besides, the error decreases as the number of observed SNeIa increases). Since distant sources are more likely to be ``demagnified'' the most probable value of the observed density parameter $\Omega_m$ is slightly smaller than its actual value. On the other hand, one may obtain some valuable information on the properties of the underlying non-linear density field from the measure of weak lensing distortions.

\end{abstract}


\section{Introduction}

Using Type Ia supernovae as standard candles, it is possible to derive the cosmological parameters $\Omega_m$ and $\Omega_{\Lambda}$ from the observed magnitude-redshift relation (e.g. Perlmutter et al.1999). However, several sources of uncertainty (e.g. redshift evolution of the luminosity of SNeIa) can affect this method. In particular, the apparent magnitude of distant supernovae can be distorted by gravitational lensing by the density fluctuations along the line of sight. In view of the importance of the measure of $\Omega_m$ and $\Omega_{\Lambda}$ from SNeIa, it is of interest to get a good estimate of the effect of weak gravitational lensing. Moreover, since the latter is directly linked to the matter distribution in the universe, one might use this effect to get some information on the large-scale structure of the universe itself. Note that contrary to the usual weak lensing statistics obtained when one considers the filtered distortions realized on large angular scales to probe the quasi-linear regime (e.g. Bernardeau et al.1997), the effects studied in this article which are relevant to SNeIa come from strongly non-linear scales ($\sim 100$ kpc).

Several authors have already studied some aspects of weak gravitational lensing in this non-linear regime, by analytical means (e.g. Frieman 1997; Kantowski 1998; Metcalf 1999; Hui 1999) or numerical simulations (e.g. Wambsganss 1997; Jain et al.1999). Here we present an analytical calculation of the probability distribution of the magnification due to weak lensing, from a model of the non-linear density field which has already been compared with numerical simulations of structure formation. Thus, we can directly express the properties of the weak lensing magnification in terms of the characteristics of the underlying density field. In particular, we recover the non-gaussian behaviour of the magnification as seen in previous numerical studies. Then we show that the the fluctuations of the magnification ($\delta \mu \sim 0.08$ at $z_s = 1$) lead to a significant uncertainty for $\Omega_m$ ($\Delta \Omega_m \ga 0.3$ for two observations at $z_s=0.5$ and $z_s=1$, but this error decreases as the number of observations increases). 

This article is organized as follows. In Sect.\ref{Density contrast probability distribution} we recall the description we use for the density field. Next, in Sect.\ref{Magnification by weak lensing} we derive the probability distribution of the magnification of distant sources by weak gravitational lensing. We present in Sect.\ref{Dependence on cosmology and redshift} the numerical results we obtain for three cosmologies (critical, open and low-density flat universes) as well as the dependence on the cosmological parameters of the amplitude of the fluctuations of the magnification. Then, in Sect.\ref{Galactic halos} we compare our results with another approach, used by several authors (e.g. Porciani \& Madau 1999), where the density field is described as a collection of smooth virialized halos. In particular, we point out the limitations of this method (it is restricted to large magnifications and it leads to some inconsistencies with some results from numerical simulations). Next, in Sect.\ref{Derivation of cosmological parameters} we show how weak lensing can affect the measure of $\Omega_m$ and $\Omega_{\Lambda}$ from SNeIa. Finally, in Sect.\ref{Bias} we present the bias due to weak lensing, that is the distortion of the luminosity function of SNeIa.

\section{Density contrast probability distribution}
\label{Density contrast probability distribution}

In order to obtain the probability distribution of the magnification of distant sources by gravitational lensing we need the properties of the density field. Hence we briefly recall here the formalism we use to characterize the density fluctuations. We shall use the same techniques to derive the statistics of the flux perturbations. It is convenient to express the probability distribution of the density contrast $\delta= (\rho-\rhob)/\rhob$ at scale $R$ and redshift $z$ (here $\rhob(z)$ is the mean universe density) in terms of the many-body correlation functions $\xi_p({\bf r}_1,...,{\bf r}_p)$. Thus we define the quantities ($p \geq 2$):
\beq
S_p = \frac{\xia_p}{\xia_2^{\; p-1}} \hspace{0.3cm} \mbox{with} \hspace{0.3cm} \xia_p = \int_V \frac{d^3r_1 ... d^3r_p}{V^p} \; \xi_p ({\bf r}_1,...,{\bf r}_p)
\label{Sp}
\eeq
where $V=4\pi/3 R^3$ is the volume of a spherical sphere of radius $R$. Next we introduce the generating function
\beq
\varphi(y)   =   \sum_{p=1}^{\infty} \frac{(-1)^{p-1}}{p!} \; S_p \;
y^{\;p}
\label{phiy}
\eeq
with $S_1=1$. Note that the parameters $S_p$ can also be written in term of the cumulants $\lag \delta_R^{\;p} \rag_c$ of the density contrast $\delta$ at scale $R$:
\beq
S_p = \frac{\lag \delta_R^{\;p} \rag_c}{\lag \delta_R^{\;2} \rag^{\;p-1}}
\label{Spcum}
\eeq
Then, one can show (White 1979; Balian \& Schaeffer 1989) that the probability distribution of the density contrast $\delta$ within spheres of size $R$ is:
\beq
P(\delta) = \inta \frac{dy}{2\pi i \xia} \; e^{[(1+\delta)y-\varphi(y)]/\xia}
\label{Phi}
\eeq
where we note $\xia_2$ as $\xia$. The relation (\ref{Phi}) provides the link of the density probability distribution with the correlation functions, hence with the cumulants of the density field. In the non-linear regime $\xia \gg 1$ it is convenient to define the function (inverse Laplace transform):
\beq
h(x) =  -\inta \frac{dy}{2 \pi i} \; e^{xy} \; \varphi(y)
\label{hphi}
\eeq
which obeys:
\beq
p \geq 1 \; : \;\; S_p = \int_0^{\infty}  x^p h(x) \; dx \hspace{0.3cm} , \hspace{0.3cm}  S_1=S_2=1  
\label{Sphx}
\eeq
From very general considerations (Balian \& Schaeffer 1989) one expects the function $\varphi(y)$ to behave as a power-law for large $y$:
\beq
y \rightarrow +\infty \; : \; \varphi(y) \sim a \; y^{1-\omega}
\hspace{0.3cm} \mbox{with} \; 0 \leq \omega \leq 1
\eeq
and to display a singularity at a small negative value of $y$:
\beq
y \rightarrow y_s^+ \; : \; \varphi(y) = - a_s \; \Gamma(\omega_s)
\; (y-y_s)^{-\omega_s}
\eeq
where we neglected less singular terms. From this behaviour of $\varphi(y)$ we have:
\beq
\left\{ \begin{array}{rl} x \ll 1 \; : & {\displaystyle  h(x) \sim
\frac{a(1-\omega)}{\Gamma(\omega)} \; x^{\omega-2} } \\ \\  x \gg 1 \; : & {\displaystyle h(x) \sim a_s \; x^{\omega_s-1} \; e^{-x/x_s} } \end{array} \right.
\label{has}
\eeq
with $x_s=1/|y_s|$. The interest of the function $h(x)$ is that in the non-linear regime for large density contrasts the density probability distribution can be written as (Balian \& Schaeffer 1989):
\beq
\xia \gg 1 \; , \; (1+\delta) \gg \xia^{\;-\om/(1-\om)} \; : \; P(\delta) = \frac{1}{\xia^{\;2}} \; h(x)
\eeq
with:
\beq
x = \frac{1+\delta}{\xia}
\eeq
Thus, the density probability distribution $P(\delta)$ shows a power-law behaviour from $(1+\delta) \sim \xia^{\;-\om/(1-\om)}$ up to $(1+\delta) \sim x_s \xia$ with an exponential cutoff above $x_s \xia$. The measure of $P(\delta)$ in numerical simulations allows one to recover $h(x)$ hence $\varphi(y)$ since (\ref{hphi}) can be inverted as:
\beq
\varphi(y) = \int_0^{\infty} \; \left( 1 - e^{-xy} \right) \; h(x) \; dx
\label{phih}
\eeq
The function $h(x)$ has been measured in the non-linear regime for various power-spectra by several authors (Valageas et al.1999; Bouchet et al.1991; Colombi et al.1997; Munshi et al.1999). In particular, although Colombi et al.(1996) found a small scale-dependence other authors found that the numerical results were consistent with $h(x)$ being scale-independent in the non-linear regime. Thus, in the following we shall use the scaling function $h(x)$ obtained by Valageas et al.(1999) for any redshift. Note that it depends on the power-spectrum and it must be obtained from numerical simulations since there is no known method to derive analytically $h(x)$. The scale-invariance of $h(x)$, hence of the coefficients $S_p$, can be interpreted as evidence for the stable-clustering ansatz (Peebles 1980):
\beq
\xi_p(\lambda {\bf r}_1,...,\lambda {\bf r}_p ;a) = a^{3(p-1)} \;
\lambda^{-\gam(p-1)} \; \hat{\xi}_p({\bf r}_1,...,{\bf r}_p)
\label{scal1}
\eeq
which was studied in details in Balian \& Schaeffer (1989). Here $a(t)$ is the scale-factor and $\gam$ is the (local) slope of the two-point correlation function. Note that for an initial linear power-spectrum which is a power-law $P(k) \propto k^n$ we have if stable-clustering is valid:
\beq
\gam = \frac{3(3+n)}{5+n}
\label{gamma}
\eeq
The interest of the formulation (\ref{Phi}) is that once $h(x)$, or $\varphi(y)$, is known the density probability distribution $P(\delta)$ can be obtained for any time and scale in the non-linear regime provided that one knows the behaviour of $\xia$ (and that indeed $h(x)$ is scale-invariant). Note that a similar technique can be used in the quasi-linear regime with a different $\varphi(y)$ obtained by perturbative calculations (Bernardeau 1994). 

Then, in order to obtain the properties of the non-linear density field we only need to model the evolution of the two-point correlation function $\xi_2$, or of the power-spectrum $P(k)$. To this order we use the fits given by Peacock \& Dodds (1996) which give the non-linear power-spectrum $P(k)$ from its linear counterpart $P_L(k)$. Note that this behaviour of $P(k)$ is consistent with the stable-clustering ansatz.

\section{Magnification by weak lensing}
\label{Magnification by weak lensing}

\subsection{Definition}
\label{Definition}

As a photon travels from a distant source towards the observer its trajectory is deflected by density fluctuations along the light path. This produces an apparent displacement of the source as well as a distortion of the image. In particular, the convergence $\kappa$ (defined as the trace of the shear matrix) will magnify (or demagnify) the source as the cross section of the beam is decreased (or increased). One can show (Bernardeau et al.1997; Kaiser 1998) that the convergence is given by:
\beq
\kappa = \frac{3}{2} \; \Omega_m \int_0^{\chi_s} d\chi \; w(\chi,\chi_s) \; \delta(\chi)
\label{kappa}
\eeq
with
\beq
w(\chi,\chi_s) = \frac{H_0^2}{c^2} \; \frac{\De(\chi) \De(\chi_s-\chi)}{\De(\chi_s)} \; (1+z)
\eeq
where $\chi$ is the radial comoving coordinate (and $\chi_s$ corresponds to the redshift $z_s$ of the source):
\beq
d\chi = \frac{ \frac{c}{H_0} \;\; dz}{\sqrt{\Omega_{\Lambda}+(1-\Omega_m-\Omega_{\Lambda})(1+z)^2+\Omega_m(1+z)^3}}
\eeq
while $\De$ is defined by:
\beq
\De(\chi) = \frac{c/H_0}{\sqrt{1-\Omega_m-\Omega_{\Lambda}}} \sinh \left( \sqrt{1-\Omega_m-\Omega_{\Lambda}} \; \chi \right)
\label{De}
\eeq
The relation (\ref{kappa}) assumes that the metric perturbations $\phi$ are small ($\phi \ll 1$) but the density fluctuations $\delta$ can be large (Kaiser 1992). The magnification $\mu$ is linked to the convergence $\kappa$ and the intensity of the shear $\gam$ by:
\beq
\mu = \frac{1}{(1-\kappa)^2 - |\gam|^2}
\label{mukappa1}
\eeq
Thus, for small values of $\kappa$ we have: 
\beq
\kappa \ll 1 \; : \; \mu = 1+2\kappa
\label{mukappa}
\eeq
and we write the flux perturbation $\delta \mu = \mu-1$ as:
\beq
\dmu = 3 \Omega_m \int_0^{\chi_s} d\chi \; w(\chi,\chi_s) \; \delta(\chi)
\label{dmu}
\eeq
Of course, because of flux conservation the mean shift of the flux over all lines of sight is zero: $\lag \dmu \rag=0$. We can see directly in (\ref{dmu}) that there is a minimum value $\mumin(z_s)$ for the magnification of a source located at redshift $z_s$:
\beq
\mumin(z_s) = 1 - 3 \Omega_m F_s(\chi_s) 
\label{mumin}
\eeq
with
\beq
F_s(\chi_s) = \int_0^{\chi_s} d\chi \; w(\chi,\chi_s)
\label{Fs}
\eeq
This corresponds to an ``empty'' beam between the source and the observer ($\delta=-1$ everywhere along the line of sight).

\subsection{Probability distribution $P(\mu)$}
\label{Probability distribution}

Next we wish to obtain the probability distribution of the magnification $\mu$ from (\ref{dmu}). To this order we simply need to derive the cumulants $\lag \dmup \rag_c$. This will provide the parameters $S_{\mu,p}$, similar to (\ref{Spcum}), and the generating function $\phimu(y)$, similar to (\ref{phiy}). However, it is convenient to introduce first a ``reduced magnification'' $\eta$ and its variation $\delta \eta$ by:
\beq
\eta = \frac{\mu-\mumin}{1-\mumin} \hspace{0.3cm} ,  \hspace{0.3cm}  \delta \eta = \eta - 1 = \frac{\dmu}{3 \Omega_m F_s}
\label{eta}
\eeq
From (\ref{dmu}) and (\ref{eta}) we obtain the cumulant of order $p$ of $\delta \eta$:
\beq
\lag \delta \eta^p \rag_c = \lag \int_0^{\chi_s} \prod_{i=1}^p d\chi_i \; \frac{w(\chi_i,\chi_s)}{F_s} \; \delta(\chi_i) \; \rag_c
\eeq
Since the correlation length (beyond which the many-body correlation functions are negligible) is much smaller than the Hubble scale $c/H(z)$ (where $H(z)$ is the Hubble constant at redshift $z$) we obtain:
\beq
\begin{array}{l} {\displaystyle \lag \delta \eta^p \rag_c = \int_0^{\chi_s} d\chi \left( \frac{w(\chi,\chi_s)}{F_s} \right)^p } \\ \\  {\displaystyle \hspace{2.5cm} \times \int_{-\infty}^{\infty} \prod_{i=2}^{p} d \chi_i \;\; \xi_p(0,\chi_2,..,\chi_p;z) }
\end{array}
\label{dmupc3}
\eeq
From (\ref{scal1}) we see that the integral over the points $\chi_i$ along the line of sight is dominated by the comoving scale $R_c(z)$ such that the local slope of the two-point correlation function is $-\gam(R_c)=-1$. This corresponds to a local linear power-spectrum index $n=-2$, see (\ref{gamma}). Indeed, for realistic power-spectra like CDM the slope of $\xi_2(x)$ decreases from 0 at small scales down to $-4$ at large scales. On the other hand, power-spectra which would be pure power-laws would lead to divergences. Since for the power-spectra we shall use $R_c(z)$ corresponds to galactic scales ($R_c(0) \sim 100$ kpc) within the highly non-linear regime ($\xia(R_c;0) \sim 2000$), we must indeed use the non-linear many-body correlation functions $\xi_p(0,\chi_2,..,\chi_p;z)$. Thus we can use the scaling laws (\ref{scal1}). Of course, at large redshifts $z > 7$ the scale $R_c(z)$ will enter the linear regime, however since we shall restrict ourselves to smaller $z$ we can always use (\ref{scal1}). Now we must estimate the contribution of the integral over the points $\chi_i$ along the line of sight. Although the points $\chi_i$ are drawn on a line and not within a sphere we shall use the approximation (compare with (\ref{Sp})):
\begin{eqnarray}
\int_{-\infty}^{\infty} \prod_{i=2}^{p} d\chi_i \; \xi_p(0,\chi_2,..,\chi_p) & \simeq & S_p  \left( \int_{-\infty}^{\infty} d\chi \; \xi_2(0,\chi) \right)^{p-1} \nonumber \\ & & 
\label{Sp2}
\end{eqnarray}
where the coefficients $S_p$ are obtained for $n=-2$ (i.e. $\gam=1$) from the numerical results described in Valageas et al.(1999). Thus, we define the quantity $\Imu(z)$ by:
\beq
\Imu(z) = \int_{-\infty}^{\infty} dx \; \xi_2(x;z)
\label{Imu1}
\eeq
It is convenient to express $\Imu$ in terms of the non-linear power-spectrum which is directly provided by the fits obtained in Peacock \& Dodds (1996). Thus, using the Fourier transform:
\beq
\xi_2({\bf x}) = \int d^3k \; e^{i {\bf k . x}} \; P(k) = \int_0^{\infty} \frac{dk}{k} \; \Delta^2(k) \; \frac{\sin (kx)}{kx}
\eeq
where we defined the power-spectrum $P(k)$ by:
\[
\lag \delta({\bf k}_1) \delta({\bf k}_2) \rag = P(k_1) \; \delta_D({\bf k}_1+{\bf k}_2) \hspace{0.2cm} , \hspace{0.2cm} \Delta^2(k) = 4 \pi k^3 P(k)
\]
we obtain (note that $\chi$, $x$ and $k$ are comoving coordinates):
\beq
\Imu(z) = \pi \int_0^{\infty} \frac{dk}{k} \; \frac{\Delta^2(k)}{k}
\label{Imu2}
\eeq
Next, using the approximation (\ref{Sp2}) in the relation (\ref{dmupc3}) we get:
\beq
\lag \delta \eta^p\rag_c = \int_0^{\chi_s} d\chi \; \left( \frac{w}{F_s} \right)^p \; S_p \; \Imu(z)^{p-1}
\label{dmup2}
\eeq
In particular, this yields for the variances $\xieta=\lag \delta \eta^2 \rag$ and $\ximu = \lag \dmu^2 \rag$ of the fluctuations of the amplification of a source located at redshift $z_s$:
\beq
\xieta = \int d\chi \left(\frac{w}{F_s}\right)^2 \Imu \hspace{0.3cm} , \hspace{0.3cm}  \ximu = (3 \Omega_m)^2 \int d\chi \; w^2 \; \Imu
\label{Ximu}
\eeq
Note that our result (\ref{Ximu}) for the {\it rms} fluctuation $\sqrt{\ximu}$ does not use the approximation (\ref{Sp2}). Moreover, the expressions (\ref{Ximu}) clearly show that most of the contribution to weak gravitational lensing effects comes from the scale $R_c(z)$ where $\Delta^2(k)/k$ is maximum. This corresponds to $\gam=1$ or $n'=-2$ where $n'$ is the slope of the {\it non-linear power-spectrum}. From (\ref{gamma}) we see that it also corresponds to $n=-2$ where $n$ is the slope of the linear power-spectrum. Then, from (\ref{dmup2}) we can define the generating function $\phieta(y)$, similar to (\ref{phiy}), by:
\beq
\phieta(y) = y + \sum_{p=2}^{\infty} \frac{(-1)^{p-1}}{p!} \; \frac{\lag\delta \eta^p\rag_c}{\xieta^{p-1}} \; y^{p}
\label{phimu}
\eeq
This yields:
\beq
\phieta(y) = \int_0^{\chi_s} d\chi \; \frac{\xieta}{\Imu} \; \varphi \left( y  \frac{w}{F_s} \frac{\Imu}{\xieta} \right)
\label{phieta}
\eeq
where $\varphi$ is the generating function of the density contrast defined in (\ref{phiy}). Next, in a fashion similar to (\ref{Phi}) since $\lag \delta \eta \rag=0$, we obtain for the probability distribution of $\eta$:
\beq
P(\eta) = \inta \frac{dy}{2\pi i \xieta} \; e^{[\eta y - \phieta(y)] / \xieta}
\label{Pphieta}
\eeq
and the {\it p.d.f.} $P(\mu)$ is given by:
\beq
P(\mu) = \frac{1}{1 - \mumin} \; P(\eta)
\label{Pphimub}
\eeq
The expansion of $\phieta(y)$ in $y=0$ verifies:
\beq
\phieta(y) = y -\frac{y^2}{2} + ...
\eeq
which implies, as it must, that:
\beq
\lag\eta\rag = 1 \hspace{1cm} \mbox{and} \hspace{1cm} \left< (\eta-\lag\eta\rag)^2 \right> = \xieta
\eeq
The relations (\ref{phieta}) and (\ref{Pphieta}) provide the probability distribution of the magnification of a source located at redshift $z_s$. Note that the generating function $\phieta(y)$ depends on $z_s$. We can also check in (\ref{Pphieta}), using (\ref{phieta}), that $P(\eta) = 0$ for $\eta \leq 0$ as it should (since for $\eta \leq 0$ we can push the integration path in (\ref{Pphieta}) towards the right, Re$(y) \rightarrow +\infty$, where the exponential vanishes). Thus, {\it the approximation (\ref{Sp2}) has preserved the fact that $P(\mu)=0$ for $\mu \leq \mumin$}. In a fashion similar to (\ref{hphi}) we can define the function:
\beq
\begin{array}{ll} {\displaystyle \heta(x)} & {\displaystyle =  -\inta \frac{dy}{2 \pi i} \; e^{xy} \; \phieta(y) } \\ \\  & {\displaystyle = \int_0^{\chi_s} d\chi \; \frac{\xieta}{\Imu} \; \frac{1}{T_s(\chi)} \; h \left( \frac{x}{T_s(\chi)} \right) }
\end{array}
\label{hphieta}
\eeq
with:
\beq
T_s(\chi) = \frac{w \; \Imu}{F_s \; \xieta}
\label{Ts}
\eeq
The function $\heta(x)$ satisfies:
\beq
p \geq 1 \; : \;\; S_{\eta,p} = \int_0^{\infty}  x^p \heta(x) \; dx \hspace{0.3cm} , \hspace{0.3cm}  S_{\eta,1}=S_{\eta,2}=1  
\label{Spheta}
\eeq
with
\beq
p \geq 2 \; : \; S_{\eta,p} = \frac{\lag\delta \eta^{\;p}\rag_c}{\xieta^{\;p-1}}
\eeq
From (\ref{hphieta}) and (\ref{has}) we can obtain the parameters which govern the asymptotic behaviour of $\heta(x)$:
\beq
\left\{ \begin{array}{l} {\displaystyle \om_{\eta} = \om \hspace{0.5cm} , \hspace{0.5cm} a_{\eta} = a \int_0^{\chi_s} d\chi \; \frac{\xieta}{\Imu} \; T_s(\chi)^{1-\om} }
\\ \\ {\displaystyle \om_{\eta,s} = \om_s - \frac{1}{2} \hspace{0.5cm} , \hspace{0.5cm} x_{\eta,s} = x_s \; T_s(\chi_c) } \\ \\ {\displaystyle a_{\eta,s} = a_s \; \frac{\xieta}{\Imu} \; \left| \frac{2\pi x_s}{T_s^{''}(\chi_c)} \right|^{1/2} \; T_s(\chi_c)^{1-\om_s} }
\end{array} \right.
\eeq
where $0<\chi_c<\chi_s$ is the point where $T_s(\chi)$ is maximum. Thus, we see that the slope of $\heta(x)$ at low $x$ is the same as $h(x)$ while the exponential cutoff is slightly modified ($T_s(\chi_c) \simeq 1$). From (\ref{Fs}) and (\ref{Ximu}) we see that $T_s(\chi) \simeq 1$ so we are led to the approximation:
\beq
\hetat(x) = h(x) \hspace{0.3cm} \mbox{and} \hspace{0.3cm} \phietat(y) = \varphi(y)
\label{phietat}
\eeq
This leads to the approximate {\it p.d.f.} for $\eta$:
\beq
\tP(\eta) = \inta \frac{dy}{2\pi i \xieta} \; e^{[\eta y - \varphi(y)] / \xieta}
\label{Pphietat}
\eeq
which is still properly normalized and satisfies $\tP(\eta)=0$ for $\eta \leq 0$. The practical advantage of the approximation (\ref{Pphietat}) is that to compute the {\it p.d.f.} $P(\eta)$ we can directly use the functions $h(x)$ and $\varphi(y)$ obtained in numerical simulations. Note that formally $\tP(\eta)$ can be expressed in terms of the probability distribution $P(\delta)$ of the density contrast $\delta$ at the scale such that $n=-2$ as:
\beq
\tP(\eta) = P( \delta \rightarrow \eta-1 ; \xia \rightarrow \xieta )
\label{PetaPdelta}
\eeq
using (\ref{Phi}). Note however that even for $\xieta \leq 1$ the probability distribution $P(\delta)$ used in (\ref{PetaPdelta}) corresponds to the highly non-linear regime. Thus in the case $\xieta \leq 1$ the probability distribution $P(\delta)$ in (\ref{PetaPdelta}) is not the one measured at the time such that $\xia=\xieta$. Nevertheless, the expressions (\ref{Pphietat}) and (\ref{PetaPdelta}) clearly show that the measure of the {\it p.d.f.} $P(\mu)$, or of $P(\kappa)$, provides a direct estimate of the {\it p.d.f.} $P(\delta)$ of the density contrast through $\varphi(y)$ (see also the study of $P(\kappa)$ in Valageas 1999b).  

As can be seen from (\ref{hphieta}) or (\ref{Pphietat}) the probability distributions $P(\eta)$ and $P(\mu)$ are non-gaussian. Indeed, they are {\it strictly zero for $\eta \leq 0$ and $\mu \leq \mumin$}, they show an {\it exponential cutoff} $\sim e^{-\eta/(x_{\eta,s} \xieta)}$ (or $\sim e^{-\mu/(x_{\eta,s} (1-\mumin) \xieta)}$) for large $\eta$ (or large $\mu$), and they have their {\it maximum at a value smaller than the mean} $\lag\eta\rag=\lag\mu\rag=1$.

We compare in details the predictions of our approach with available numerical results from N-body simulations (Jain et al.1999) for the convergence $\kappa$ smoothed on small angular scales ($\theta \sim 1'$) in Valageas (1999b). This comparison shows that our approach (which can be extended to finite smoothing windows in a straightforward fashion) provides very good results for all three cosmologies we consider here (e.g., see Fig.4 and Fig.5 in Valageas 1999b). Moreover, we show in that paper that the approximation (\ref{phietat}) gives reasonable results which are quite close to the more accurate expression (\ref{phieta}). Indeed, we shall check below that the correction to the third moment (for instance) due to (\ref{Pphietat}) is quite small and well within the errorbars of the value of $S_3$ obtained by counts-in-cells calculations from numerical simulations. Note that the evaluation of the probability distribution of the magnification $\mu$ by ray tracing through N-body simulations would of course suffer from the same uncertainty, which affects the moments of the density distribution itself (because of numerical inaccuracy).

\subsection{Scaling functions $\varphi(y)$ and $h(x)$}
\label{Scaling functions}

In order to perform numerical calculations, we need to choose the function $h(x)$, or equivalently the parameters $S_p$ defined in (\ref{Sp}). As we explained above, most of the contributions to the weak lensing effects come from the scales where the local slope of the linear power-spectrum is $n=-2$, see also Fig.2 in Valageas (1999b). Hence we use in the following the scaling function $h(x)$ obtained from numerical simulations by Valageas et al.(1999) for the case of a critical universe with an initial linear power-spectrum which is a power-law $n=-2$:
\beq
h(x) = \frac{a(1-\omega)}{\Gamma(\omega)} \; \frac{x^{\omega-2}}{(1+bx)^{c}}
\; \exp(-x/x_s)    
\label{fithx}
\eeq
with:
\[
a=1.71 \; , \; \omega=0.3 \; , \; x_s=13 \; , \; b= 5 \;\; \mbox{and} \;\; c=0.6
\]
\[ 
\mbox{hence} \;\; \omega_s = \omega -1 -c =-1.3
\]
In fact, the curvature of the CDM power-spectrum may slightly change the parameters $S_p$ from the value they would have for a pure power-law $P(k)$. However, in order to improve meaningfully this approximation one would need to measure the parameters $S_p$ realized on a line rather than in a sphere, see (\ref{Sp2}). Thus, we think our approach is the best analytical tool one can currently build. The scaling function $h(x)$ shown in (\ref{fithx}) defines the generating function $\varphi(y)$ through (\ref{phih}). In particular, one obtains (see Gradshteyn \& Ryzhik 1965, \S 9.211, p.1058):
\beq
\varphi'(y) = a (1-\om) b^{-\om} \; \psi \left(\om,2+\om_s;\frac{y-y_s}{b}\right)
\eeq
and
\beq
\begin{array}{l} {\displaystyle \varphi(y) = a \; b^{1-\om} \left[ \; \psi \left(\om-1,1+\om_s;\frac{y-y_s}{b}\right) \right. } \\ \\ {\displaystyle \left. \hspace{3cm} - \psi \left(\om-1,1+\om_s;\frac{-y_s}{b}\right) \right] } \end{array}
\eeq
where $\psi$ is Kummer's function which can be expressed in terms of the difference between two confluent hypergeometric functions $ _{1}F_{1}$. Next, to obtain $\tP(\eta)$ from (\ref{Pphietat}) one simply needs to perform an integration in the complex plane. In order to make the integral (\ref{Pphietat}) converge sufficiently fast, it is convenient to define the integration path by the constraint ${\cal I}m[\eta y - \varphi(y)] = 0$. However, in practice it is sufficient to use ${\cal I}m[\eta y - a (y-y_s)^{1-\om}] = 0$ (we replace $\varphi(y)$ by a power-law with the right behaviour for $y \rightarrow +\infty$ and the right location of the singularity at $y_s$) which gives for the integration path:
\beq
y = y_s + \rho \; e^{i\theta} \;\;\;\;\;\; \mbox{with} \;\;\;\;\;\; \rho = \left[ \frac{a}{\eta} \; \frac{\sin(1-\om)\theta}{\sin\theta} \right]^{1/\om}
\eeq
Note that it is better to define $\varphi(y)$ from $h(x)$ rather than trying to use a fit for $\varphi(y)$ itself. Indeed, from (\ref{Sp}) and (\ref{phiy}) we see that:
\beq
p \geq 1 \; : \hspace{0.6cm} (-1)^{p-1} \varphi^{(p)}(0) = S_p > 0
\eeq
and moreover, using Schwarz' inequality and the scalar product $\lag f|g \rag=\int f(x) g(x) x h(x) dx$, one can see that the coefficients $S_p$ must obey:
\beq
S_{p+q} \; S_{p-q} \geq S_p^2
\eeq
These constraints are automatically verified if one defines $\varphi(y)$ from $h(x)$. If one uses a fit for $\varphi(y)$ which does not obey these constraints one may get negative probabilities (since in this case $h(x)$ has to be negative in some range).

The fit (\ref{fithx}) for the function $h(x)$ was obtained for a critical universe. In the case of a low density universe, we use the same function although there are no numerical results available (from counts-in-cells statistics) to validate (or invalidate) this choice. However, we note that the fact the dependence on cosmology of the two-point correlation function is accurately given by the simple term described in Peacock \& Dodds (1996) suggests that the structure of the non-linear clustering pattern is the same for low $\Omega_m$ as for a critical universe, once the effect of the slow-down of the linear growth factor is taken into account. Indeed, most highly non-linear structures formed when the universe was close to critical ($\Omega_m(z) \simeq 1$) since at later times the slow-down of the linear growth factor prevents additional new structures to form. Of course, this break in the hierarchy of scales which successively turn non-linear may also lead with time to some difference with the case of a critical universe (at least for the scales which were the last to collapse). Detailed numerical studies are needed to investigate more precisely this point. However, for reasonable cosmologies $\Omega_m \ga 0.1$ our model provides a good approximation as shown by a direct comparison of $P(\kappa)$ with results from N-body simulations, as described in Valageas (1999b). 

Note that, except for this possible dependence of the parameters $S_p$, all our results are valid for any realistic power-spectrum such that $n<-2$ on small scales, $n>-2$ on large scales and the scale where $n=-2$ is non-linear. In particular, note that for such power-spectra all moments of the convergence $\kappa$ and of the magnification $\mu$ converge: there is no need to introduce a cutoff at small scales.

\section{Dependence on cosmology and redshift}
\label{Dependence on cosmology and redshift}

Using the results obtained in the previous sections we can compute the probability distribution $P(\mu)$ of the magnification for various cosmologies. We shall mainly consider three cases: a CDM critical density universe (SCDM), a low-density open universe (OCDM) and a low-density flat universe with a non-zero cosmological constant ($\Lambda$CDM). The main cosmological parameters of these scenarios are described in Tab.\ref{table1}. Here $\Gamma$ is the usual shape parameter of the power-spectrum and we use the fit given by Bardeen et al.(1986) for $P(k)$. These cosmologies are those we use in Valageas (1999b) to compare our predictions with available results from N-body simulations. The reader is refered to that paper for a detailed discussion of the accuracy of our approach and for an extension of our model to finite smoothing windows.

\begin{table}
\begin{center}
\caption{Cosmological models}
\label{table1}
\begin{tabular}{|r|lll|}\hline

  & SCDM & OCDM &  $\Lambda$CDM \\ \hline

$\Om$ & 1 & 0.3 & 0.3 \\
$\Ol$ & 0 &  0 &   0.7  \\

$H_0$ [km/s/Mpc] & 50 & 70 & 70 \\

$\sigma_8$ & 0.6 & 0.85 & 0.9 \\

$\Gamma$ & 0.5 & 0.21 & 0.21 \\ \hline

\end{tabular}
\end{center}
\end{table}

\subsection{Fluctuations of the magnification}
\label{Fluctuations of the magnification}

\begin{figure}

{\epsfxsize=8 cm \epsfysize=5.4 cm \epsfbox{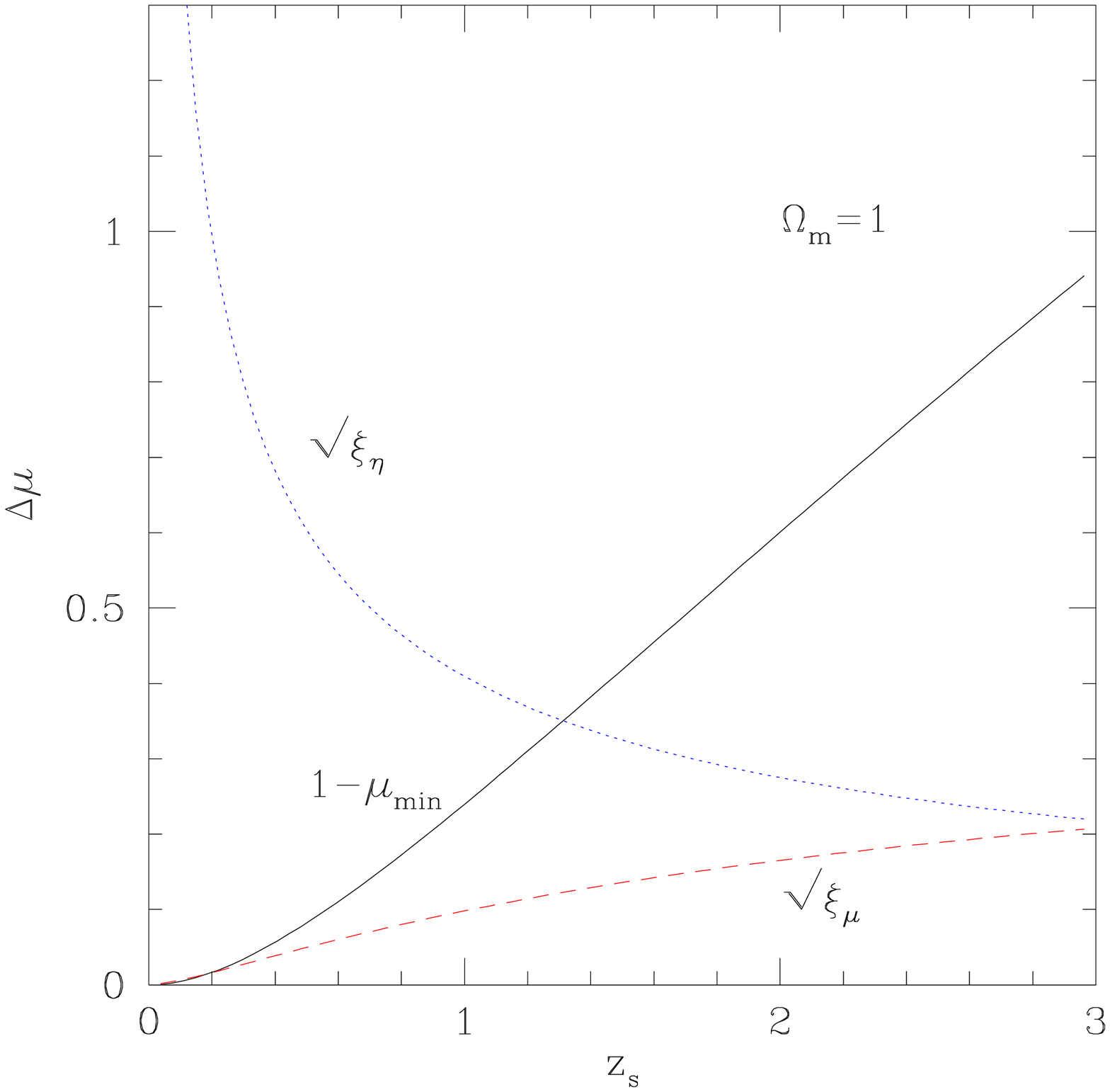} }
{\epsfxsize=8 cm \epsfysize=5.4 cm \epsfbox{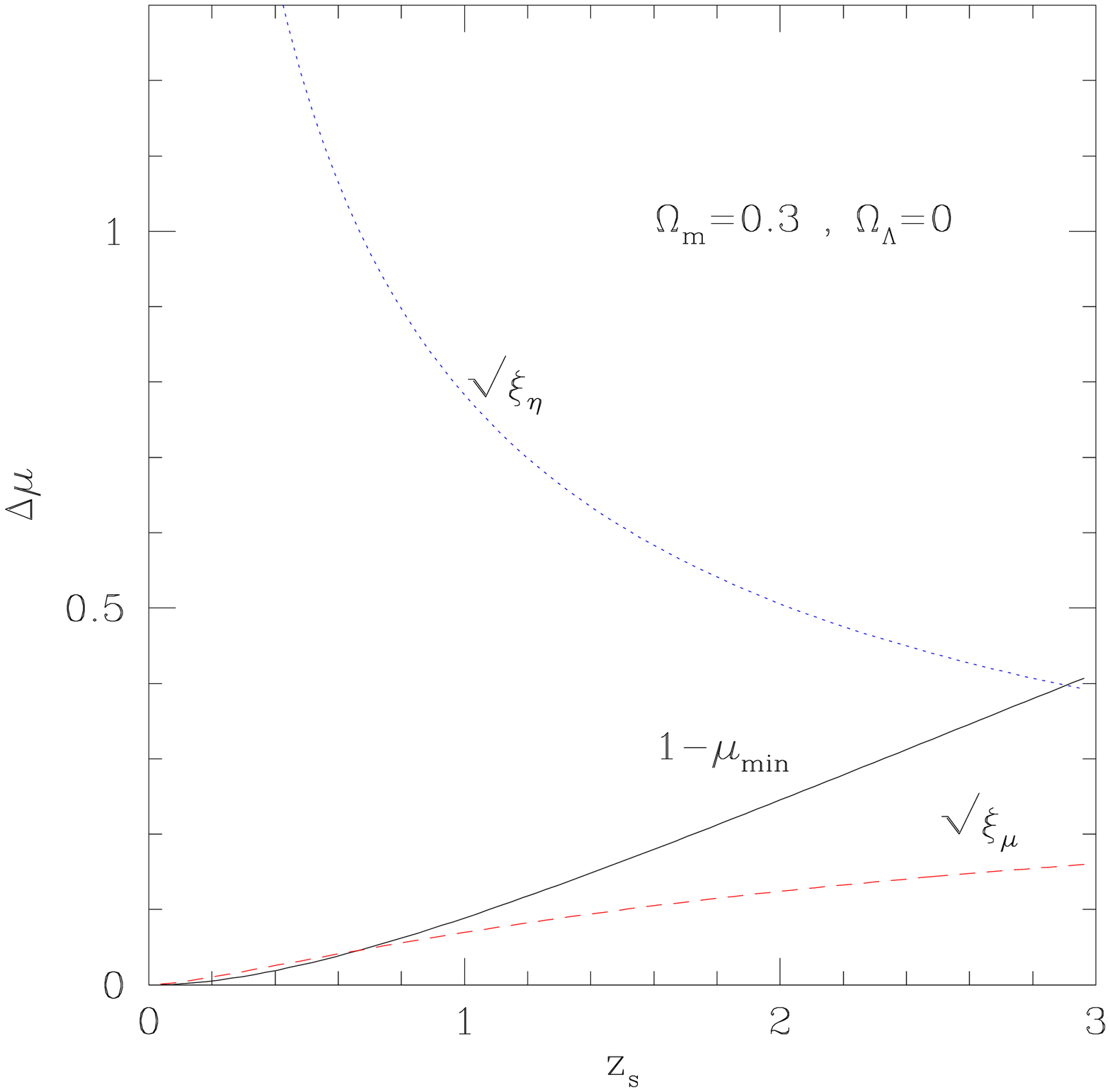} }
{\epsfxsize=8 cm \epsfysize=5.4 cm \epsfbox{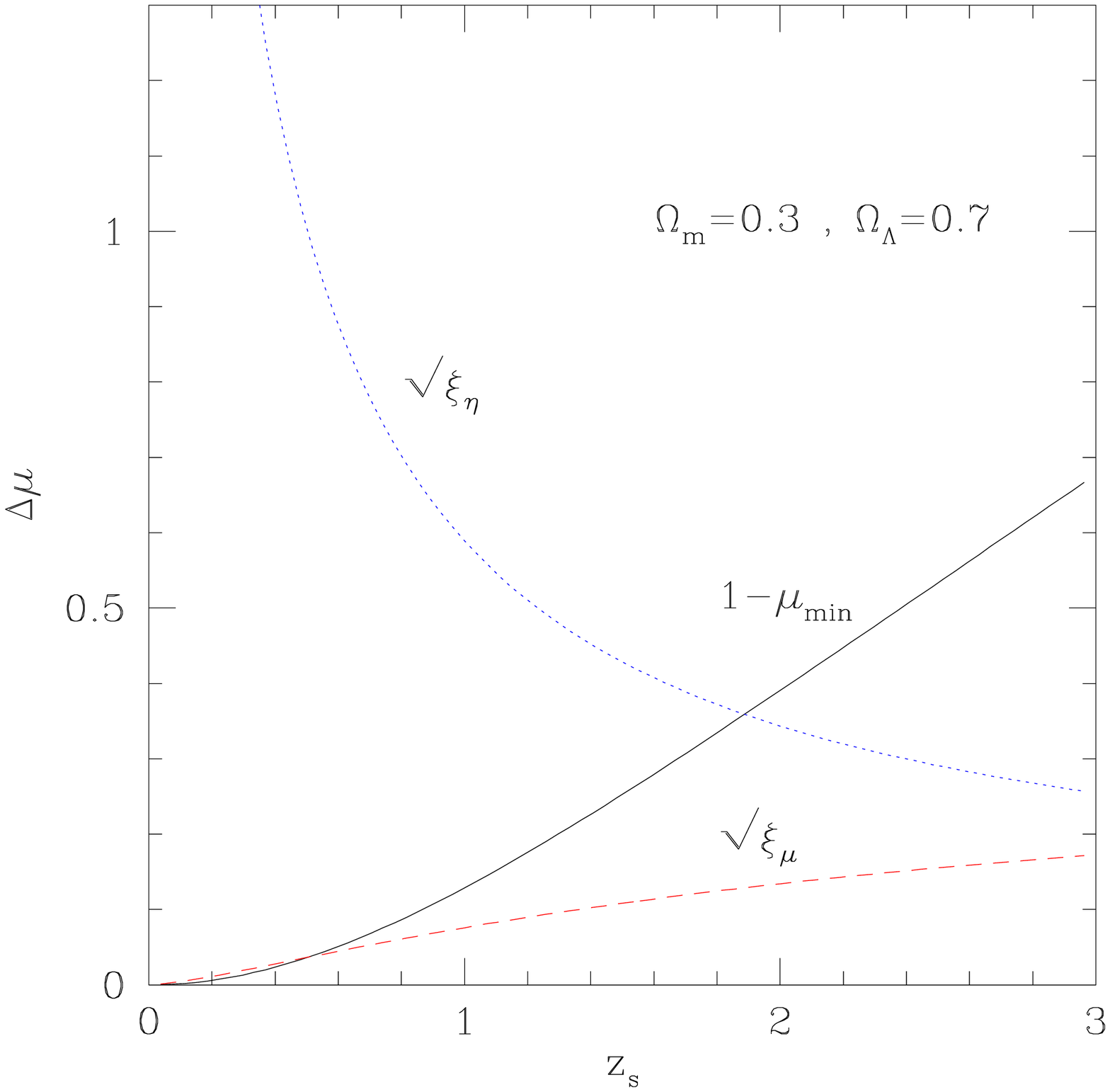} }

\caption{The fluctuations of the magnification $\mu$ for critical, open and low-density flat universes, for a source located at redshift $z_s$. The solid lines show $1-\mumin$ where $\mumin$ given by (\ref{mumin}) is the minimum value of the magnification. The dashed lines present the variance $\surd \ximu$ of the magnification, from (\ref{Ximu}). The dotted lines show the variance $\surd \xieta$ of $\eta$ from (\ref{Ximu}).}

\label{figXi}

\end{figure}

First, we present in Fig.\ref{figXi} the redshift evolution of the amplitude of the fluctuations of the magnification $\mu$ of distant sources located at $z_s$. The increase with $z_s$ of the interval $(1-\mumin)$ between the mean $\lag\mu\rag=1$ and the minimum value $\mumin$ is due to the more extended line of sight which gives more room for the weak lensing effects. This deviation is smaller for the low-density universes than for the critical case because of the factor $\Omega_m$ which enters (\ref{mumin}): the difference of matter between the mean and $0$ is smaller as it is proportional to $\Omega_m$ at low $z$. It is larger for the flat cosmology (with the same $\Omega_m$) because of the detailed dependence on $\Omega_{\Lambda}$ of the factor $F_s$ (see also Bernardeau et al.1997): indeed the distances $\chi$ and $\De$ are larger at fixed $z$ which gives more room for weak lensing effects at a given $z_s$. On the other hand, the variance $\sqrt{\xieta}$ of the ``reduced magnification'' $\eta$ decreases at larger redshift because the integration in (\ref{dmu}) over the successive ``slabs'' of matter leads $P(\mu)$ to be ``closer'' to a gaussian, in a fashion similar to the central limit theorem (although the latter does not apply here since the probability distribution of the magnification due to each slab evolves with $z$ and $z_s$). It diverges for $z_s \rightarrow 0$ where the number of (highly non-linear) density fluctuations which intersect the line of sight declines. Note however that for large $\mu$ the probability distribution $P(\mu)$ is always very different from a gaussian, whatever the value of $\ximu$ and $\xieta$, since it shows a simple exponential cutoff rather than a gaussian cutoff. As a consequence, at low $z_s$ the variance $\sqrt{\ximu}$ of the magnification $\mu$ becomes of the order of, and even larger than, $(1-\mumin)$ (thus $P(\mu)$ is strongly non-gaussian and sharply peaked close to the minimum $\mumin$) while at large redshift $\sqrt{\ximu}$ becomes significantly smaller than $(1-\mumin)$ (thus the peak of $P(\mu)$ gets closer to the mean $\lag\mu\rag=1$). In particular, we can check from (\ref{mumin}) and (\ref{Ximu}) that:
\beq
z_s \rightarrow 0 \; : \;  (1-\mumin) \propto z_s^2 \; , \; \ximu \propto z_s^3 \;\; \mbox{and} \;\; \xieta \propto z_s^{-1}
\label{zs}
\eeq
and:
\beq
z_s \rightarrow \infty \; : \;\;\; \ximu \mbox{ is finite and  } \xieta \propto (1+z_s)^{-2} 
\eeq
We show in Valageas (1999b) that our results agree with the values obtained by Jain et al.(1999) using ray tracing through N-body simulations, for smoothing angles $\theta \ga 0.1'$. In fact, since our prediction for the variance $\ximu$ only relies on the weak lensing approximation and on the fits for the non-linear power-spectrum given by Peacock \& Dodds (1996) this agreement mainly shows that both sets of simulations are consistent. As seen in Fig.1 in Valageas (1999b), the variance $\ximu$ obtained without smoothing ($\theta=0$), which we consider here, is slightly larger than the value reached at $\theta=0.1'$ (since a finite smoothing removes the power from small scales). For instance, for the SCDM case we get $\ximu \simeq 0.1$ for $\theta=0$ (as in Fig.\ref{figXi} here) and $\ximu \simeq 0.08$ for $\theta=0.1'$.

\subsection{Probability distributions}
\label{Pdf}

\begin{figure}

{\epsfxsize=8 cm \epsfysize=5.4 cm \epsfbox{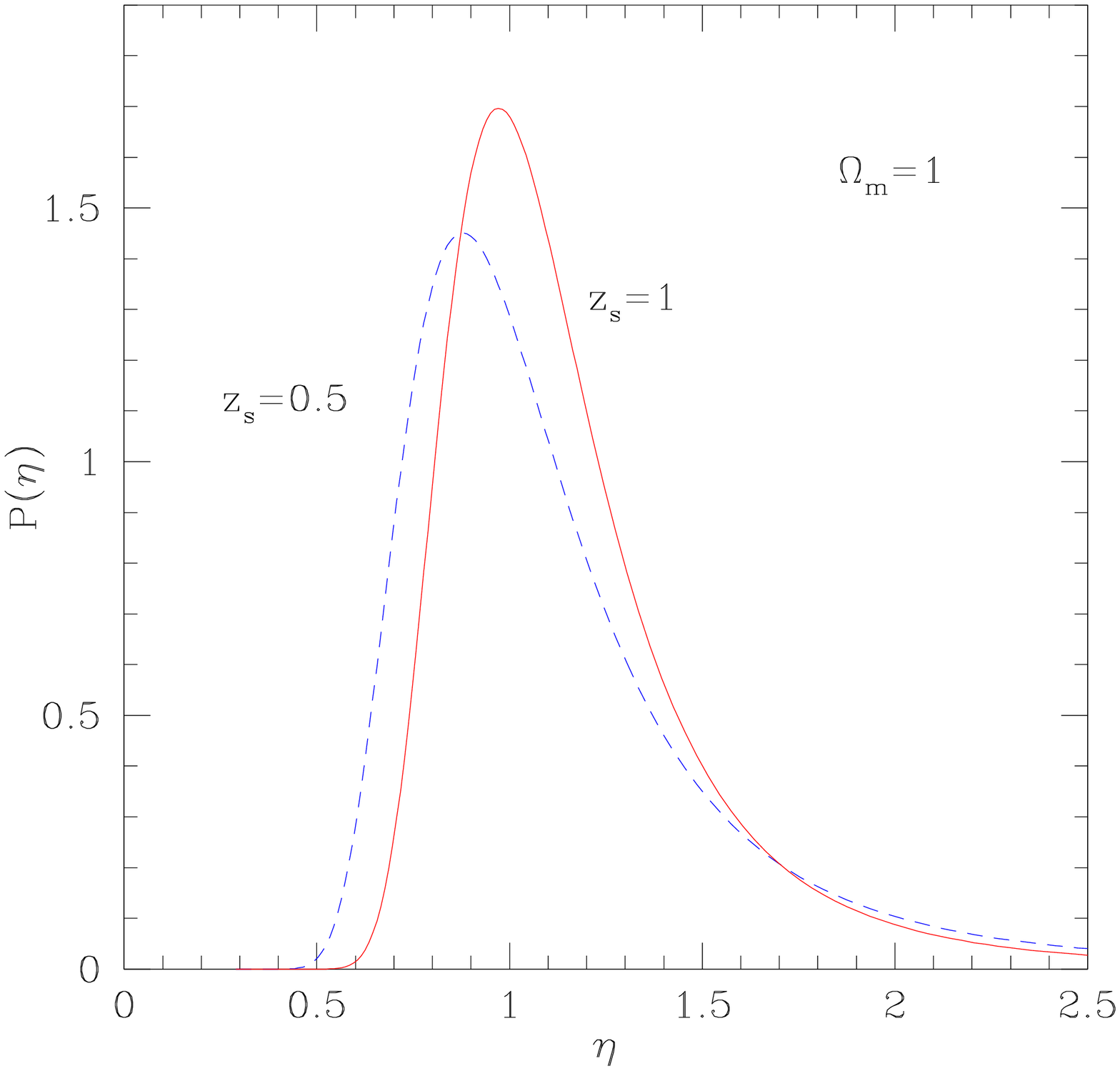} }
{\epsfxsize=8 cm \epsfysize=5.4 cm \epsfbox{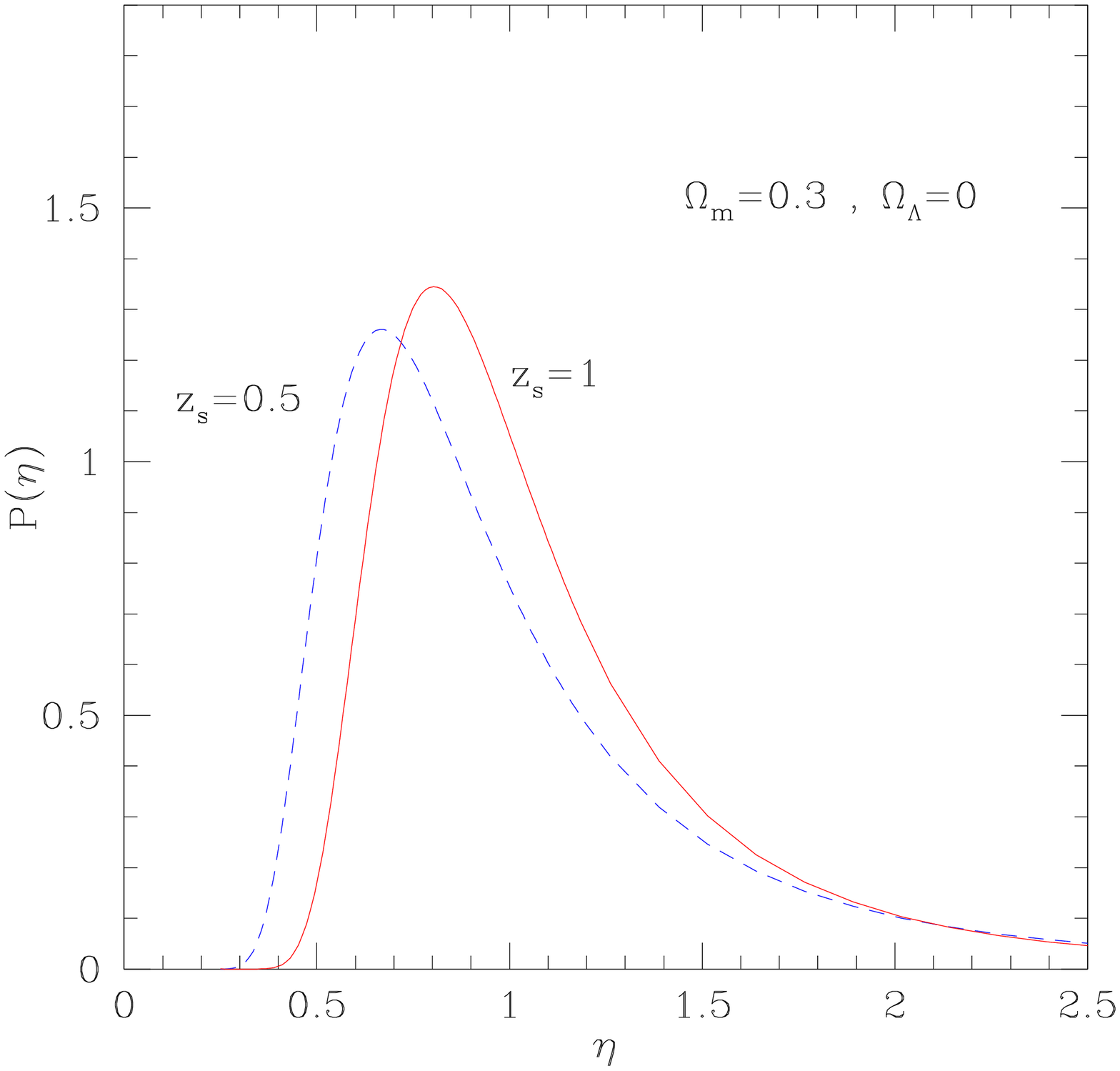} }
{\epsfxsize=8 cm \epsfysize=5.4 cm \epsfbox{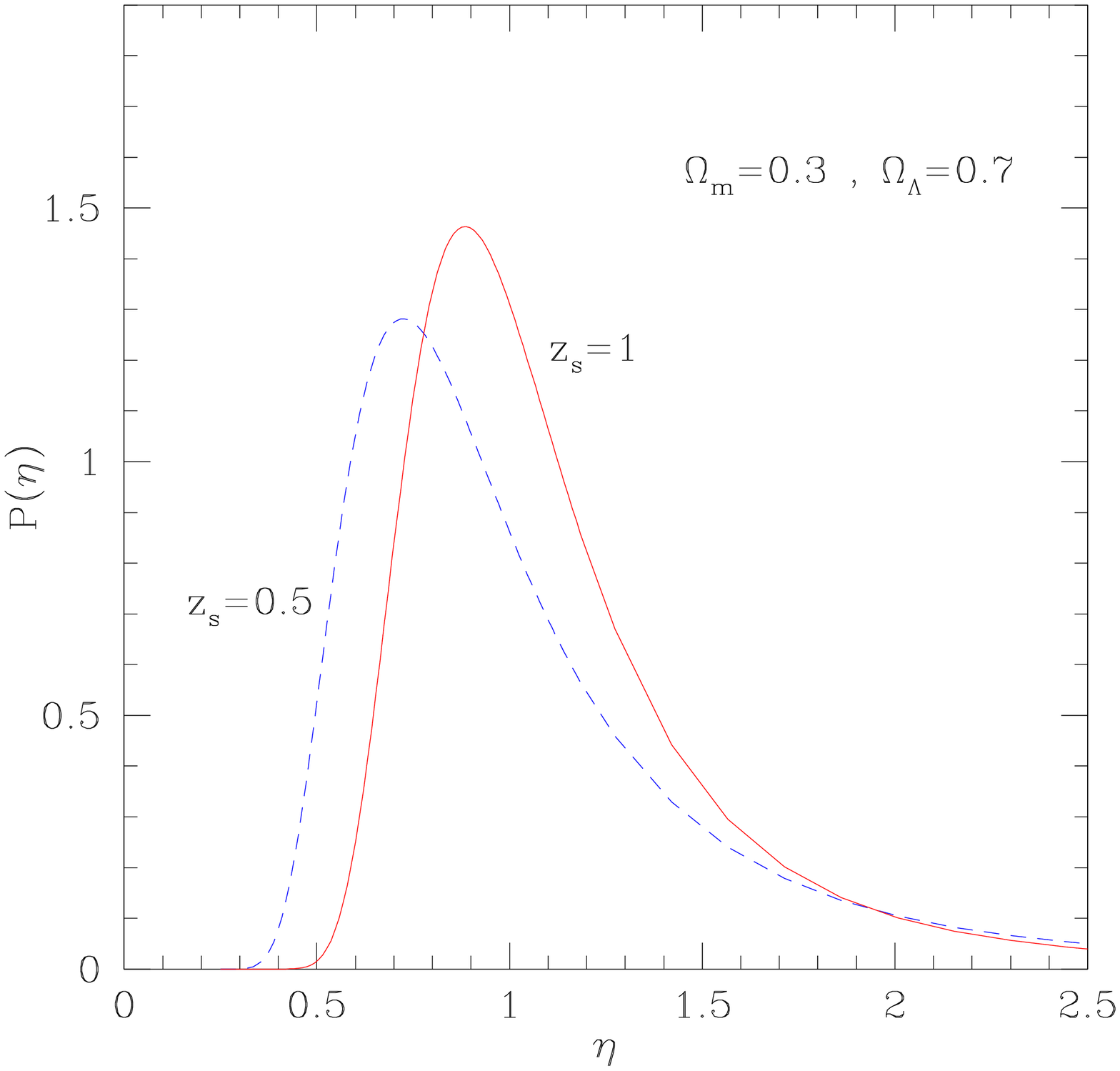} }

\caption{The probability distribution $\tP(\eta)$ of the ``reduced magnification'' $\eta$, from (\ref{Pphietat}). The solid lines correspond to $z_s=1$ and the dashed lines to $z_s=0.5$.}

\label{figPeta}

\end{figure}

We display in Fig.\ref{figPeta} the probability distribution $\tP(\eta)$ of the ``reduced magnification'' $\eta$, from (\ref{Pphietat}). As explained in Sect.\ref{Magnification by weak lensing} it is strongly non-gaussian with a maximum below the mean $\lag\eta\rag=1$ and a power-law tail followed by a simple exponential cutoff at large $\eta$. Its is more strongly peaked around its maximum which is closer to the mean $\lag\eta\rag=1$ for the $\Lambda$CDM model, and even more for the SCDM scenario, following the decrease of $\sqrt{\xieta}$ shown in Fig.\ref{figXi} (but of course this depends on the cosmological parameters one chooses). Note that even for very small variance $\sqrt{\ximu}$ of the magnification ($\sqrt{\ximu} \la 0.1$, see Fig.\ref{figXi}) the probability distribution of the magnification displays clear deviations from gaussianity, as explained above. In particular, the normalized magnification $\eta$ clearly shows the shape of the probability distribution, which appears squeezed towards the mean $\lag\mu\rag=1$ when displayed as a function of $\mu$.

\begin{figure}

{\epsfxsize=8 cm \epsfysize=5.4 cm \epsfbox{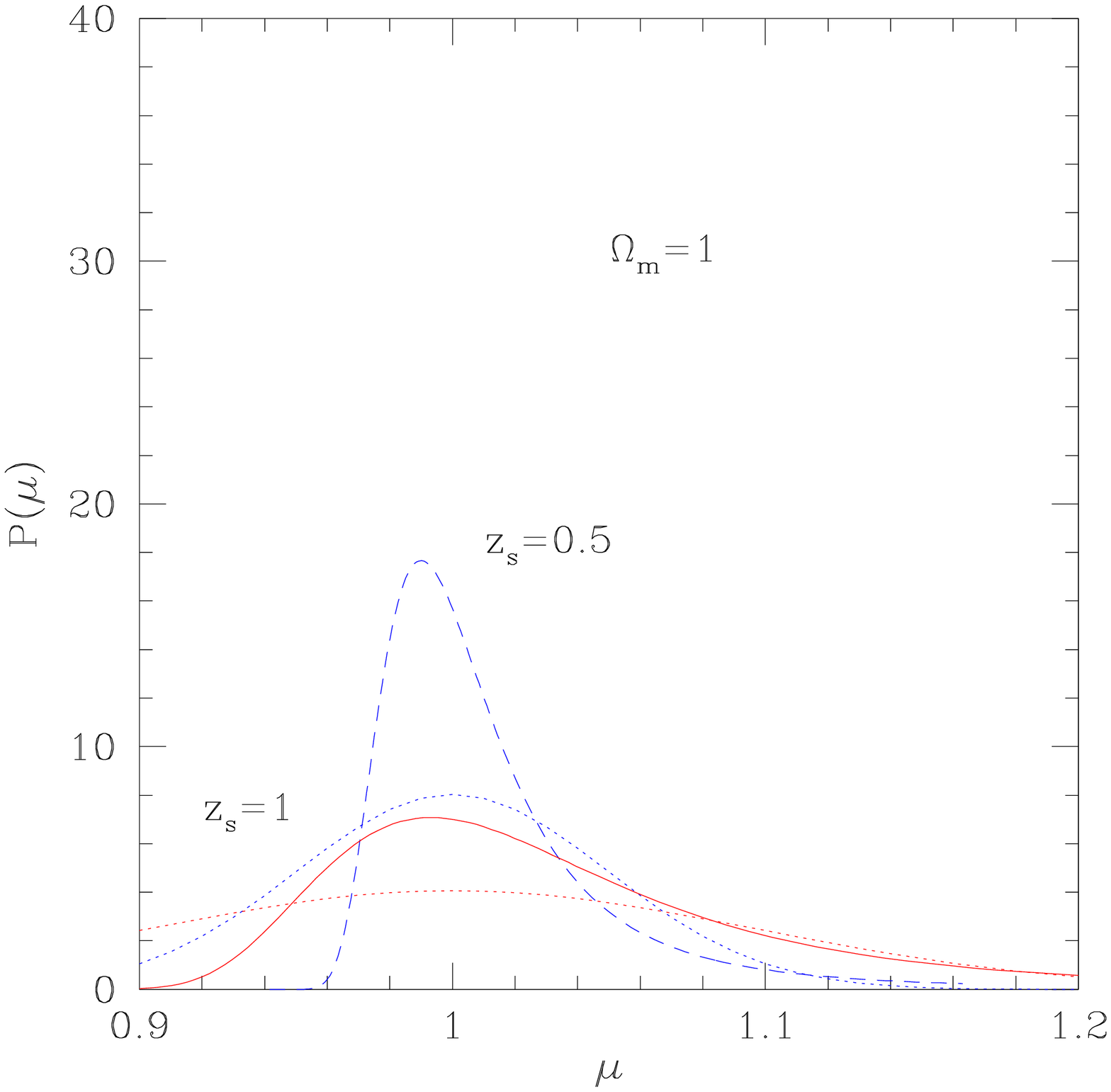} }
{\epsfxsize=8 cm \epsfysize=5.4 cm \epsfbox{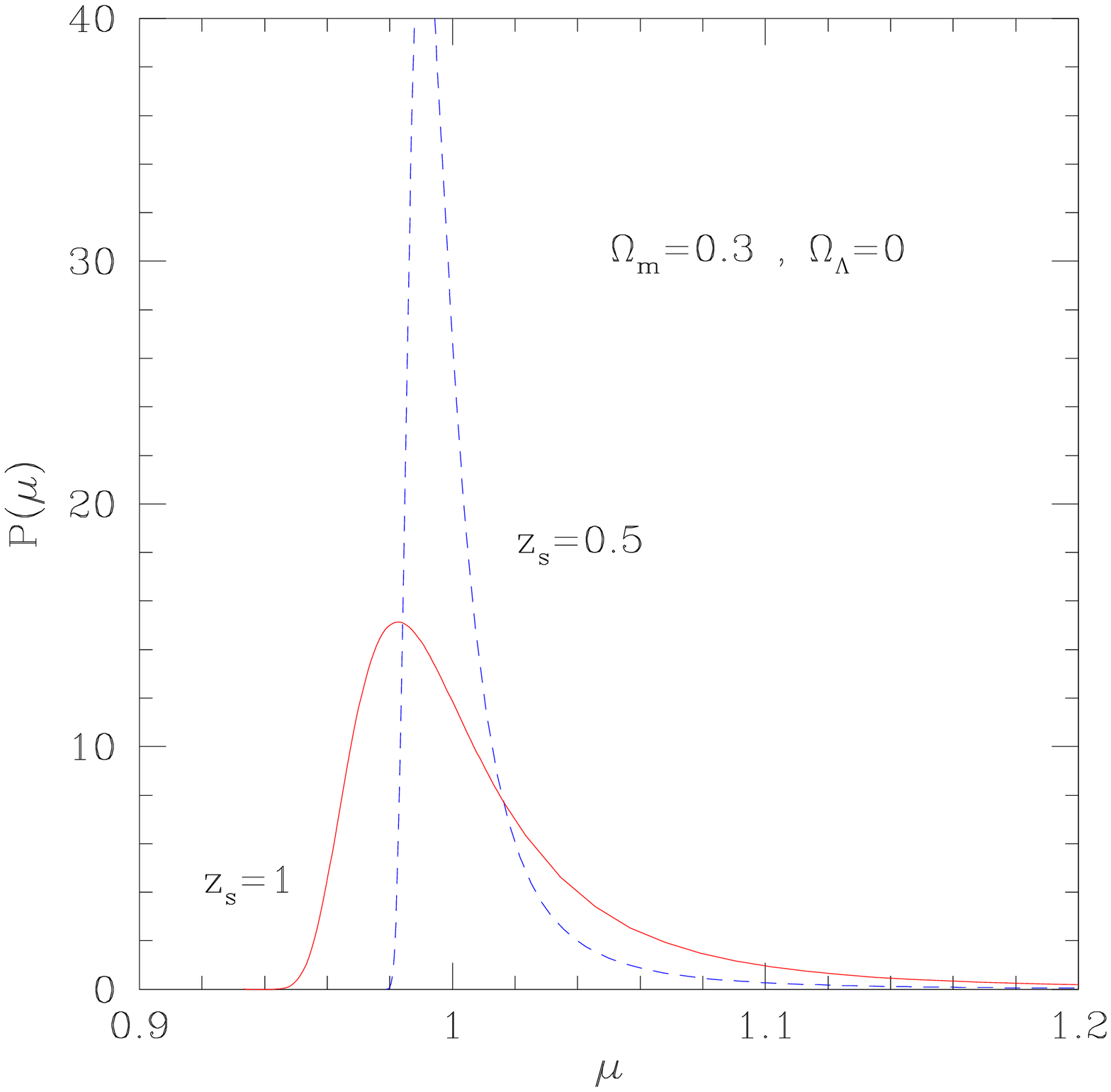} }
{\epsfxsize=8 cm \epsfysize=5.4 cm \epsfbox{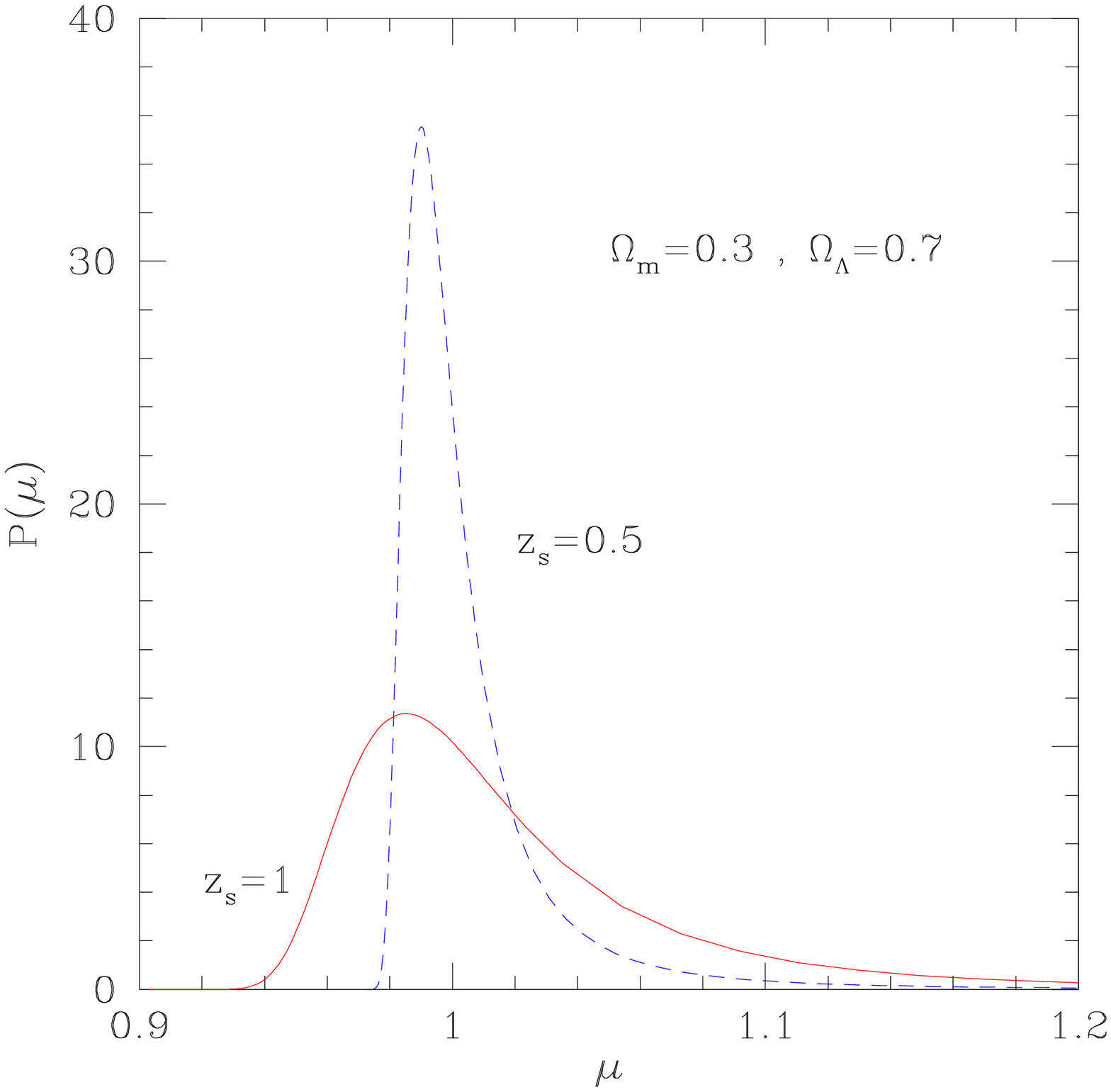} }

\caption{The probability distribution $P(\mu)$ of the magnification $\mu$, from (\ref{Pphietat}) and (\ref{Pphimub}). The solid lines correspond to $z_s=1$ and the dashed lines to $z_s=0.5$. For the critical density universe we also show the gaussians (dotted lines) which correspond to the same variance $\ximu$. These gaussians have a peak at the mean $\lag\mu\rag=1$ and the larger redshift $z_s=1$ corresponds to the larger variance and to the lower height of the maximum.} 

\label{figPmu}

\end{figure}

This is shown in Fig.\ref{figPmu} where one can check that at low redshift $P(\mu)$ tends to a Dirac $\delta_D(\mu-1)$. Due to the smaller value of $(1-\mumin)$ and $\sqrt{\ximu}$ for low density universes $P(\mu)$ is much more sharply peaked around its maximum than for a critical cosmology. One can clearly see the asymmetry due to the lower cutoff at $\mumin$ and the extended large $\mu$ tail, as well as the shift of the maximum of $P(\mu)$ below the mean $\lag\mu\rag=1$. This is similar to the behaviour obtained from numerical simulations (e.g. Wambsganss et al.1997). We display below in Fig.\ref{figlPmu} the curve $\log[P(\mu)]$ which shows even more clearly the non-gaussian features of $P(\mu)$. For the critical density universe we also show in Fig.\ref{figPmu} the gaussians (dotted lines) which correspond to the same variance $\ximu$. We can see that the probability distribution $P(\mu)$ is indeed very different from a gaussian, at both redshifts. Thus, it would be quite meaningless to model $P(\mu)$ as a gaussian. For the sake of completeness, we note here how one can obtain the approximate behaviour of the locations $\etamax$ and $\mumax$ of the peak of the probability distributions $P(\eta)$ and $P(\mu)$. From (\ref{Pphietat}) and (\ref{has}) one can show that for large variance $\xieta$ the location $\etamax$ of the maximum of $P(\eta)$ is given by (Balian \& Schaeffer 1989; Valageas \& Schaeffer 1997):
\beq
\xieta \gg 1 \; : \; \etamax \simeq a^{1/(1-\om)} \; \xieta^{-\om/(1-\om)}
\eeq
which implies:
\beq
\xieta \gg 1 \; : \; \mumax \simeq \mumin + (1-\mumin) a^{1/(1-\om)} \; \xieta^{-\om/(1-\om)}
\eeq
Thus, for large $\xieta$ the peak of the probability distribution of the magnification gets very close to the minima $\eta=0$ and $\mumin$. For smaller values of the variance $\xieta$ one may write an Edgeworth expansion of the probability distribution $P(\eta)$ (e.g. Bernardeau \& Kofman 1995):
\beq
P(\eta) \simeq \frac{1}{\sqrt{2\pi\xieta}} \; e^{-\nu^2/2} \; \left[ 1 + \sigma_{\eta} \frac{S_3}{6} He_3(\nu) \right]
\eeq
where $He_3(\nu)=\nu^3-3 \nu$ is the Hermite polynomial of order $3$, $\nu= (\eta-1)/\sigma_{\eta}$ and $\sigma_{\eta}=\sqrt{\xieta}$. This gives for the location of the peak of the probability distributions:
\beq
\xieta \ll 1 \; : \; \etamax \simeq 1 - \frac{S_3}{2} \xieta
\eeq
and
\beq
\xieta \ll 1 \; : \; \mumax \simeq 1 - \frac{S_3}{2} (1-\mumin) \xieta
\eeq
Thus, for both large and small redshift, that is for small and large $\xieta$, the peak $\mumax$ tends to the mean $1$ (but for different reasons). Hence the deviation $(1-\mumax)$ is maximum for an intermediate redshift of order unity.

\begin{figure}

{\epsfxsize=8 cm \epsfysize=5.4 cm \epsfbox{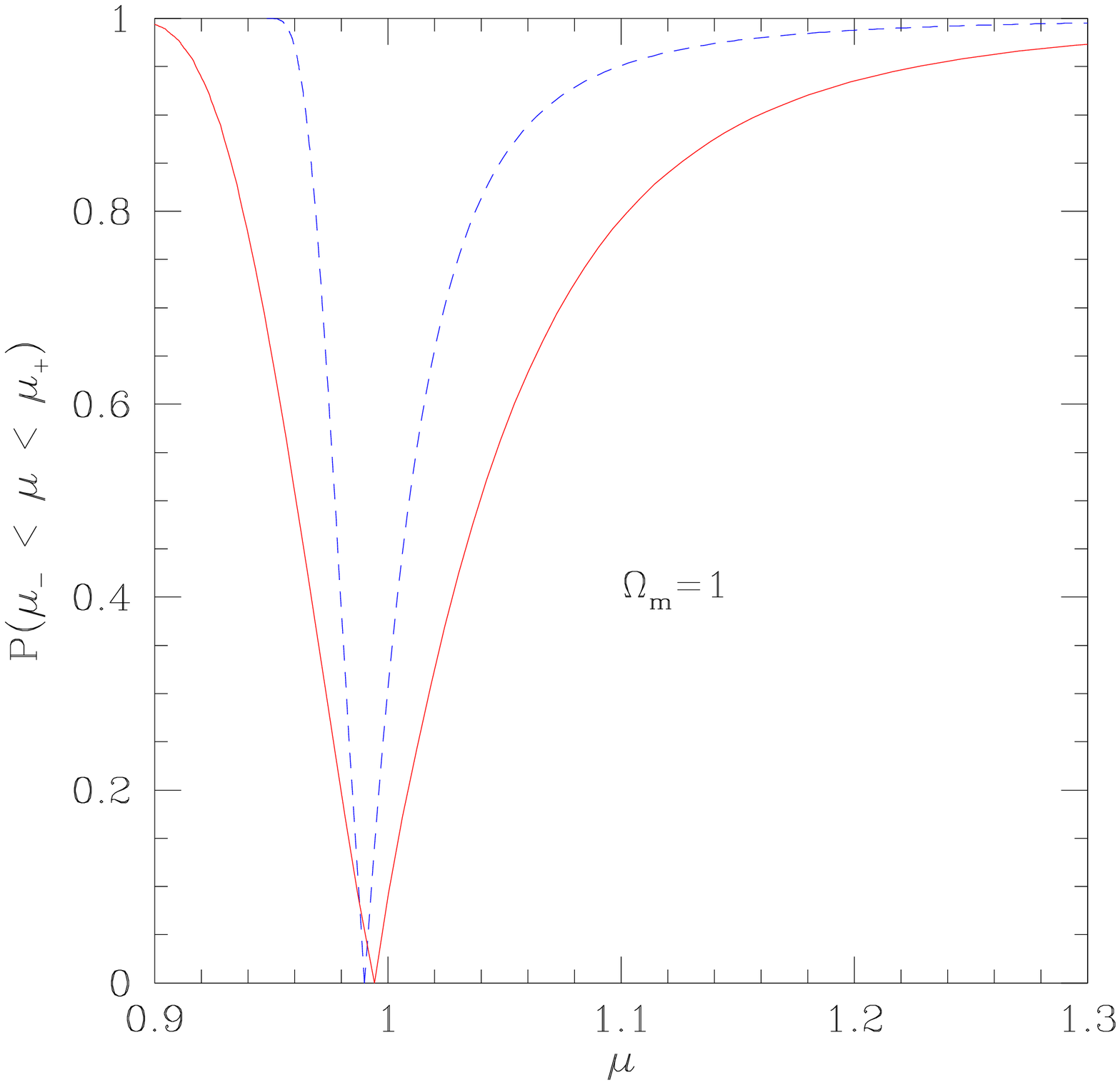} }
{\epsfxsize=8 cm \epsfysize=5.4 cm \epsfbox{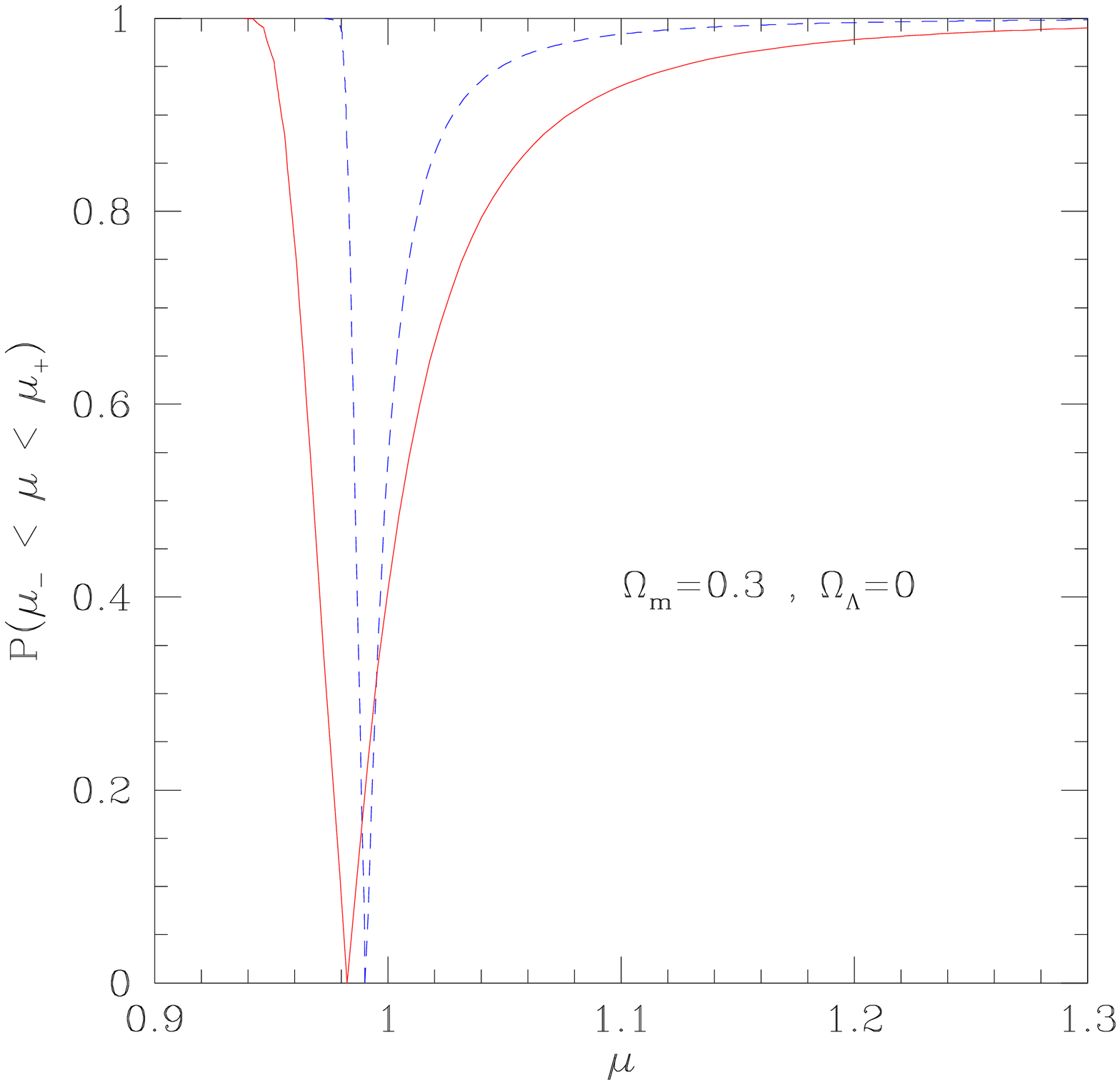} }
{\epsfxsize=8 cm \epsfysize=5.4 cm \epsfbox{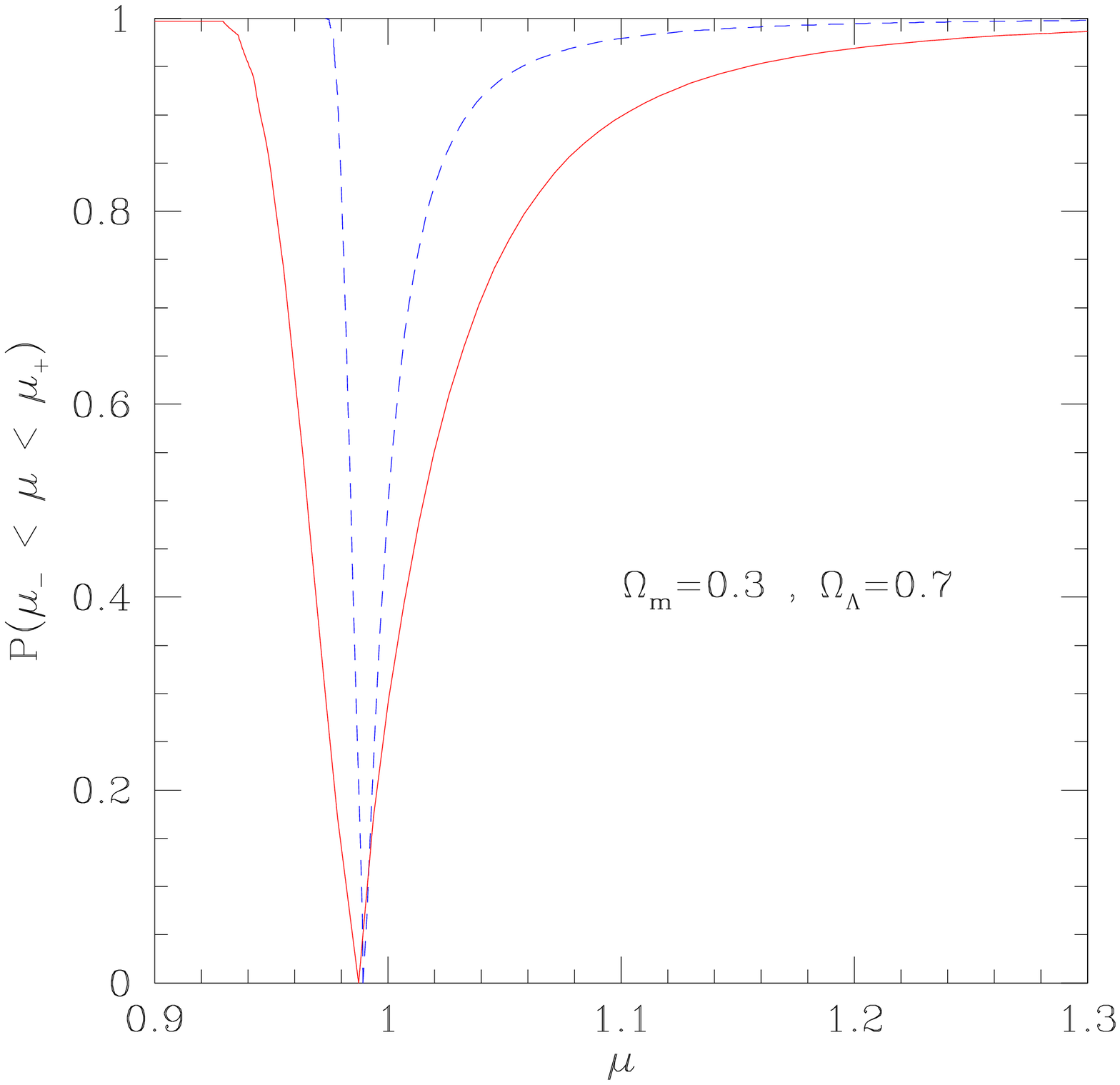} }

\caption{Intervals of confidence for the magnification $\mu$. The curves show the probability $P(\mu_{-}<\mu<\mu_{+})$ that the magnification $\mu$ is within the minimal interval $[\mu_{-},\mu_{+}]$. The solid lines correspond to $z_s=1$ and the dashed lines to $z_s=0.5$.}

\label{figPsig}

\end{figure}

From the probability distribution $P(\mu)$ one can obtain intervals of confidence for the magnification $\mu$. Thus, for $0 \leq \cP \leq 1$ we define $[\mu_{-},\mu_{+}]$ as the minimal interval (i.e. with the smallest length) such that $\mu \in [\mu_{-},\mu_{+}]$ with probability $\cP$. These intervals contain the location $\mumax$ of the peak of $P(\mu)$ and obey:
\beq
P(\mu_{-}) = P(\mu_{+}) \hspace{0.5cm} , \hspace{0.5cm} P(\mu_{-}<\mu<\mu_{+}) = \cP
\label{Psig}
\eeq
They are displayed in Fig.\ref{figPsig} for the redshifts $z_s=0.5$ and $z_s=1$: any horizontal line of ordinate $\cP$ intersects the curves at the points $\mu_{-}$ and $\mu_{+}$. For $\cP \rightarrow 0$ the length of the interval goes to 0 as $\mu_{-}$ and $\mu_{+}$ tend to $\mumax$. For $\cP=1$ we have $\mu_{-}=\mumin$ and $\mu_{+}=\infty$. Thus, Fig.\ref{figPsig} clearly shows the range of $\mu$ one can expect, as well as the asymmetry of the underlying probability distribution. In particular, the large $\mu$ tail of $P(\mu)$ is clearly seen. Moreover, one can check that there is a non-negligible probability to have $\mu<1$ on a line-of-sight ($\mu_{+}<1$).

\subsection{Skewness}
\label{Skewness}

\begin{figure}

{\epsfxsize=8 cm \epsfysize=5.4 cm \epsfbox{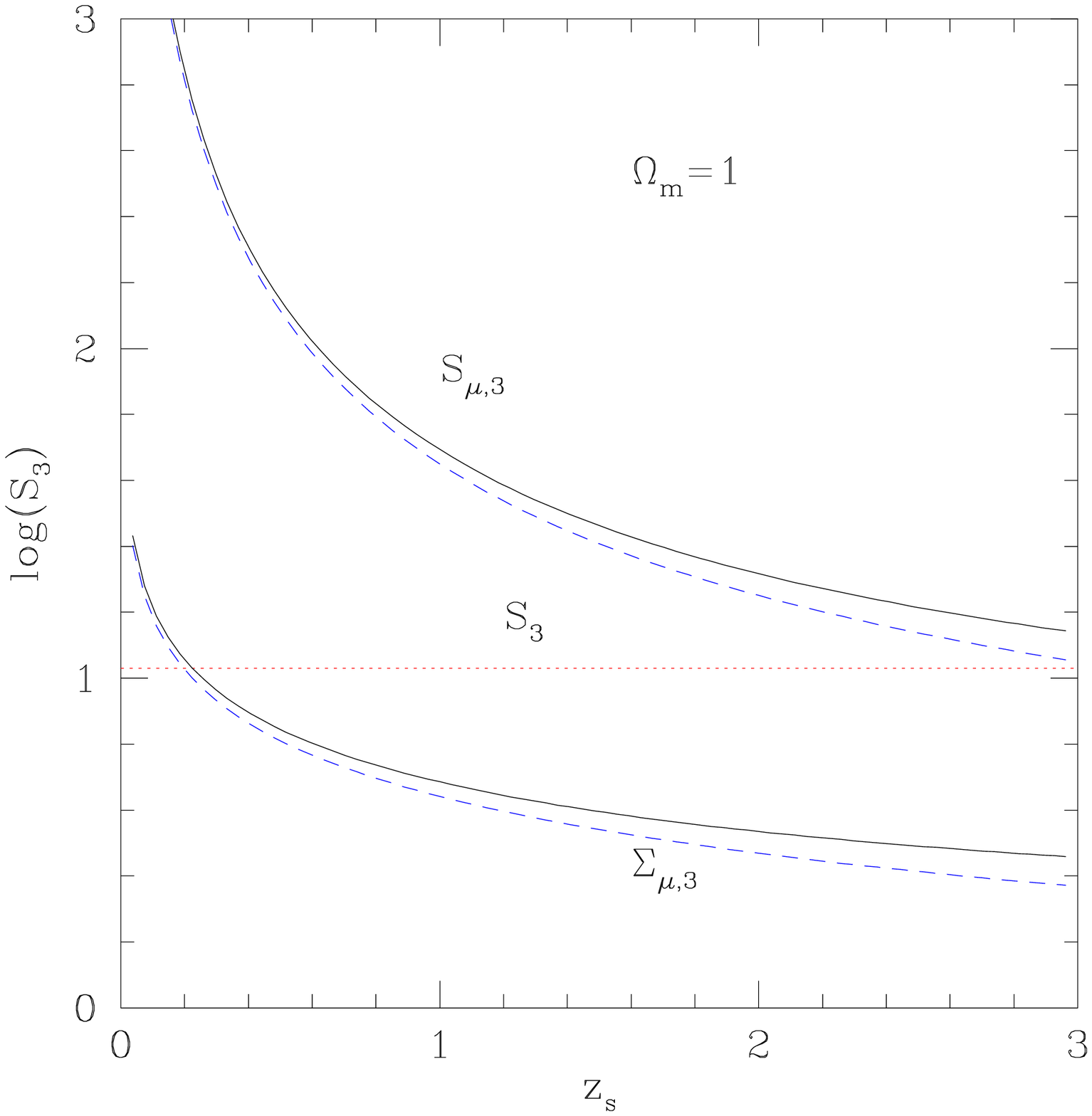} }
{\epsfxsize=8 cm \epsfysize=5.4 cm \epsfbox{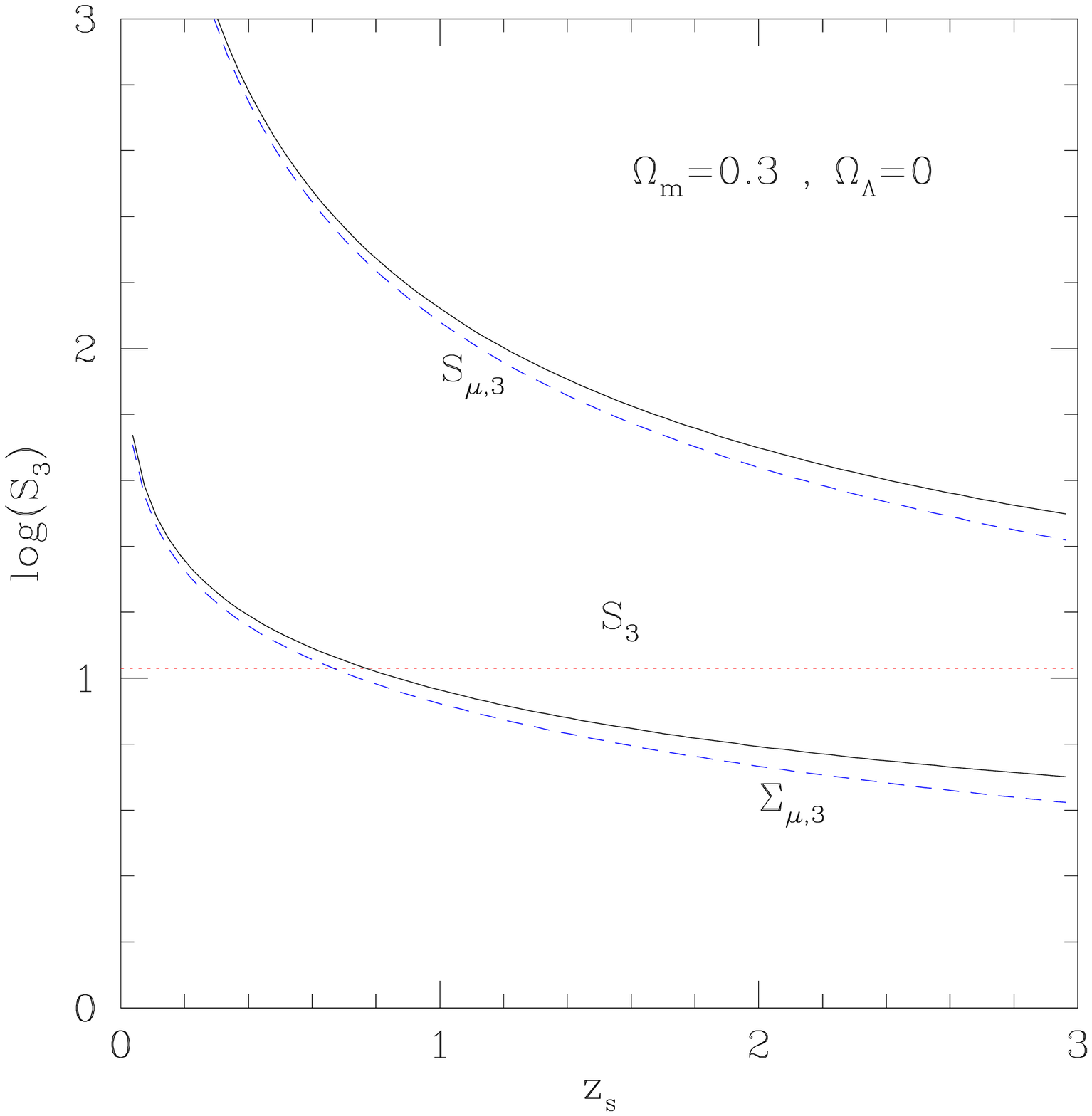} }
{\epsfxsize=8 cm \epsfysize=5.4 cm \epsfbox{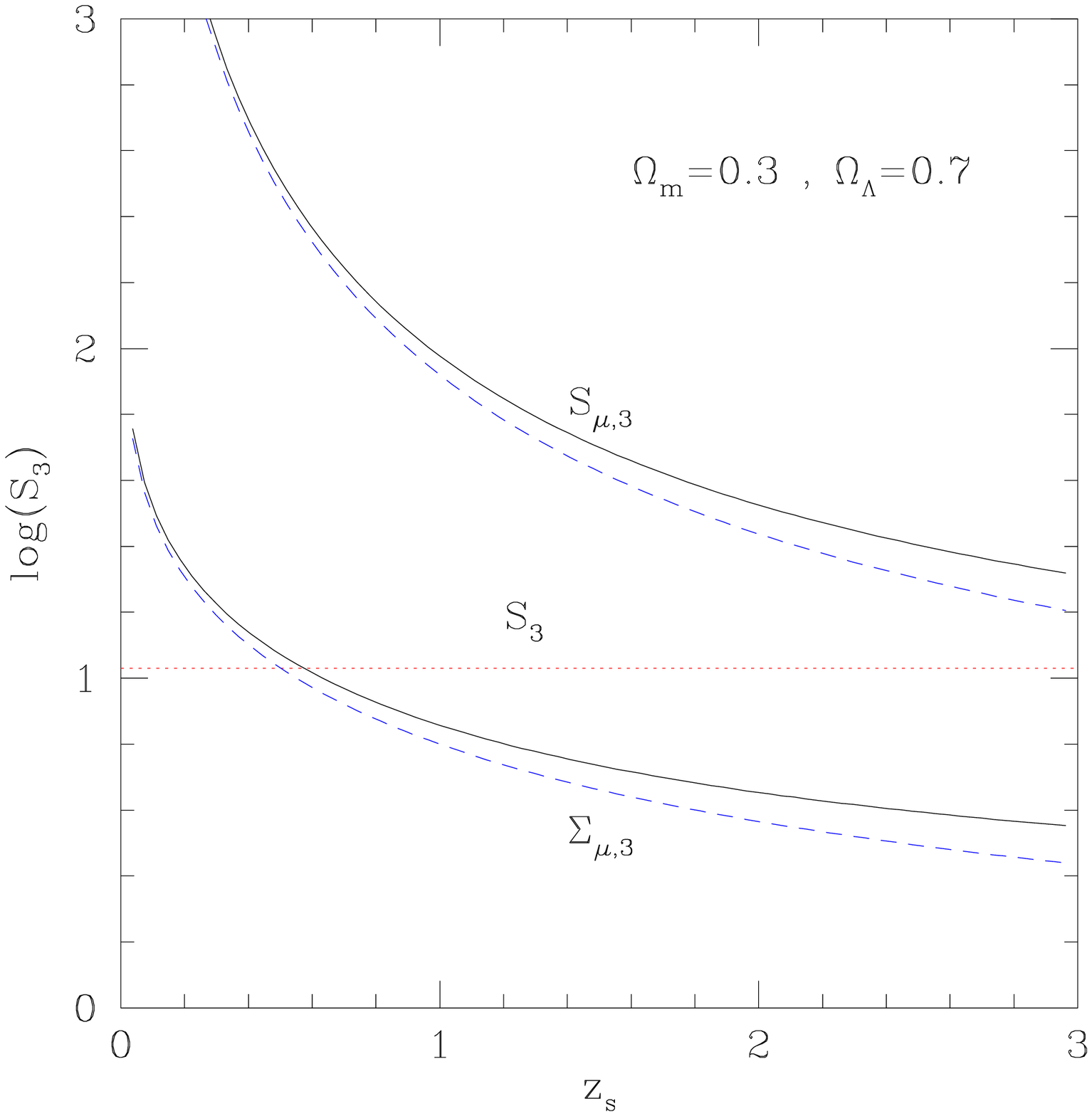} }

\caption{The parameters which define the third moment of the probability distribution for the magnification $\mu$ ($S_{\mu,3}$ and $\Sigma_{\mu,3}$) and the ``reduced magnification'' $\eta$ or the density contrast ($S_3$). The solid lines show the value of $\log(S_{\mu,3})$ and $\log(\Sigma_{\mu,3})$ obtained from (\ref{Smuex}) while the dashed curves correspond to the approximation (\ref{Smuap}). At larger redshift $P(\mu)$ gets closer to a gaussian so that $S_{\mu,3}$ and $\Sigma_{\mu,3}$ decrease while $S_3$ is independent of $z_s$.}

\label{figS3z}

\end{figure}

Finally, in Fig.\ref{figS3z} we display the third moment of the probability distribution of the magnification. From (\ref{Sp}) the parameter $S_3$ (skewness) is given by:
\beq
S_3 =  \frac{\xia_3}{\xia^{\;2}} = \frac{\lag\delta_R^{\;3}\rag}{\lag\delta_R^{\;2}\rag^2}
\eeq
since $\lag\delta_R^{\;3}\rag=\lag\delta_R^{\;3}\rag_c$. It measures the third moment of the probability distribution of the density contrast $\delta_R$ realized in spherical cells of radius $R$. As explained in Sect.\ref{Density contrast probability distribution} it is constant with time in the non-linear regime at the scale $R_c(z)$ such that the local slope of the linear power-spectrum is $n=-2$. From the results obtained by Valageas et al.(1999) from numerical simulations we have $S_3 \simeq 10.7$. From (\ref{Pphietat}), $S_3$ also measures the third order moment of $\eta$:
\beq
\frac{\lag\delta \eta^{\;3}\rag}{\xieta^2} \simeq S_3
\eeq
From the change of variable (\ref{eta}) we also  obtain:
\beq
\begin{array}{ll} {\displaystyle  S_{\mu,p} \equiv \frac{\lag\dmu^{\;p}\rag_c}{\lag\ximu\rag^{p-1}} } & {\displaystyle = (3\Omega_m F_s)^{2-p} S_{\eta,p} } \\ \\ & {\displaystyle \simeq (3\Omega_m F_s)^{2-p} S_p } \end{array}
\label{Smuap}
\eeq
This clearly shows the dependence on cosmology of the coefficients $S_{\mu,p}$. Moreover, from (\ref{dmup2}) we can also derive the parameters $S_{\mu,p}$ without the approximation (\ref{phietat}). This leads to:
\beq
S_{\mu,p} = (3\Omega_m F_s)^{2-p} \; S_p \; \int_0^{\chi_s} d\chi \; \left( \frac{w}{F_s} \right)^p \; \left( \frac{\Imu}{\xieta} \right)^{p-1} 
\label{Smuex}
\eeq
Finally, we also define the parameter $\Sigma_{\mu,3}$ by:
\beq
\Sigma_{\mu,3} = \frac{\lag\dmu^{\;3}\rag}{\lag\ximu\rag^{3/2}} = S_{\mu,3} \;\; \ximu^{1/2}
\eeq
which may be seen as a more convenient measure of the deviation of $P(\mu)$ from a gaussian. We can check in Fig.\ref{figS3z} that the error introduced by the approximation (\ref{phietat}) is quite small. In particular, it is negligible as compared to the inaccuracy due to the results of numerical simulations of structure formation. Indeed, while Valageas et al.(1999) get $S_3 \simeq 10.7$, Colombi et al.(1997) obtain $S_3 \simeq 10.2$ and Munshi et al.(1999) find $S_3 \simeq 6$. Thus, the approximation (\ref{phietat}) is quite sufficient in view of the accuracy of the scaling function $h(x)$ obtained from numerical results. However, one should note that the functions $h(x)$ measured in simulations are reasonably close in the range where they have been tested against numerical data. Indeed, the uncertainty which affects the parameters $S_p$ (and increasingly so for large $p$) comes from the large density tail of the probability distribution of the density contrast while the behaviour of $x^2 h(x)$ around its maximum (i.e. at $x \sim 1$) is fairly well constrained, see Valageas et al.(1999) for a detailed discussion. Note that the measure of $S_{\mu,p}$ from ray-tracing in N-body simulations would of course bear the same inaccuracy, which is due to the dispersion of the properties of the non-linear density field itself obtained from different numerical simulations. 

However, we can note that at $z_s=1$ we get $S_{\mu,3} = 49$, $131$ and $95$ for the critical density, open and low-density flat universes, while Jain et al.(1999) obtain $S_{\mu,3} = 43$, $107$ and $75$ (we have $S_{\mu,3} = S_{\kappa,3} /2$). For the critical density universe both values agree quite well, while for the low-density universes our values are somewhat larger than those obtained by these authors (although they show the same trends). This could be due to the slow rise of the skewness with smaller smoothing angle (due to the fact that $S_3$ is larger for non-linear scales than for quasi-linear scales, see Colombi et al.1997). Indeed, as seen in Jain et al.(1999) the skewness may not have reached its asymptotic value at $\theta = 1'$ yet. Moreover, as discussed in Valageas (1999b) the errorbars on the measure of $S_{\mu,3}$ from numerical simulations may be larger than the dispersion of the estimator used to compute the skewness may suggests because two {\it p.d.f.} with a rather different skewness can still agree very well, as shown by the good agreement of our prediction for $P(\kappa)$ with the results from these N-body simulations. In particular, although this comparison may suggest a small dependence of $S_3$ on $\Omega_m$ the dispersion of numerical results (which provide values for $S_3$ which can vary by a factor $1.8$ as seen above) prevents us from drawing definite conclusions. Of course, it would be interesting to measure both $S_3$ and $S_{\mu,3}$ in the same numerical simulation to directly check the accuracy of our relation (\ref{Smuex}). On the other hand, the main improvement to our calculation would be to directly measure (or obtain from first principles) the parameters $S_p$ defined by the integrals on the line-of-sight of the many-body correlation functions, i.e. the {\it l.h.s.} terms in (\ref{Sp2}), rather than the ratios of the averages defined in (\ref{Sp}). Then one could still apply our method and simply use the new function $\varphi(y)$ (or $h(x)$) defined by these new parameters $S_p$ as in (\ref{phiy}).

As explained above, we can check in Fig.\ref{figS3z} that at larger redshift the probability distribution of the magnification becomes ``closer'' to a gaussian as the parameters $S_{\mu,3}$ and $\Sigma_{\mu,3}$ decrease. However, it is interesting to note that these parameters can be fairly large (higher than $S_3$) and even diverge for $z \rightarrow 0$. Thus, at low $z$ the probability distribution of the magnification is strongly non-gaussian. In particular, we obtain:
\beq
z_s \rightarrow 0 \; : \; S_{\mu,3} \propto z_s^{-2} \;\; \mbox{and} \;\; \Sigma_{\mu,3} \propto z_s^{-1/2}
\eeq
and
\beq
z_s \rightarrow \infty \; : \; S_{\mu,3} \propto (1+z_s)^{-1} \;\; \mbox{and} \;\; \Sigma_{\mu,3} \propto (1+z_s)^{-1}
\eeq
Thus, if the intrinsic magnitude dispersion of the sources is sufficiently small, one might observe these non-gaussian features and check that they agree with the usual models of the density field. Moreover, from (\ref{Smuex}) we see that one could get an estimate of the properties of the underlying density field itself (e.g. its skewness $S_3$) from $P(\mu)$. However, as seen in Fig.\ref{figPmu} the width of the probability distribution $P(\mu)$ is quite small at low $z$ which would make such a study rather difficult, so that intermediate redshifts $z_s \sim 1$ may provide better results.

\subsection{Influence of cosmological parameters}
\label{Influence of cosmological parameters}

\begin{figure}

{\epsfxsize=8 cm \epsfysize=10 cm \epsfbox{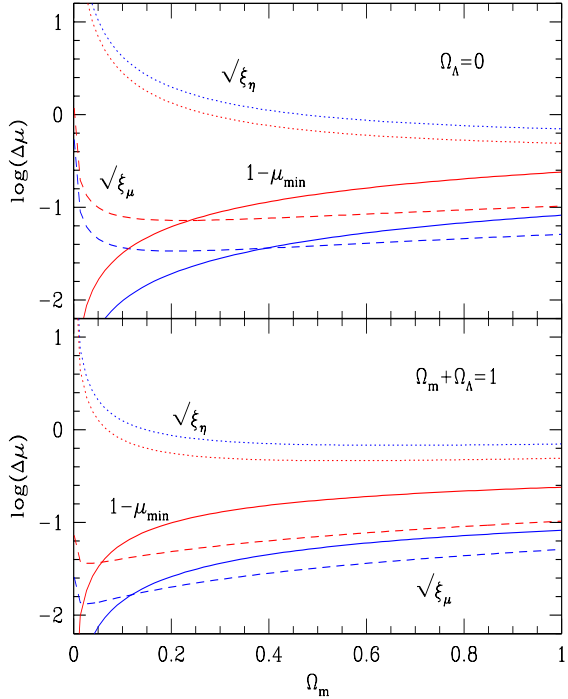} }

\caption{The dependence on $\Omega_m$ and $\Omega_{\Lambda}$ of the fluctuations of the magnification $\mu$ of a source located at redshift $z_s=0.5$ and $z_s=1$. The solid lines show $\log(1-\mumin)$, the dashed lines present the variance $\log(\surd \ximu)$ of the magnification and the dotted lines show the variance $\log(\surd \xieta)$ of $\eta$. The higher redshift corresponds to smaller $\xieta$ and larger $\ximu$ and $(1-\mumin)$.}

\label{figXiopfl}

\end{figure}

We show in Fig.\ref{figXiopfl} the dependence on $\Omega_m$ and $\Omega_{\Lambda}$ of the fluctuations of the magnification $\mu$ of a source located at redshift $z_s=0.5$ and $z_s=1$. We vary the normalization $\sigma_8$ of the power-spectrum with $\Omega_m$ as $\sigma_8 \propto \Omega_m^{-1/2}$ which roughly accounts for the change of $\sigma_8$ needed in order to reproduce as well as possible large-scale structure observations (abundance of clusters, velocity fields) with different cosmologies (of course, some cosmologies match such observations better than others, whatever their choice of $\sigma_8$). Note that the shape of $P(k)$ also depends on $\Omega_m$ through $\Gamma = \Omega_m h$. We can see in Fig.\ref{figXiopfl} that $(1-\mumin)$ increases for larger $\Omega_m$. This is due to the factor $\Omega_m$ which appears in (\ref{mumin}). It translates the fact that for higher $\Omega_m$ there is more matter in the universe hence there is more room for deviations from the mean $\lag\mu\rag=1$, for instance in the case where $\delta=-1$ everywhere along the line of sight (``empty beam'') which leads to $\mumin$. In particular, we have:
\beq
\Omega_m \rightarrow 0 \; : \; (1-\mumin) \propto \Omega_m
\eeq
for both flat and open universes, since for an empty universe all beams are empty. The variances $\sqrt{\xieta}$ and $\sqrt{\ximu}$ increase for $\Omega_m \rightarrow 0$ because of the rise of the two-point correlation function which compensates the slower growth of the linear growth factor, see Peacock \& Dodds (1996). The amplitude of the fluctuations of the density contrast, hence of the magnification $\mu$, is smaller for a flat universe than for the open case with the same $\Omega_m$ because of the detailed form of this linear growth factor. Overall, the variation of the quantities $(1-\mumin)$, $\sqrt{\xieta}$ and $\sqrt{\ximu}$ is rather small over the usual range of cosmological parameters $0.2 \leq \Omega_m \leq 1$. Note that the moments of the probability distribution $P(\mu)$ also depend on the cosmological parameters $\Omega_m$ and $\Omega_{\Lambda}$ (mainly on $\Omega_m$), as can be seen from (\ref{Smuap}). In particular, we have:
\beq
S_{\mu,3} \simeq \frac{S_3}{1-\mumin}
\eeq
so that the variation of $S_{\mu,3}$ can be directly seen in Fig.\ref{figXiopfl} from the evolution of $(1-\mumin)$.

\section{Galactic halos}
\label{Galactic halos}

In the previous sections, we have obtained the probability distribution of the magnification due to weak lensing by assuming the density field follows a specific scaling model described by (\ref{scal1}), which is consistent with numerical results (Valageas et al.1999; Munshi et al.1999; Colombi et al.1997). More precisely, although the parameters $S_p$ and the function $h(x)$ may show some slow scale-dependence through the variation of the local power-spectrum index $n$ they should not explicitely depend on scale (i.e. on $\sigma^2$) so that the approximation (\ref{Sp2}) is valid. In turn, such a picture of the non-linear density field implies that non-linear mass condensations can be described as an infinite hierarchy of increasingly small and overdense substructures (which follow the rise at small scales of $\xia$), as discussed in Valageas et al.(1999) and Valageas (1999a). On the other hand, it is customary to model collapsed objects as smooth halos with a density profile which follows a power-law or a curved profile as those obtained by Navarro et al.(1996) or Hernquist (1990). Moreover, within galactic cores where baryonic matter plays a dominant role one may expect a smooth density profile. Hence in this section we consider the case where collapsed halos are described by a power-law density profile:
\beq
\rho(r) = (1+\Delta_c) \; \rhoa \; \left(\frac{r}{R}\right)^{-\beta}
\label{beta}
\eeq
which defines the slope $\beta$ (with $0<\beta<3$). The factor $(1+\Delta_c)$ is the overdensity at the virial radius $R$ of the object. Recently a similar description was used by Porciani \& Madau (1999), with $\beta=2$ (isothermal sphere), with the Press-Schechter mass function (Press \& Schechter 1974) of just-virialized objects. Here we intend to compare the results obtained from such a model to those we obtained in the previous sections.

First, we note that such a description of the density field leads to a power-law tail at high $\mu$ for the probability distribution of the magnification. Indeed, as shown in Valageas (1999a) the $p-$point correlation functions implied by (\ref{beta}) follow in the highly non-linear regime the power-law behaviour:
\beq
p \geq 2 \; : \; \xia_p(R) \propto R^{-\gam_p} \hspace{0.4cm} \mbox{with} \hspace{0.4cm} \gam_p = p \beta -3
\eeq
From (\ref{dmupc3}) we see that:
\beq
\lag\dmu^{\;p}\rag_c \sim \lag \delta \eta^{\;p}\rag_c \sim R^{p-1} \; \xia_p \sim R^{-p(\beta-1)+2}
\eeq
where $R$ is the upper or lower cutoff of the integrals in (\ref{dmupc3}). Thus, if $\beta>1$ and there is no lower cutoff (which would correspond to the radius of galactic cores where $\beta < 1$) the cumulants $\lag\dmu^{\;p}\rag_c$ and the moments $\lag\mu^{\;p}\rag$ diverge for $p>2/(\beta-1)$. This is inconsistent with an exponential cutoff for $P(\mu)$ while it is the natural outcome of a power-law tail given by:
\beq
\mbox{large } \mu \; : \; P(\mu) \sim \mu^{-1-2/(\beta-1)}
\label{tail1}
\eeq
We also note from the definition (\ref{Sp}) that the description (\ref{beta}) means that the parameters $S_p$ are scale-dependent:
\beq
S_p \propto R^{-(p-2)(3-\beta)}
\eeq
For reasonable values of $\beta$ ($\beta \leq 2$) this implies a strong growth of the coefficients $S_p$ ($p \geq 3$) at small scales which contradicts numerical results as discussed in Valageas (1999a). Note on the other hand that $\beta<2$ leads to a small slope $\gam<1$ for the two-point correlation function. Thus, we think this is not a good model for the dark matter density field. However, we briefly present below the results obtained by such a description as it may be valid for baryonic cores of collapsed objects and it allows us to compare our previous results with other approaches. We restrict ourselves to $1<\beta<3$.

Following the method used in Sect.\ref{Magnification by weak lensing}, in order to derive the probability distribution $P(\mu)$ we first obtain the cumulants $\lag\dmu^{\;p}\rag_c$. We divide the line-of-sight from the observer to the redshift $z_s$ of the source into small physical elements $\Delta l_i$ so that they contain at most one halo and from (\ref{dmu}) we write for each realization the flux perturbation as:
\beq
\hdmu = \sum_i \sum_{\alpha} \; \hn_{i,\alpha} \; \dmu_{i,\alpha}
\eeq
where $\hn_{i,\alpha}=1$ (resp. $\hn_{i,\alpha}=0$) if there is (resp. there is not) a halo of type $\alpha$ within the length element $i$ while $\dmu_{i,\alpha}$ is the contribution to the magnification of the source by a halo $\alpha$ at redshift $z_i$. From (\ref{dmu}) we write:
\beq
\dmu_{i,\alpha} = 3\Omega_m \; (1+z) w(\chi,\chi_s) \; \Ib (1+\Delta_c) \; b \left(\frac{b}{R}\right)^{-\beta}
\eeq
where:
\beq
\Ib = 2 \; \int_0^{\infty} \frac{du}{(1+u^2)^{\beta/2}}
\eeq
comes from the integration along the line-of-sight through the halo. Here we made the approximation that the impact parameter $b$ of the line-of-sight is much smaller than the radius $R$ of the halo. The factor $(1+z)$ comes from the fact that here $b$ and $R$ (and $l$ below) are physical scales while $\chi$ is a comoving scale. Note that the index $\alpha$ denotes both the mass (or radius) of the halos and the impact parameter. Then, if we neglect the correlations between the collapsed objects we obtain in the continuous limit, using $\hn_{i,\alpha} = \hn_{i,\alpha}^2$:
\beq
\begin{array}{l} 
{\displaystyle  \lag\dmu^{\;p}\rag_c = \int_0^{l_s} dl \int \frac{dM}{M} \int_0^R db \; 2\pi b \; (1+z)^3 \; \eta(M) }   \\  \\  {\displaystyle  \hspace{0.8cm} \times \left[ 3\Omega_m \; (1+z) w(\chi,\chi_s) \; \Ib (1+\Delta_c) \; b \left(\frac{b}{R}\right)^{-\beta} \right]^p }
\end{array}
\eeq
where we defined the comoving mass function of halos of mass $M$ to $M+dM$ as $\eta(M) dM/M$. We also used:
\beq
\lag\hn_{i,\alpha}\rag = \Delta l_i \; db \; 2\pi b \; (1+z)^3 \; \eta(M) \frac{dM}{M}
\eeq
Some of the integrals over the impact parameter diverge for $\beta>1$ but we could add for these intermediate steps of the calculation an ad-hoc cutoff which we would put to 0 later on. In a fashion similar to (\ref{phimu}) we can define the generating function $\bphimu(y)$ and we get:
\beq
\begin{array}{l} 
{\displaystyle  \bphimu(y) = y + \int d\chi \int \frac{dM}{M} \int db \; B \; \bximu } \\ \\ {\displaystyle  \hspace{3cm} \times \left[ e^{-yA/\bximu} -1 + \frac{yA}{\bximu} \right] }
\end{array}
\eeq
with:
\beq
\left\{ \begin{array}{l} {\displaystyle A =  3\Omega_m \; (1+z) w(\chi,\chi_s) \; \Ib (1+\Delta_c) \; b \left(\frac{b}{R}\right)^{-\beta} } \\ \\ {\displaystyle B = (1+z)^2 \; 2\pi b \; \eta(M) } \end{array} \right.
\eeq
and the variance of the magnification is:
\beq
\bximu = \int_0^{\chi_s} d\chi \int_0^{\infty} \frac{dM}{M} \int_0^R db \; B \; A^2
\eeq
Next, we define:
\beq
\left\{ \begin{array}{l} {\displaystyle \lag n_h\rag = \int d\chi \int \frac{dM}{M} \int db \; B } \\ \\ {\displaystyle 1-\muminb = \int d\chi \int \frac{dM}{M} \int db \; B \; A } \end{array} \right.
\label{nbeta}
\eeq
and:
\beq
\bmu = \mu - \muminb
\eeq
Thus, $\lag n_h\rag$ is the mean number of halos along the line of sight while $\muminb$ is the minimum value of the magnification $\mu$. Indeed, using (\ref{Phi}) we obtain:
\beq
P(\bmu) = e^{-\lag n_h\rag} \inta \frac{dy}{2\pi i} e^{\bmu y + \int d\chi \int \frac{dM}{M} \int db B e^{-yA} }
\label{Pbmu3}
\eeq
which shows that $P(\bmu)=0$ for $\bmu < 0$. For large $\mu$ such that $P(\bmu) e^{\lag n_h\rag} \ll 1$ we can develop the exponential and only keep the first two terms which leads for $\bmu>0$ to:
\beq
P(\bmu) = e^{-\lag n_h\rag} \int d\chi \int \frac{dM}{M} \int db \; B \; \delta_D(\mu-A)
\label{Pmub}
\eeq
where $\delta_D$ is Dirac's function. Of course, we could have written (\ref{Pmub}) directly, as it corresponds to the case where there is only one object along the line of sight. In practice $\lag n_h\rag \; \ll 1$ and we obtain after integration over $b$:
\beq
\begin{array}{l} {\displaystyle P(\bmu) = \frac{2\pi}{\beta-1} \;\;  \bmu^{-\frac{\beta+1}{\beta-1}} \;\; \int d\chi \int \frac{dM}{M} \; R^2 \; \eta(M) }  \\ \\ {\displaystyle  \hspace{0.6cm} \times (1+z)^2 \left[ 3 \Omega_m \; (1+z) w(\chi,\chi_s) \; \Ib (1+\Delta_c) \; R \right]^{\frac{2}{\beta-1}} } \end{array}
\eeq
We can check that we recover the power-law behaviour of (\ref{tail1}). Next, in order to compare with (\ref{Pphietat}) we need to estimate the halo mass function $\eta(M) dM/M$. As discussed in details in Valageas \& Schaeffer (1997) and Valageas \& Schaeffer (1999), from the description of the non-linear density field presented in Sect.\ref{Density contrast probability distribution} one can write the comoving mass function of halos defined by the density contrast $\Delta(M,z)$ (which may depend both on $M$ and $z$) as:
\beq
\eta(M) \frac{dM}{M} = \frac{\rhob_0}{M} \; x^2 h(x) \; \frac{dx}{x}  \hspace{0.5cm} ,  \hspace{0.5cm} x = \frac{1+\Delta(M,z)}{\xia(R,z)}
\label{etah}
\eeq
where $\rhob_0$ is the present mean universe density and $h(x)$ is the function obtained from counts-in-cells statistics defined in (\ref{hphi}). A comparison of (\ref{etah}) with numerical results is described in Valageas et al.(1999). In order to get a direct comparison with (\ref{Pphietat}) we now make the approximation:
\beq
\xia(R;z) \simeq \xia_{c0} \; \frac{R_{c0}}{R} \; (1+z)^{-3}
\eeq
This is possible because the halos we consider here correspond to galactic masses hence to the scale $R_c(z)$ as noticed in Sect.\ref{Magnification by weak lensing}, where the local index of the initial linear power-spectrum is $n=-2$. Here $R_{c0}= R_c(0)$ and $\xia_{c0}=\xia(R_{c0};0)$. Thus we eventually get:
\beq
\begin{array}{l} {\displaystyle P(\bmu) = \bmu^{-\frac{\beta+1}{\beta-1}} \;\; \frac{3 (3\Omega_m\Ib)^{\frac{2}{\beta-1}}}{2(\beta-1)}  \; S_{\frac{2}{\beta-1}} \; ( \xia_{c0} R_{c0})^{\frac{3-\beta}{\beta-1}} } \\ \\ {\displaystyle \hspace{1.9cm} \times \int d\chi \; (1+z)^{\frac{2(\beta-3)}{\beta-1}} \; w(\chi,\chi_s)^{\frac{2}{\beta-1}} } \end{array}
\label{Pbmu2}
\eeq
where we defined:
\beq
S_{\frac{2}{\beta-1}} = \int_0^{\infty}  \; x^{\frac{2}{\beta-1}} \; h(x) \; dx
\eeq
as in (\ref{Sphx}). In a similar fashion, we obtain from (\ref{nbeta}):
\beq
\begin{array}{l} {\displaystyle \lag n_h\rag = \frac{3a}{4\Gamma(\om)}  ( \xia_{c0} R_{c0} )^{-\om} \left( \frac{4 (\beta-1) kT_{min}}{\mu_m m_p \Omega_m c^2} \right)^{\frac{\om-1}{2}} } \\ \\ {\displaystyle \hspace{1.2cm} \times \left( \frac{H_0}{c} \right)^{1-\om} \int d\chi \; (1+\Delta_c)^{\frac{\om-1}{2}} \; (1+z)^{\frac{3\om+1}{2}} } \end{array}
\label{nbeta1}
\eeq
where the parameters $a$ and $\om$ describe the small $x$ behaviour of $h(x)$, see (\ref{has}), $\Delta_c \sim 177$ is the density contrast of the smallest collapsed halos we consider at the threshold $T_{min}$ and $\mu_m$ is the mean molecular weight. Indeed, the number of halos along the line of sight is dominated by the contribution of the smallest objects (the multiplicity function diverges at small mass if there is no cutoff). We use $T_{min}= 3 \; 10^4$ K, it corresponds to inefficient cooling and photo-heating by the UV background. However, note that $\lag n_h\rag$ does not enter the probability distribution (\ref{Pbmu2}) as long as it is small (the exponential factor in (\ref{Pmub}) can be put to unity) which is the case in practice. Thus, our results do not depend on the cutoff $T_{min}$. We also get:
\beq
1-\muminb = \frac{3\Ib}{2(3-\beta)} \; (1-\mumin)
\eeq
where $\mumin$ is the exact minimum value of the magnification obtained in (\ref{mumin}). As expected, we recover the factor $(1-\mumin)$ since the physical process is the same. The term $3\Ib/(2(3-\beta))$ of order unity comes from the approximations involved in our calculation (assumption of small impact parameter and approximation $(1+\Delta) \simeq \Delta$). From the results obtained in the previous section, we know that $\mumin \simeq 1$ so that we still have $\muminb \simeq 1$ and $\bmu \simeq \mu-1$. Moreover, the probability distribution (\ref{Pbmu2}) is only valid for $P(\bmu) \ll 1$ as explained above, that is for $\bmu \ga 1$, hence the value of $\muminb$ plays no role as long as it remains small as compared to 1.

\begin{figure}

{\epsfxsize=8 cm \epsfysize=5.4 cm \epsfbox{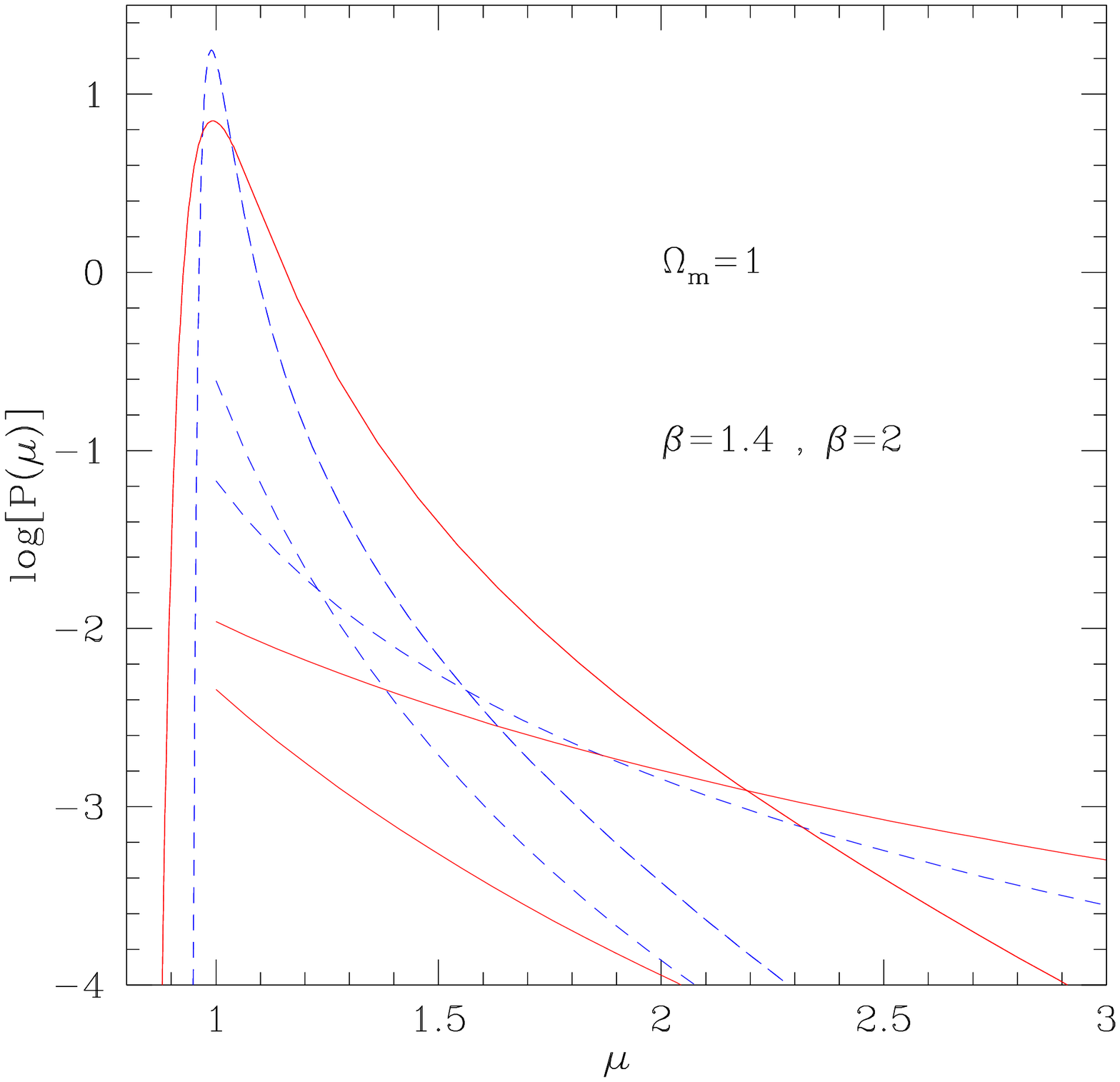} }
{\epsfxsize=8 cm \epsfysize=5.4 cm \epsfbox{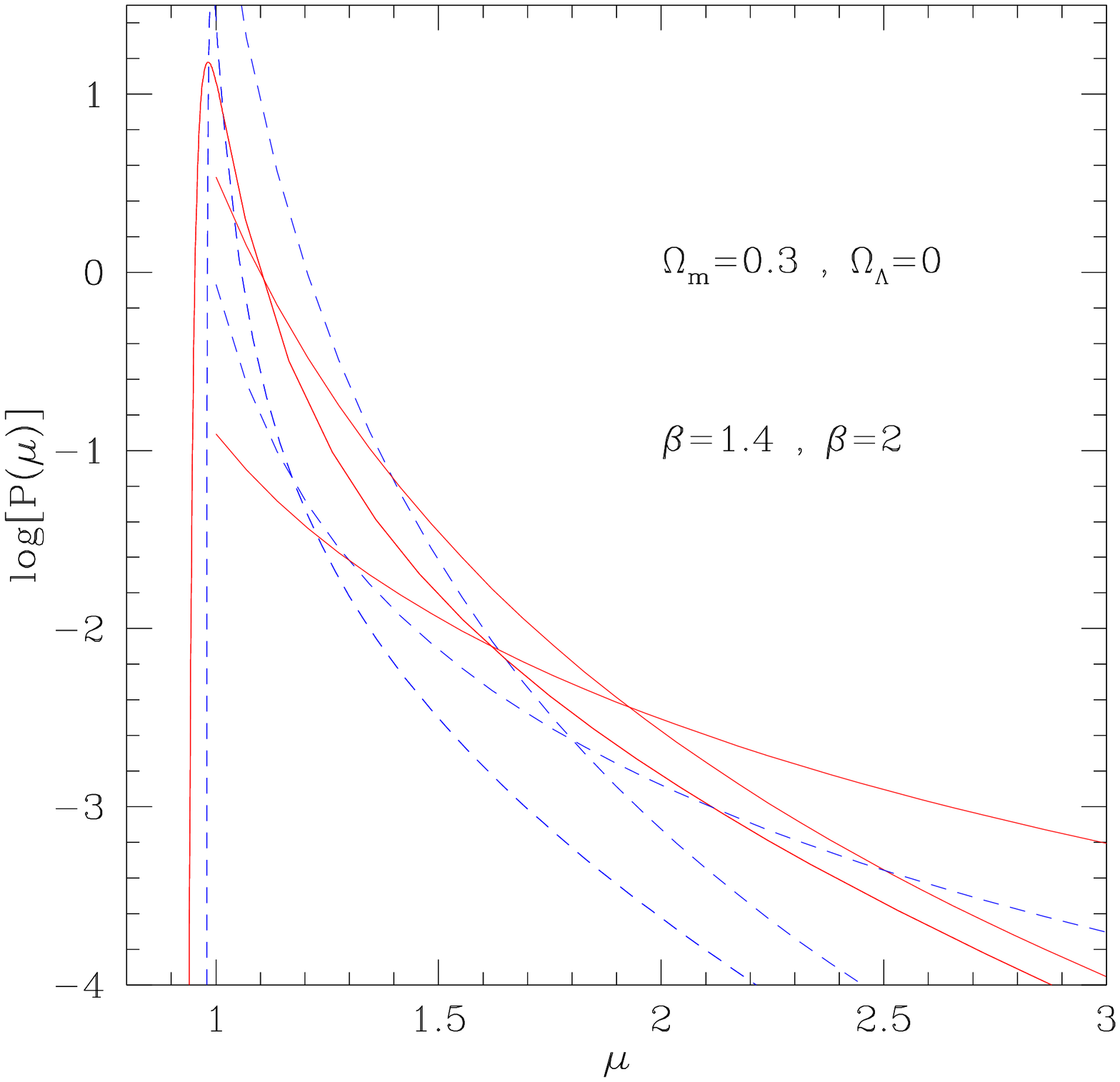} }
{\epsfxsize=8 cm \epsfysize=5.4 cm \epsfbox{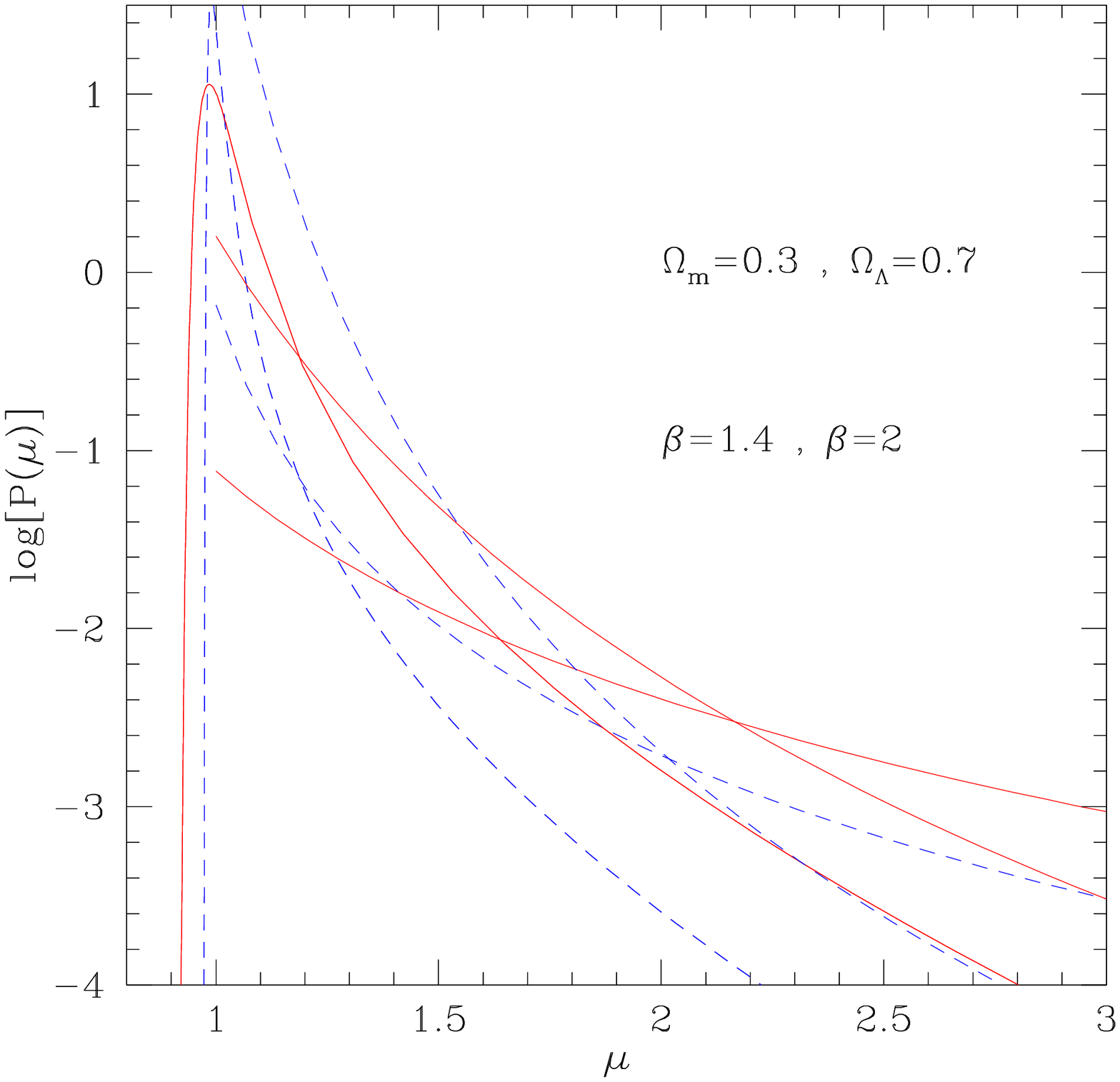} }

\caption{The probability distribution $P(\mu)$ of the magnification of distant sources located at $z_s=0.5$ (dashed lines) and at $z_s=1$ (solid lines). The curves which start at $\mu=1$ with no falloff at low $\mu$ correspond to the formulation (\ref{Pbmu2}) where the density field is described as a collection of virialized halo with a power-law density profile (\ref{beta}) with $\beta=1.4$ (steeper falloff) or $\beta=2$ (smoother falloff). The curves which show a peak at $\mu < 1$ and a falloff at lower $\mu$ are from (\ref{Pphietat}).}

\label{figlPmu}

\end{figure}

We compare in Fig.\ref{figlPmu} the probability distribution $P(\mu)$ obtained from (\ref{Pbmu2}) for $\beta=1.4$ and $\beta=2$ with the results obtained in Sect.\ref{Magnification by weak lensing} from (\ref{Pphietat}). Note that we have defined throughout the magnification $\mu$ by (\ref{mukappa}). For large values of the convergence $\kappa \ga 1$ the approximation (\ref{mukappa}) breaks down and $\mu$ should be obtained from (\ref{mukappa1}). However, since in this article we are mostly interested in the regime $\kappa \ll 1$ we always define $\mu$ by (\ref{mukappa}). Thus, $P(\mu)$ can also be understood as $P(\kappa)$ with the change of variable (\ref{mukappa}). Then, large values of the magnification $\mu \geq 3$ correspond to $\kappa \geq 1$ and to strong lensing events with multiple images. We can see in Fig.\ref{figlPmu} that the probability distribution $P(\mu)$ from (\ref{Pbmu2}) is of the same order as the results from (\ref{Pphietat}) in the range $1.5 < \mu < 3$. Indeed, we count the same mass and the same collapsed halos. Our results are similar to those obtained by Porciani \& Madau (1999) who got a probability of a few $10^{-4}$ to have a strong lensing event for a source at $z_s=1$ with $\beta=2$. However, we can see that (\ref{Pbmu2}) is quite sensitive to the slope $\beta$ of the halos. Indeed, for $\beta=2$ we get $P(\mu) \sim \mu^{-3}$ while for $\beta=1.4$ we have $P(\mu) \sim \mu^{-6}$. 
On the other hand, below the exponential cutoff the prescription developped in Sect.\ref{Magnification by weak lensing} leads to $P(\mu) \sim \mu^{-1.7}$ down to $P(\mu) \sim \mu^{-2.8}$. However, since $\ximu$ is small this pure power-law regime does not really appear as the corrections due to the peak at $\mu \la 1$ and the exponential cutoff are not negligible. Nevertheless, we can clearly see the extended large $\mu$ tail of $P(\mu)$ and its non-gaussian behaviour. In particular, we recover the trend seen in numerical simulations (e.g. Wambsganss et al.1997). As we explained above the formulation (\ref{Pbmu2}) cannot describe the regime $\mu \la 1$ while for large magnifications it predicts a higher probability $P(\mu)$ since it leads to a power-law tail instead of an exponential cutoff. This is directly linked to the strong growth with $p$ of the $p-$point correlation functions at small scales implied by this model, which does not seem to be compatible with numerical simulations as argued in Valageas (1999a). On the other hand, if we only use (\ref{Pbmu2}) to obtain the weak lensing effect due to inner galactic halos where baryonic matter dominates the density field we would get lower probabilities $P(\mu)$ since the matter content and size of these objects would be smaller, so that up to $\mu \leq 3$ the probability distribution (\ref{Pphietat}) would dominate. Thus, the formalism developped in Sect.\ref{Magnification by weak lensing} is better suited to obtain the magnification of distant sources by weak lensing effects.

\begin{figure}

{\epsfxsize=8 cm \epsfysize=5.4 cm \epsfbox{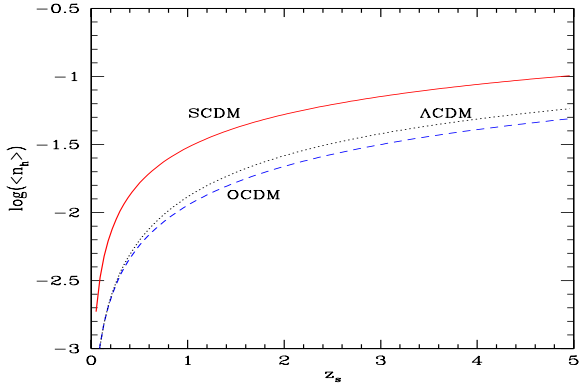} }

\caption{The mean number of halos $\lag n_h\rag$ along the line of sight up to the redshift $z_s$ of the source. The solid curve is for a critical universe, the dashed curve for an open cosmology and the dotted line for a flat low-density universe.}

\label{fignb}

\end{figure}

We show in Fig.\ref{fignb} the mean number of halos $\lag n_h\rag$ along the line of sight up to the redshift $z_s$ of the source, from (\ref{nbeta1}). We use $T_{min} =3 \; 10^4$ K. We can check that $\lag n_h\rag \ll 1$ which justifies the approximation (\ref{Pbmu2}). Thus, large magnifications $\mu$ come from the deflection of the light ray by a single object. Of course, as explained above this approach cannot handle low $\mu$ close to $\mumin$. Indeed, these lines of sight do not cross totally empty regions but low-density patches of matter so that $P(\mu) \rightarrow 0$ for $\mu \rightarrow \mumin$, as in Fig.\ref{figPmu}, while the formulation (\ref{Pmub}) leads to a Dirac $e^{-\lag n_h\rag} \delta_D(\bmu)$ in $\bmu=0$.

\section{Derivation of cosmological parameters}
\label{Derivation of cosmological parameters}

In practice, one uses the Type Ia supernovae as standard candles in order to derive the cosmological parameters $\Omega_m$ and $\Omega_{\Lambda}$ from the observed relation redshift $\leftrightarrow$ luminosity distance (e.g. Perlmutter et al.1999). However, even if these sources are perfect candles with no intrinsic dispersion nor instrumental noise, the weak lensing effects discussed in the previous sections will introduce some dispersion and some bias as the supernovae will be randomly magnified by the density fluctuations located along the line of sight. Thus, the luminosity distance measured by the observer will differ from the actual one by a small deviation which we can relate to the magnification $\mu$ by:
\beq
\sqrt{\mu} = \left[ \frac{\De(z_s)}{\Deo(z_s)} \right] (\Omega_{m,obs},\Omega_{\Lambda,obs})
\label{muOm}
\eeq
where $\Deo(z_s)$ is the observed distance defined in (\ref{De}) seen as a function of $\Omega_{m,obs}$ and $\Omega_{\Lambda,obs}$ which are thus determined by the observation (if there is no distortion by weak lensing: $\mu=1$, these parameters are equal to the actual cosmological parameters $\Omega_m$ and $\Omega_{\Lambda}$). Here we assume the absolute magnitude of the supernovae is known, for instance from local or low redshifts observations. We used the fact that the luminosity distance $d_L(z)$ obeys: $d_L(z) = (1+z) \De(z)$. Thus, for a given redshift $z_s$ of the source and a peculiar cosmology $(\Omega_m,\Omega_{\Lambda})$ of the actual universe we can obtain ``confidence regions'' in the $(\Omega_{m,obs},\Omega_{\Lambda,obs})$ plane for the observed parameters $\Omega_{m,obs}$ and $\Omega_{\Lambda,obs}$. This means that for any $\cP$ with $0 \leq \cP \leq 1$, the ``observed universe'' has a probability $\cP$ to lie within the domain $\RP$. 
Thus, for any point $(\Omega_{m,obs},\Omega_{\Lambda,obs})$ we calculate through (\ref{muOm}) the weak lensing magnification $\mu$ which is needed so that the observer would measure $(\Omega_{m,obs},\Omega_{\Lambda,obs})$ in an actual universe defined by $(\Omega_m,\Omega_{\Lambda})$. Then, from the intervals of confidence (\ref{Psig}), displayed in Fig.\ref{figPsig}, we obtain the probability $\cP$ (defined by the constraint that either $\mu=\mu_{-}$ or $\mu=\mu_{+}$) that such a magnification is realized, hence that such a cosmology is derived from observations. In this way we obtain confidence regions $\RP$ in the $(\Omega_{m,obs},\Omega_{\Lambda,obs})$ plane. Of course, from (\ref{muOm}) a measure at a single redshift only provides a value for the distance $\De_{obs}$ (hence for the deceleration parameter $q_0$ at low redshift $z_s \rightarrow 0$). Hence the parameters $\Omega_{m,obs}$ and $\Omega_{\Lambda,obs}$ are only constrained to lie on a line in the $(\Omega_{m,obs},\Omega_{\Lambda,obs})$ plane. As a consequence, the regions $\RP$ are unbounded strips in the $(\Omega_{m,obs},\Omega_{\Lambda,obs})$ plane.

\begin{figure}

{\epsfxsize=8 cm \epsfysize=5.4 cm \epsfbox{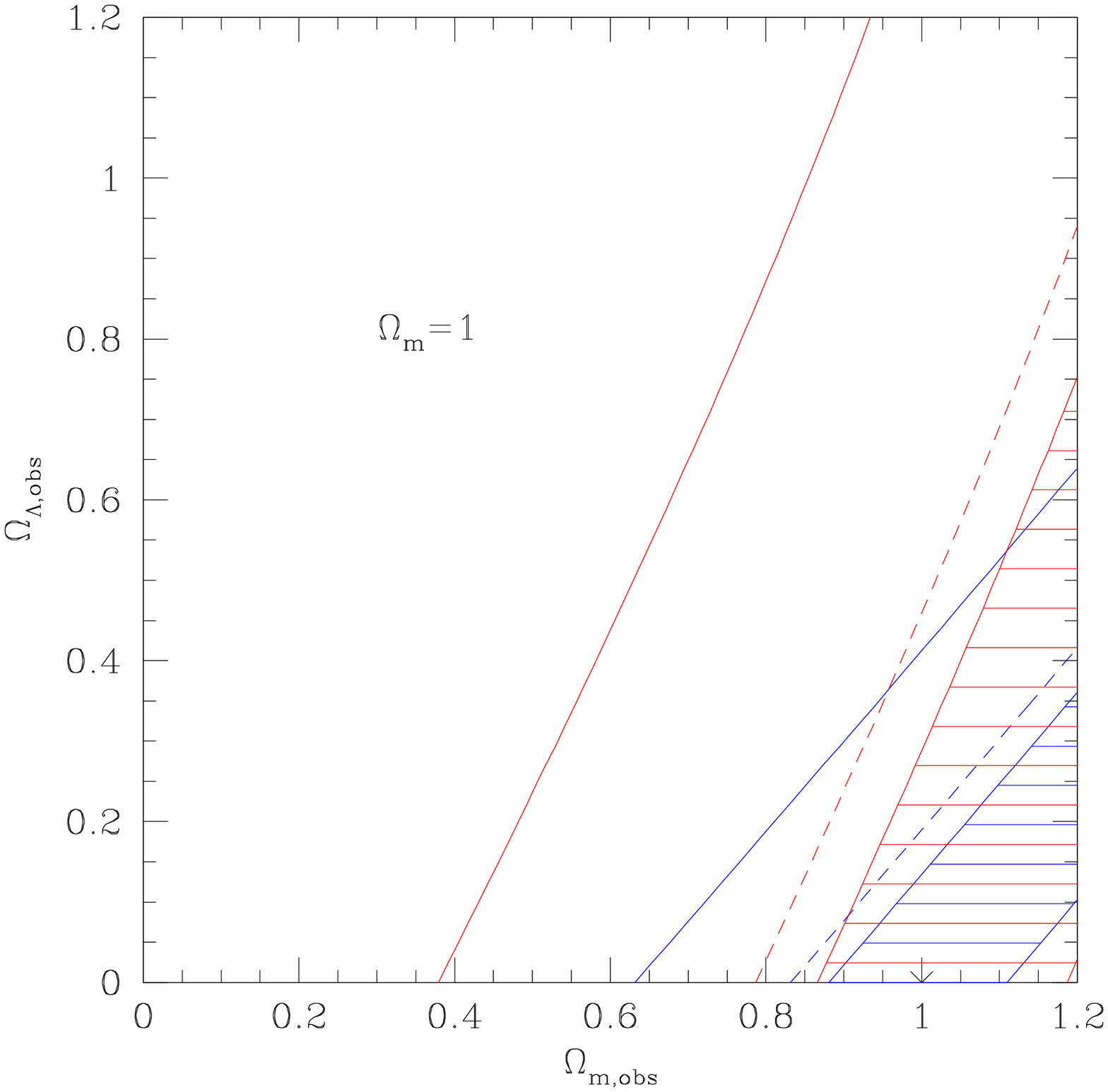} }
{\epsfxsize=8 cm \epsfysize=5.4 cm \epsfbox{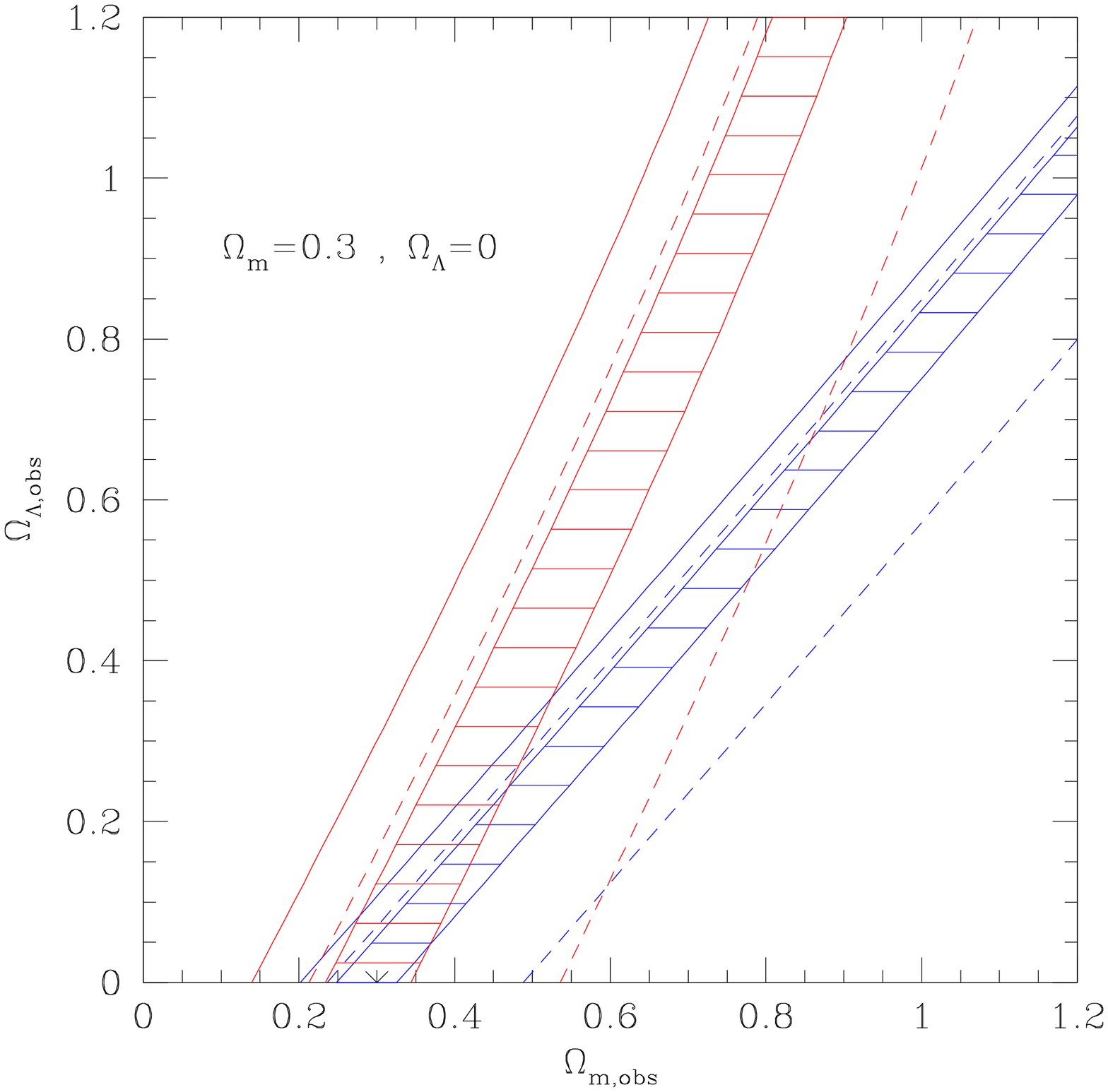} }
{\epsfxsize=8 cm \epsfysize=5.4 cm \epsfbox{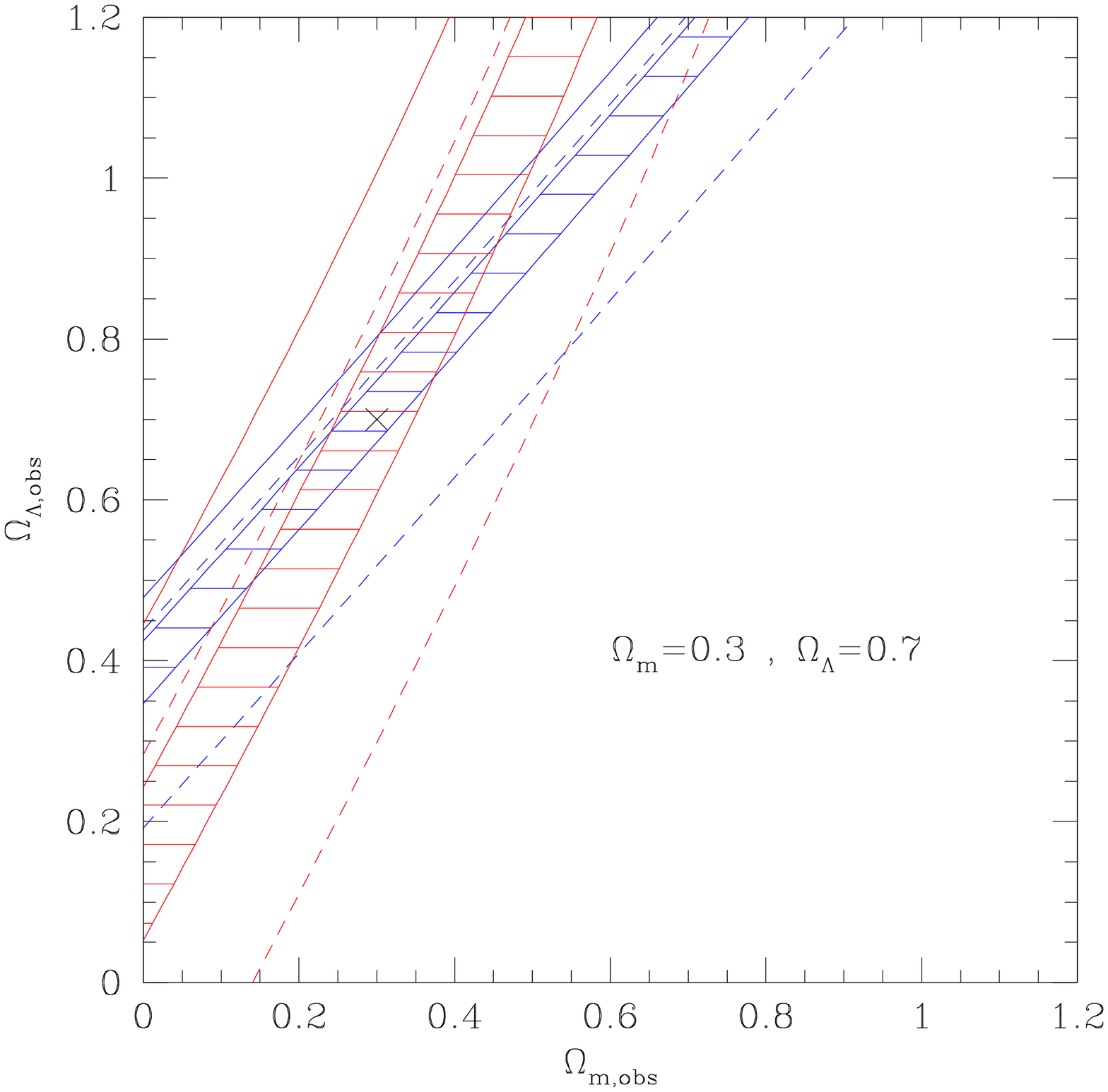} }

\caption{The ``confidence regions'' $\RP$ for the observed parameters $\Omega_{m,obs}$ and $\Omega_{\Lambda,obs}$, for three actual cosmologies and the redshifts $z_s=0.5$ and $z_s=1$. The dashed region within two solid lines corresponds to $\cP=68\%$: within $68\%$ of the cases the cosmology will be observed to lie within this domain in the $(\Omega_{m,obs},\Omega_{\Lambda,obs})$ plane. The region within the two dashed lines corresponds to $\cP=95\%$. The solid line $\cLmin$ on the left corresponds to the lower bound $\mumin$: the observed parameters $(\Omega_{m,obs},\Omega_{\Lambda,obs})$ cannot lie to the left of this curve. The boundary lines obtained for a larger redshift are steeper (i.e. closer to the vertical).}

\label{figcont}

\end{figure}

We display in Fig.\ref{figcont} these ``confidence regions'' $\RP$ we obtain for the three cosmologies we have studied in details in the previous sections, for the redshifts $z_s=0.5$ and $z_s=1$. At a given redshift $z_s$, we show the domains $\RP$ defined by $\cP=68 \%$ (dashed region within two solid lines) and by $\cP=95 \%$ (within the two dashed lines). This means that from one observation at a given redshift the observer will conclude with a probability of $68\%$ that the cosmological parameters $(\Omega_{m,obs},\Omega_{\Lambda,obs})$ lie in the dashed domain. We also display the boundary line $\cLmin$ given by $\mumin$ within (\ref{muOm}) (solid line on the left side). Thus, because of the lower bound $\mumin$ the observed parameters $(\Omega_{m,obs},\Omega_{\Lambda,obs})$ cannot lie to the left of the line $\cLmin$ (of course, here we only consider the effects of weak lensing). The cross is the point $(\Omega_m,\Omega_{\Lambda})$. For a given redshift, all these boundary lines are parallel. Moreover, their slope (when seen as a function $\Omega_{\Lambda,obs}$ of $\Omega_{m,obs}$) increases with $z_s$. Indeed, as emphasized by Perlmutter et al.(1997) the parameters $\Omega_{m,obs}$ and $\Omega_{\Lambda,obs}$ enter $\De_{obs}(z)$ with different powers of $(1+z)$ so that the observed distance $\Deo(z)$ is not a function of $q_{obs}$ (except in the limit $z_s \rightarrow 0$) but of a combination of $\Omega_{m,obs}$ and $\Omega_{\Lambda,obs}$ which varies with $z_s$. This implies the drift with $z_s$ of the slope of the boundary lines we defined above. Of course, this is the reason why observations can simultaneously constrain $\Omega_m$ and $\Omega_{\Lambda}$ (provided one has a finite range of source redshifts). 
The sign of the slope of these strips translates the fact that for the same $\Omega_m$ and redshift a flat universe corresponds to a larger distance $\De(z)$ than an open geometry. Of course, the strong asymmetry of the probability distribution $P(\mu)$, and of the intervals $[\mu_{-},\mu_{+}]$, leads to asymmetric regions $\RP$. Thus, as $\cP$ increases the strip $\RP$ grows but it is bounded on the left by $\cLmin$ while it can extend to infinity to the right. Indeed, as can be seen from (\ref{muOm}) larger $\mu$ corresponds to larger $\Omega_{m,obs}$ and smaller $\Deo$. Thus, the most likely values of $(\Omega_{m,obs},\Omega_{\Lambda,obs})$, corresponding to $\mumax<1$, lie slightly to the left of the point $(\Omega_m,\Omega_{\Lambda})$, while for large $\cP$ the domain $\RP$ shows an extended tail towards large $\Omega_{m,obs}$. The strips are larger for the critical universe which had a higher dispersion for the probability distribution $P(\mu)$, see Fig.\ref{figPmu}. We note that although the spread due to weak lensing cannot make a critical universe appear as $\Omega_m=0.3$ (due to the cutoff $\mumin$) while a low-density universe $\Omega_m=0.3$ has a negligible probability to appear as $\Omega_m=1$, the effect of the weak lensing is not negligible. Thus, two observations at redshifts $z_s=0.5$ and $z_s=1$ only determine $\Omega_m$ within an interval $\Delta \Omega_m \ga 0.3$. For instance, for the low-density flat universe (lower panel) we have:
\beq
\!\!\!\!\left\{ \begin{array}{rl} \cP=68\% : & 0.17 < \Omega_{m,obs} < 0.45 \; , \; 0.52 < \Omega_{\Lambda,obs} < 0.9 \\  \\  \cP=95\% : & 0 < \Omega_{m,obs} < 0.7 \; , \; 0.25 < \Omega_{\Lambda,obs} < 1.2  \end{array} \right.
\label{intcosmo}
\eeq
Note that if both observations (at $z_s=0.5$ and $z_s=1$) are independent the intersection of the regions labelled $\cP=68\%$, used in (\ref{intcosmo}), corresponds to a probability $0.68^2 = 46\%$. These uncertainties are larger for both other cosmologies we consider in Fig.\ref{figcont}. Of course, by averaging over many observations at a given redshift one diminishes this inaccuracy (one expects that $\Delta \Omega_{m,obs}$ roughly decreases as $1/\sqrt{N}$). Moreover, we can check in the figure that observations can unambiguously discriminate between $\Omega_m=0.3$ and $\Omega_m=1$. In the case of a low-density universe one can also clearly discriminate between $\Omega_{\Lambda}=0$ and $\Omega_{\Lambda}=1-\Omega_m$.

\begin{figure}

{\epsfxsize=8 cm \epsfysize=5.4 cm \epsfbox{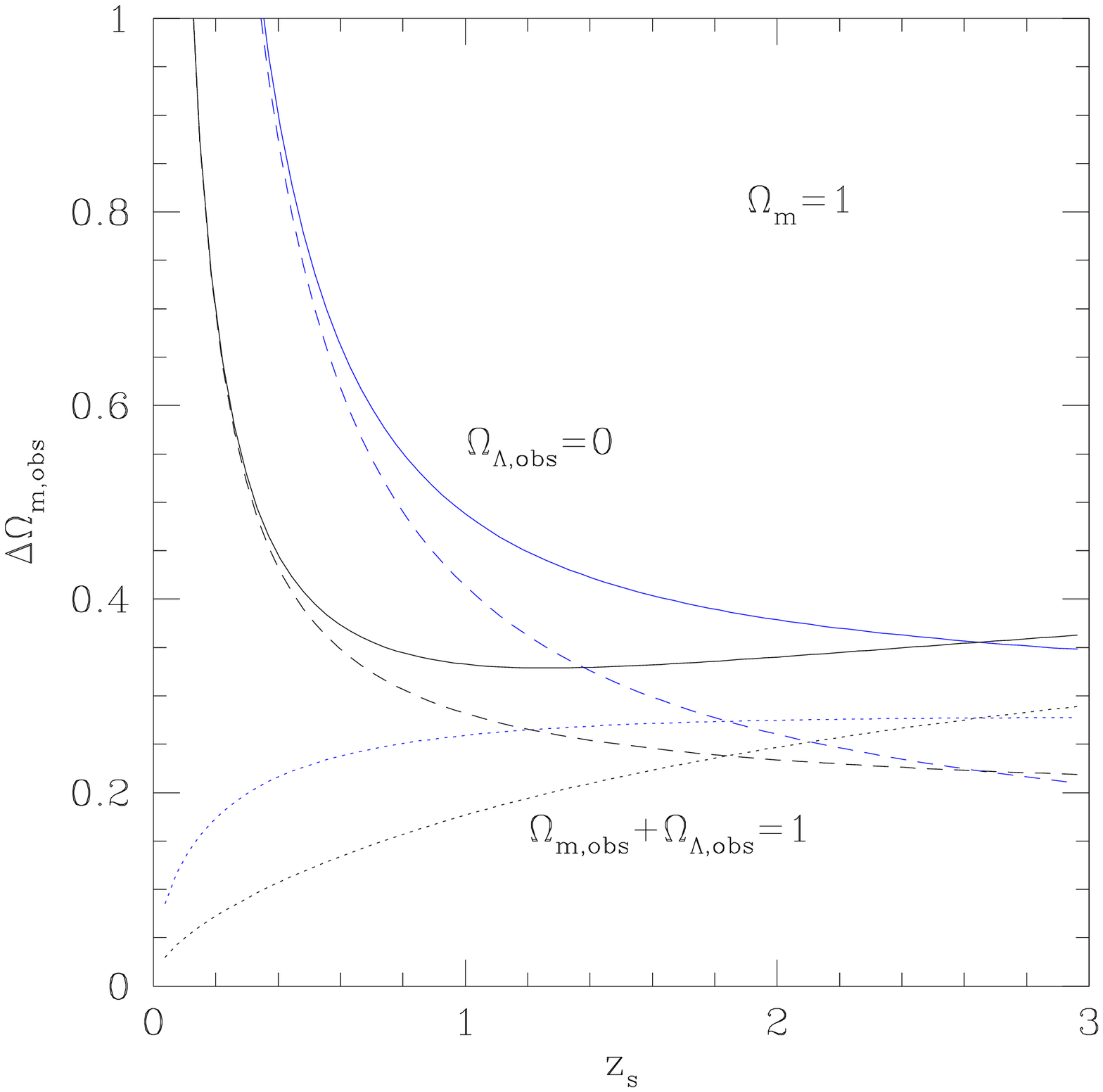} }
{\epsfxsize=8 cm \epsfysize=5.4 cm \epsfbox{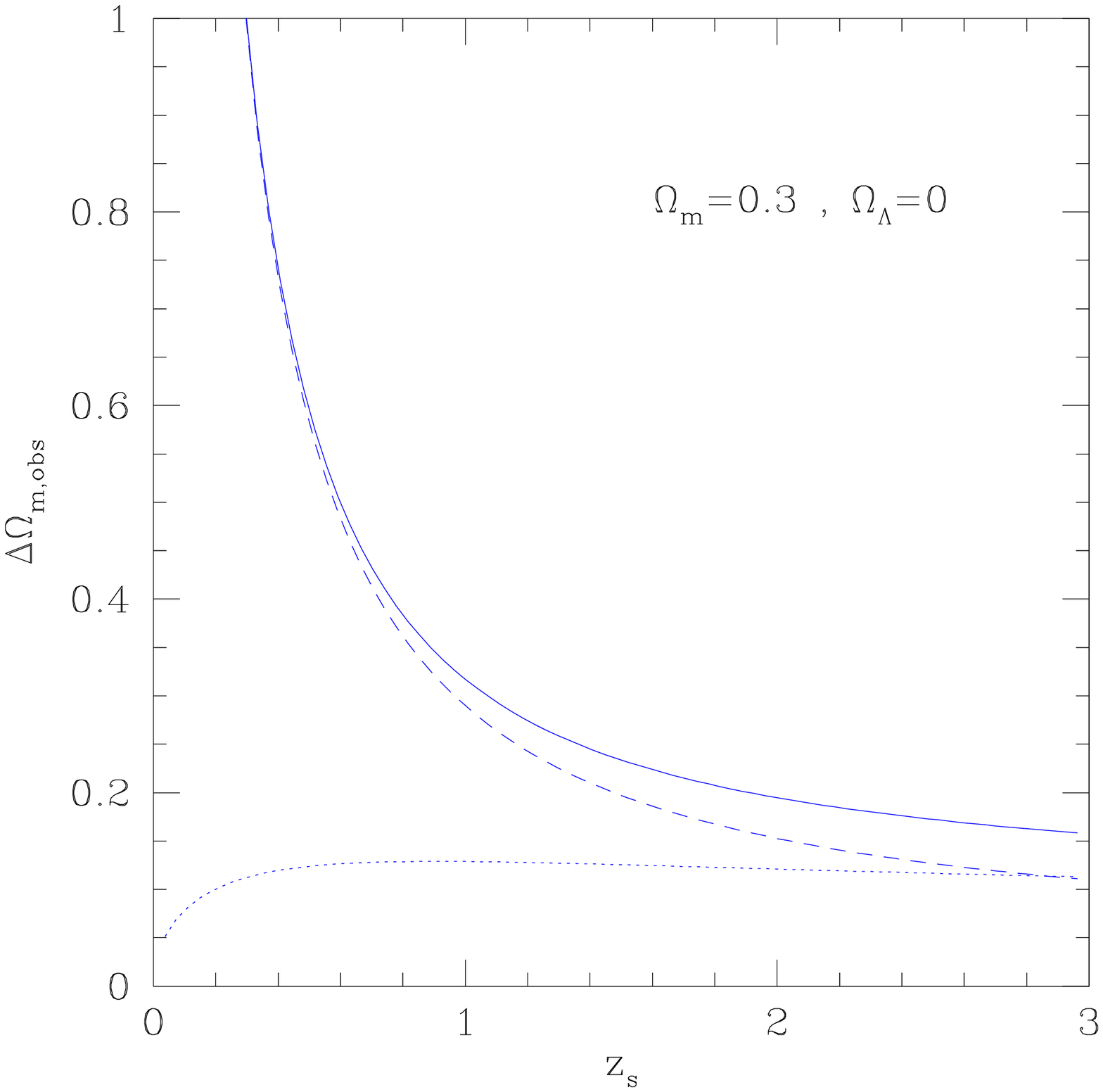} }
{\epsfxsize=8 cm \epsfysize=5.4 cm \epsfbox{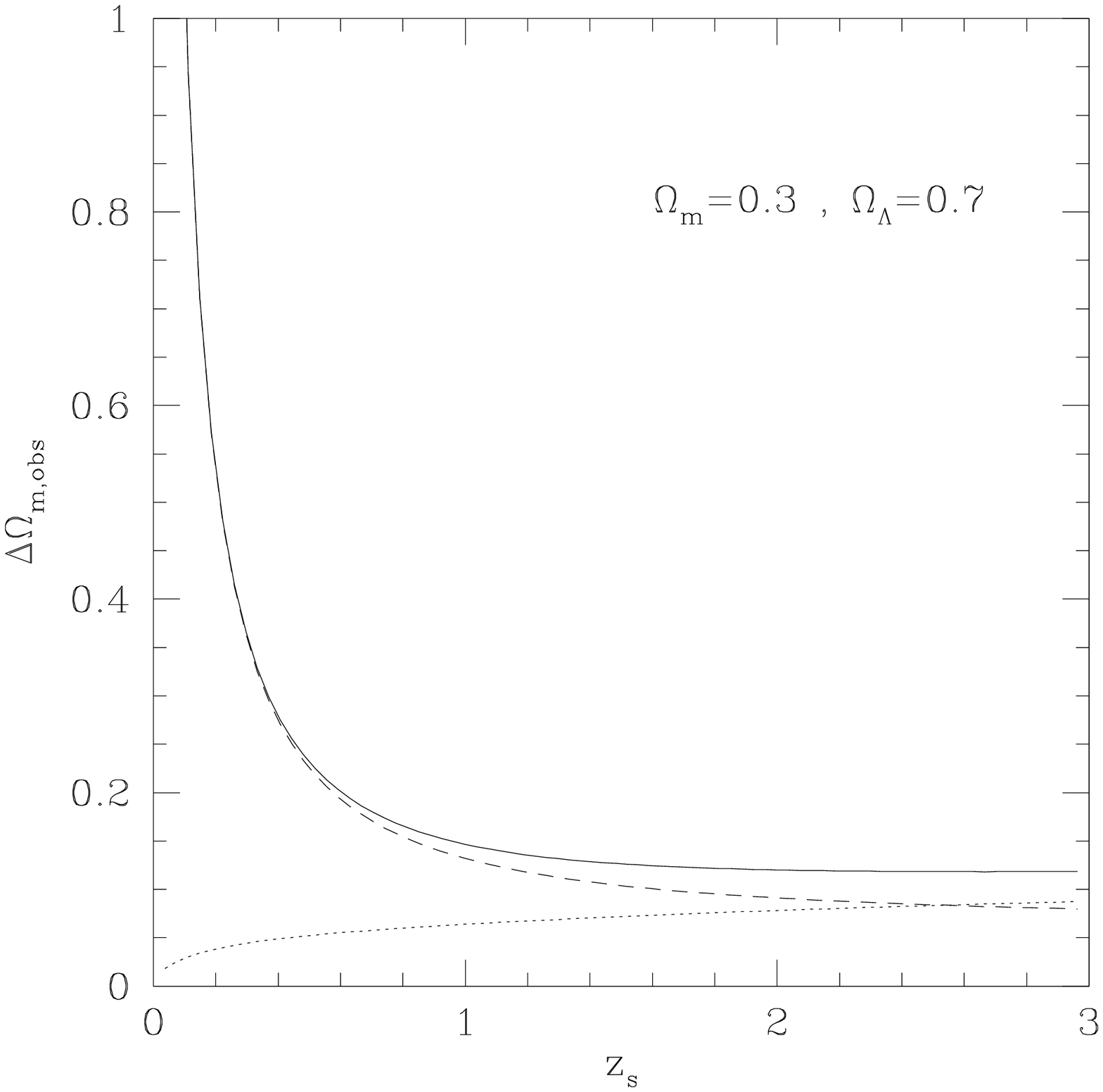} }

\caption{The uncertainty $\Delta \Omega_{m,obs}$ of the observed parameter $\Omega_m$ from sources at redshift $z_s$, for three cosmologies. The solid lines take into account both the intrinsic dispersion $\sigma_B$ of supernovae and the fluctuation $\ximu$ of the magnification by weak lensing. The dashed curves show the influence of $\sigma_B$ alone (case with $\ximu=0$) while the dotted curves represent the influence of weak lensing alone (case with $\sigma_B=0$). For an actual universe which is critical ($\Omega_m=1$) the lower curves correspond to the assumption of a flat universe and the upper curves to $\Omega_{\Lambda}=0$ universes. For low-density universes we assume the correct cosmology (flat or open).}

\label{figDOmz}

\end{figure}

In order to compare the effect of weak gravitational lensing on the derivation of the cosmological parameters with the inaccuracy due to the intrinsic magnitude dispersion $\sigma_B$ of the sources, we show in Fig.\ref{figDOmz} the quantity:
\beq
\Delta \Omega_{m,obs} = \left| \frac{d\Omega_{m,obs}}{dM_B} \right| \; \left( \sigma_B^2 + \Delta M_{Blensing}^2 \right)^{1/2}
\eeq
Here $M_B$ is the absolute magnitude of the supernova and $\Delta M_{Blensing}^2$ is the magnitude dispersion due to weak lensing. Thus, we have:
\beq
\Delta \Omega_{m,obs} = \frac{\mbox{ln}10}{2.5} \; \left. \frac{d\Omega_{m,obs}}{d\mu} \right|_{\mu=1} \; \left[ \sigma_B^2 + \left( \frac{2.5}{\mbox{ln}10} \right)^2 \ximu \right]^{1/2}
\label{DeltaOm}
\eeq
where $d\Omega_{m,obs}/d\mu$ is obtained from (\ref{muOm}). For low-density universe we eliminate $\Omega_{\Lambda,obs}$ in (\ref{muOm}) by assuming the right cosmology (flat or $\Omega_{\Lambda,obs}=0$) so that $\Deo$ is a function of $\Omega_{m,obs}$ alone. In the case of a critical universe we display the results we obtain when we assume $\Omega_{\Lambda,obs}=0$ or $\Omega_{\Lambda,obs}=1-\Omega_{m,obs}$. We use for all redshifts $\sigma_B=0.17$ mag for the lightcurve-width-corrected luminosity dispersion of SNeIa, see Perlmutter et al.(1999). The solid lines in Fig.\ref{figDOmz} show $\Delta \Omega_{m,obs}$ from (\ref{DeltaOm}). The dashed curves show the inaccuracy on $\Omega_{m,obs}$ due to $\sigma_B$ alone while the dotted curves show the effect of $\ximu$ alone. We can see that for low-density universes the error due to the intrinsic dispersion $\sigma_B$ dominates up to $z_s \leq 2$ while for a critical universe the weak lensing contribution already dominates for $z_s \geq 1.4$. Note also that the inaccuracy of the measure of $\Omega_{m,obs}$ is much larger for a critical universe, which has a non-negligible probability to appear as an open universe with $\Omega_m=0.5$ or a flat universe with $\Omega_m=0.6$ for $z_s \leq 1$. In particular, note that going to high redshift $z_s > 1$ does not increase the accuracy of the determination of $\Omega_m$ by much. Thus, the minimum dispersion of $\Omega_{m,obs}$ is $\Delta \Omega_{m,obs} \sim 0.15$ for low-density universe and $\Delta \Omega_{m,obs} \sim 0.4$ for a critical universe. Of course, one can reduce the uncertainties by observing many supernovae. Note that this analysis does not take into account the non-gaussian behaviour of the probability distribution of the magnification $\mu$. This was studied in Fig.\ref{figcont}. Moreover, as can be seen in Fig.\ref{figcont} the uncertainty on $\Omega_{m,obs}$ due to weak gravitational lensing is larger than the estimate (\ref{DeltaOm}) because of the degeneracy in the plane $(\Omega_{m,obs} , \Omega_{\Lambda,obs})$. Indeed, although observations at two different redshifts $z_s=0.5$ and $z_s=1$ remove this degeneracy the intersection of the domains $\RP$ remains elongated along an axis roughly parallel to $\Omega_{\Lambda,obs} = \Omega_{m,obs}$. Hence the uncertainty on $\Omega_{m,obs}$, given by the projection onto the $\Omega_{m,obs}$-coordinate of the length of this region along this axis, is larger than the value obtained from (\ref{DeltaOm}) shown in Fig.\ref{figDOmz} which corresponds to a cut of the domain $\RP$ along the axis $\Omega_{\Lambda,obs} = 0$ or $\Omega_{\Lambda,obs} = 1- \Omega_{m,obs}$. This is why the values obtained in (\ref{intcosmo}) are higher than those one would derive from (\ref{DeltaOm}).

\section{Bias}
\label{Bias}

The random magnification of distant sources by weak lensing induces some bias in any observed sample. For instance, the fraction of gravitationally lensed SNeIa increases for bright magnitudes due to the falloff of the luminosity function of the sources which means that the contribution from lower luminosity SNeIa which have been amplified by weak lensing becomes more important. We assume that the luminosity function of the SNeIa at redshift $z$ is a gaussian, as a function of the absolute magnitude $M_B$:
\beq
\Phi(M_B;z) = \frac{\Phi_0(z)}{\sqrt{2\pi} \sigma_B} e^{-(M_B-M_{B0}(z))^2 / (2\sigma_B^2)}
\label{gauss}
\eeq
Here $M_{B0}(z)$ is the mean absolute magnitude while $\sigma_B$ is the intrinsic magnitude dispersion of the sources. Next, we define the bias $B(<M_B,z)$ as the ratio of the number of SNeIa observed with a magnitude brighter than $M_B$, at redshift $z$, by the number of SNeIa which actually are brighter than this luminosity threshold:
\[
B(<M_B,z) = \frac{ \int_{-\infty}^{\infty} dM_B' \Phi(M_B',z) \int_{10^{(M_B'-M_B)/2.5}}^{\infty} d\mu P(\mu) } { \int_{-\infty}^{M_B} dM_B' \Phi(M_B',z) }
\]
where $P(\mu)$ is the probability distribution of the magnification. Using (\ref{gauss}) we obtain:
\beq
B(<m,z) = \int_0^{\infty} d\mu P(\mu) \frac{ \mbox{erfc} \left( - \frac{m}{\sqrt{2}} - \frac{2.5}{\sqrt{2} \sigma_B} \log \mu \right) } { \mbox{erfc} \left( - \frac{m}{\sqrt{2}} \right) }
\label{bias1}
\eeq
where we defined the ``reduced magnitude'' $m$ by:
\beq
m= \frac{M_B - M_{B0}}{\sigma_B}
\label{mreduc}
\eeq
and $\mbox{erfc}(x) = 2/\sqrt{\pi} \int_x^{\infty} dt \; e^{-t^2}$ is the complementary error function. We can see from (\ref{bias1}) that:
\beq
\left\{ \begin{array}{l} {\displaystyle m \rightarrow \infty \; : \; B(<m,z) \rightarrow 1 } \\ \\ {\displaystyle m \rightarrow -\infty \; : \; B(<m,z) \rightarrow \infty } \end{array} \right.
\eeq
Indeed, for $m \rightarrow \infty$  one recovers all SNeIa (the survey is complete) while for $m \rightarrow -\infty$ the local slope of the luminosity function is increasingly steep so that the effect of weak lensing gets larger.

\begin{figure}

{\epsfxsize=8 cm \epsfysize=5.4 cm \epsfbox{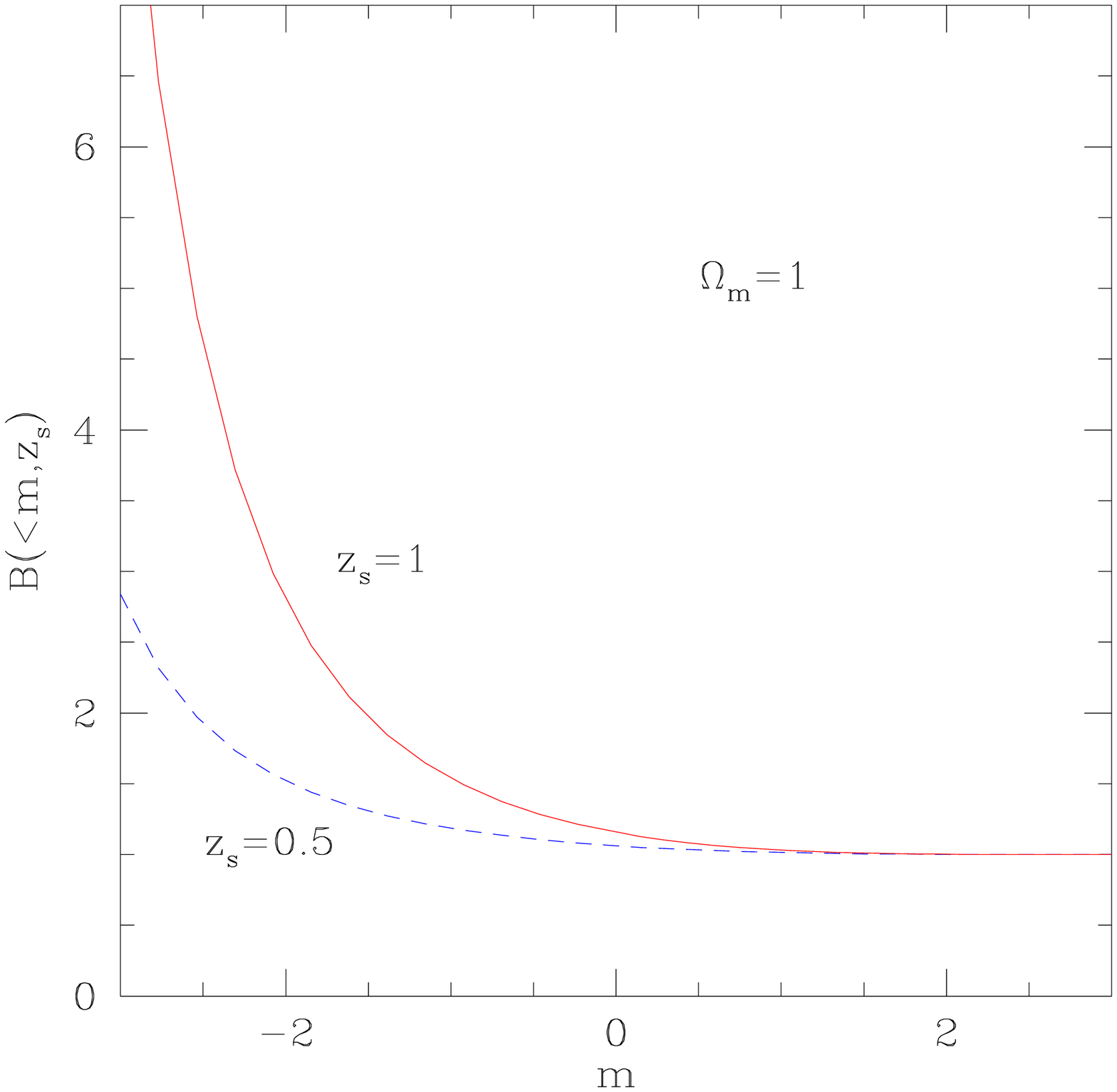} }
{\epsfxsize=8 cm \epsfysize=5.4 cm \epsfbox{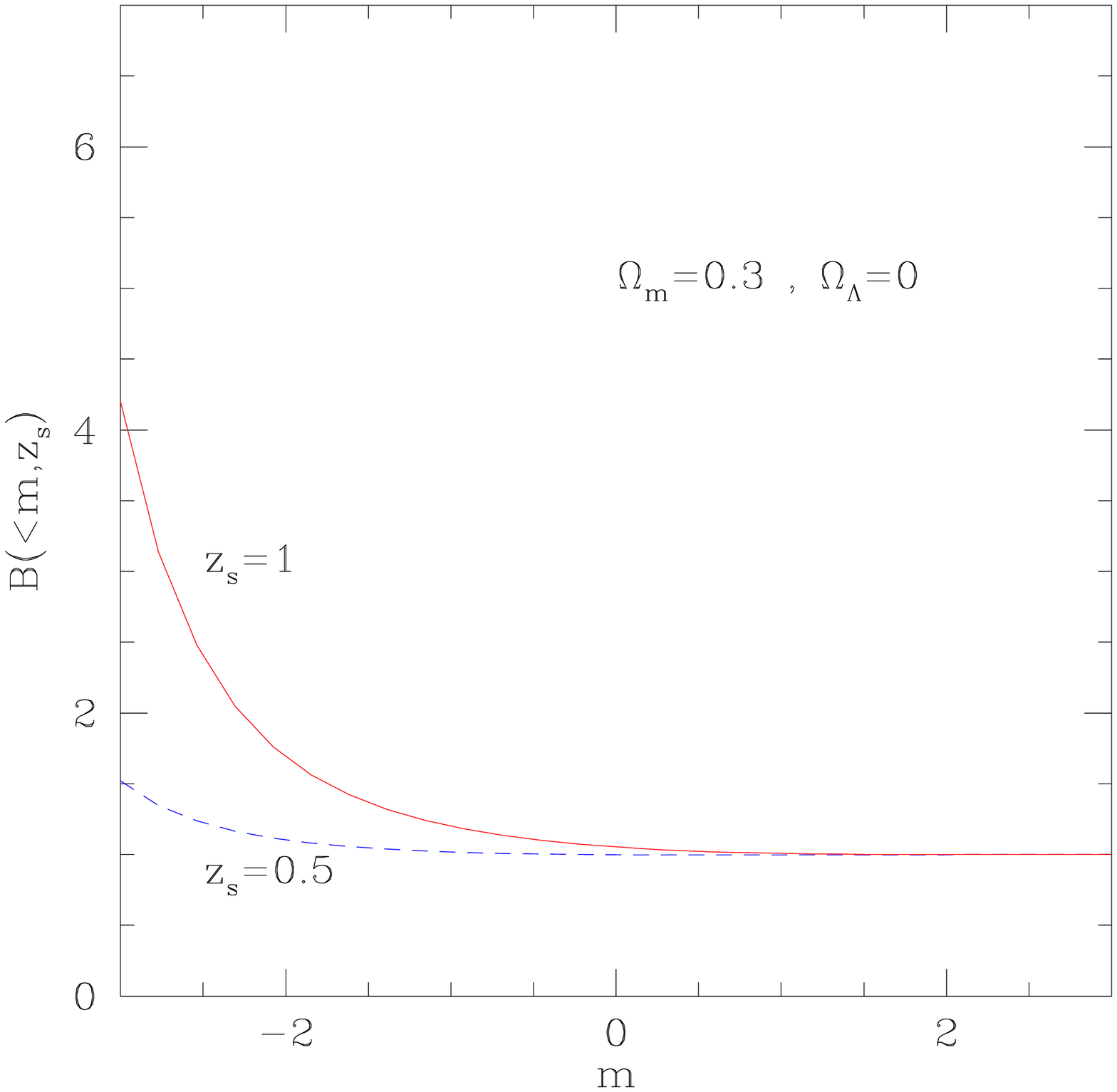} }
{\epsfxsize=8 cm \epsfysize=5.4 cm \epsfbox{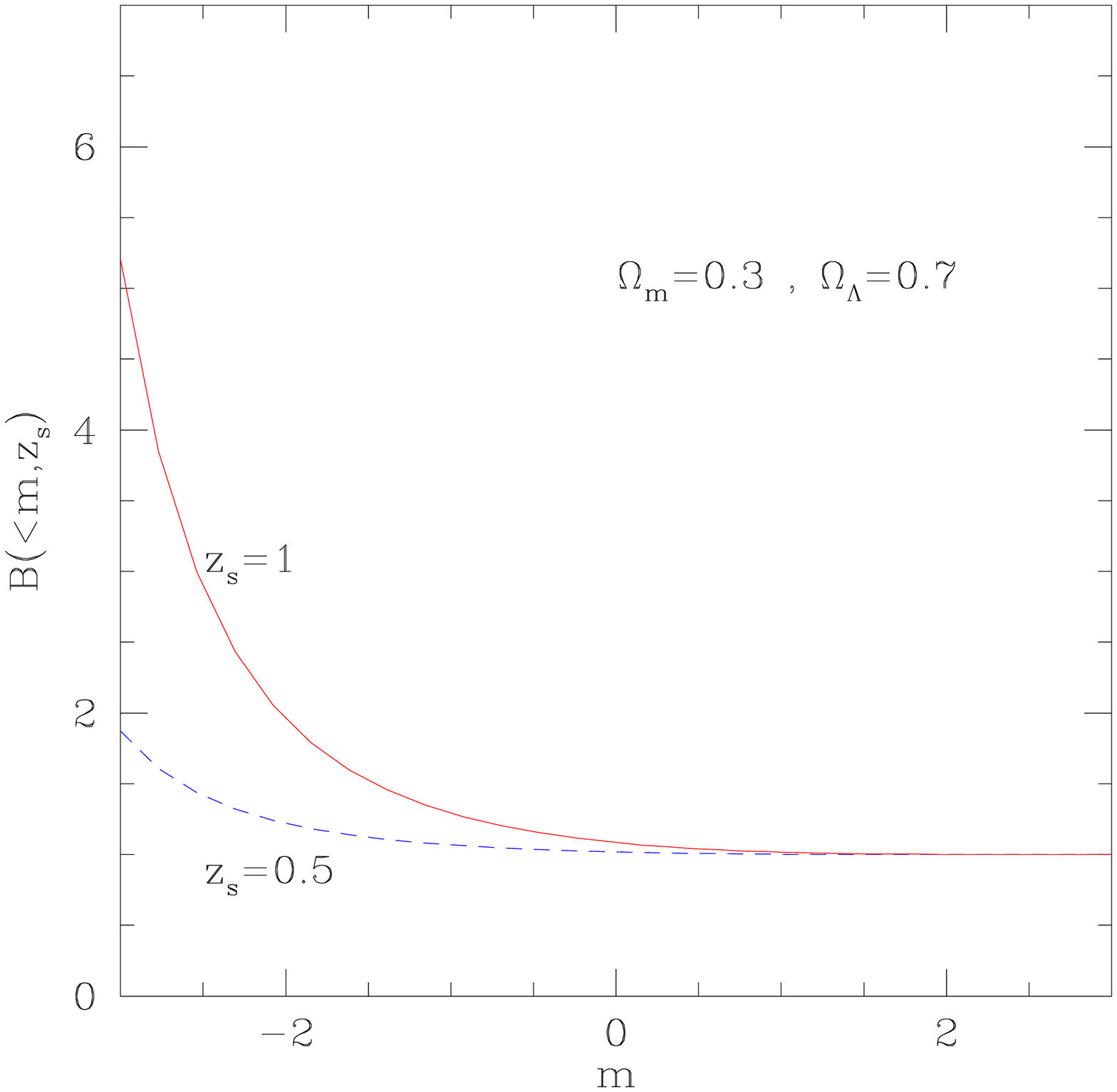} }

\caption{The bias $B(<m,z_s)$ for $z_s=0.5$ (dashed lines) and $z_s=1$ (solid lines).}

\label{figbias}

\end{figure}

We present in Fig.\ref{figbias} our results for the source redshifts $z_s=0.5$ and $z_s=1$. We can see that the bias is larger at higher redshift. Indeed, as can be seen from Fig.\ref{figPmu} the large $\mu$ tail of the probability distribution $P(\mu)$ is more important at higher $z$ because of the larger variance $\ximu$. This implies that the effect of weak lensing is higher (it is more likely for SNeIa to be ``amplified'' by density fluctuations) hence $B(<m,z_s)$ is larger. Note that the bias is not negligible: we already get $B(<m,z_s) \sim 2$ for $m=-2$ at $z_s=1$, which only corresponds to a ``2 sigma'' deviation from the mean $M_{B0}$ for the magnitude of the observed supernova.

These distortions due to weak lensing lead to an apparent luminosity function $\Phi_L$ of the sources:
\beq
\Phi_L(M_B;z) = \int_0^{\infty} d\mu \; P(\mu) \; \Phi(M_B + 2.5 \log \mu ;z)
\eeq
Thus, the mean magnitude $\lag M_B\rag_{M_{Bth}}$ of a survey limited by the absolute magnitude threshold $M_{Bth}$ is:
\[
\begin{array}{l} {\displaystyle \lag M_B\rag_{M_{Bth}} - M_{B0} =  \frac{-1}{\int d\mu P(\mu) \mbox{erfc} \left( - \frac{m_{th} \sigma_B + 2.5 \log \mu}{\sqrt{2}\sigma_B} \right) } } \\ \\ {\displaystyle \hspace{0.5cm} \times \int d\mu P(\mu) \; \Biggl \lbrace \sqrt{\frac{2}{\pi}}  \sigma_B \; \exp \left[ - \left( \frac{m_{th} \sigma_B + 2.5 \log \mu}{\sqrt{2}\sigma_B} \right)^2 \right] } \\ \\  {\displaystyle  \hspace{2.3cm} + 2.5 \; \log \mu \; \mbox{erfc} \left( - \frac{m_{th} \sigma_B + \; 2.5 \log \mu}{\sqrt{2}\sigma_B} \right) \Biggl \rbrace } \end{array}
\]
\beq 
\label{meanMag}
\eeq
The first term (with the exponential) mainly corresponds to the fact that the survey is limited by the upper magnitude $M_{Bth}$ which implies that $\lag M_B\rag_{M_{Bth}} < M_{Bth}$. It vanishes for $M_{Bth} \rightarrow \infty$ when the survey is complete. The second term (with the prefactor $2.5 \log \mu$) does not go to zero even if $M_{Bth} \rightarrow \infty$: it is due to the distortion of the luminosity function which implies that the mean of $\Phi_L$ is no longer $M_{B0}$. Indeed, we obtain at the lowest order in $\ximu$:
\beq
\begin{array}{ll} {\displaystyle \lag M_B\rag_{\infty} } & {\displaystyle  = M_{B0} - 2.5 \int d \mu \; P(\mu) \; \log \mu } \\ \\  & {\displaystyle \simeq M_{B0} + \frac{2.5}{2 \; \mbox{ln} 10} \; \ximu } \end{array}
\label{MBmeanL}
\eeq
and:
\beq
\begin{array}{l} {\displaystyle  \Phi_L(M_B) \simeq \Phi(M_B) \Biggl \lbrace 1 + \frac{2.5}{2 \; \mbox{ln} 10} \; \frac{M_B-M_{B0}}{\sigma_B^2} \; \ximu } \\ \\ {\displaystyle \hspace{1.cm}  + \left( \frac{2.5}{\sqrt{2} \; \mbox{ln} 10} \right)^2 \left( \frac{(M_B-M_{B0})^2}{\sigma_B^4} - \frac{1}{\sigma_B^2} \right) \; \ximu \Biggl \rbrace } \end{array}
\label{PhiL0}
\eeq
As we can check in (\ref{PhiL0}) the dispersion due to weak lensing increases the bright and faint tails of the luminosity function. However, this distortion is not symmetric, even at the lowest order in $\ximu$, because $\lag \log\mu\rag <0$. Hence the faint part of $\Phi(M_B)$ shows a larger increase than the very bright part and $\lag M_B\rag_{\infty} > M_{B0}$. As can be seen from (\ref{PhiL0}) and Fig.\ref{figbias} the effect of weak lensing is more important if the survey is not complete, hence the dependence on $\ximu$ (i.e. on the redshift $z_s$ of the source) is larger. In particular, we have:
\beq
\begin{array}{l} {\displaystyle  m_{th} \ll -1 \; , \; \ximu \ll \left| \frac{\sigma_B}{2.5 m_{th}} \right|^2 \; : } \\ \\ {\displaystyle \lag M_B\rag_{M_{Bth}} \simeq M_{Bth} + \left( \frac{2.5}{\mbox{ln}10} \right)^2 \; \frac{m_{th}}{2\sigma_B} \; \ximu } \end{array}
\eeq

\begin{figure}

{\epsfxsize=8 cm \epsfysize=5.4 cm \epsfbox{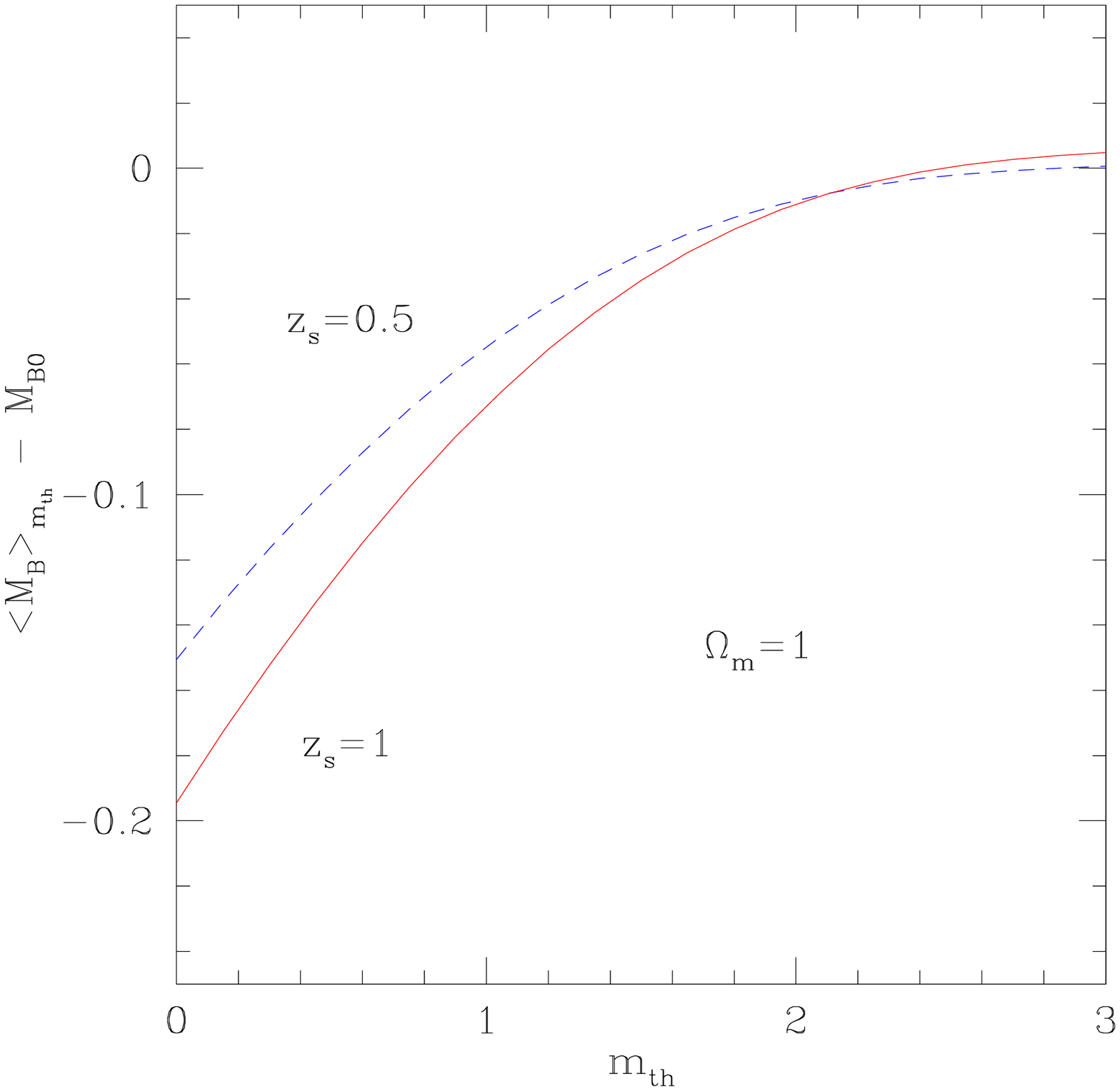} }
{\epsfxsize=8 cm \epsfysize=5.4 cm \epsfbox{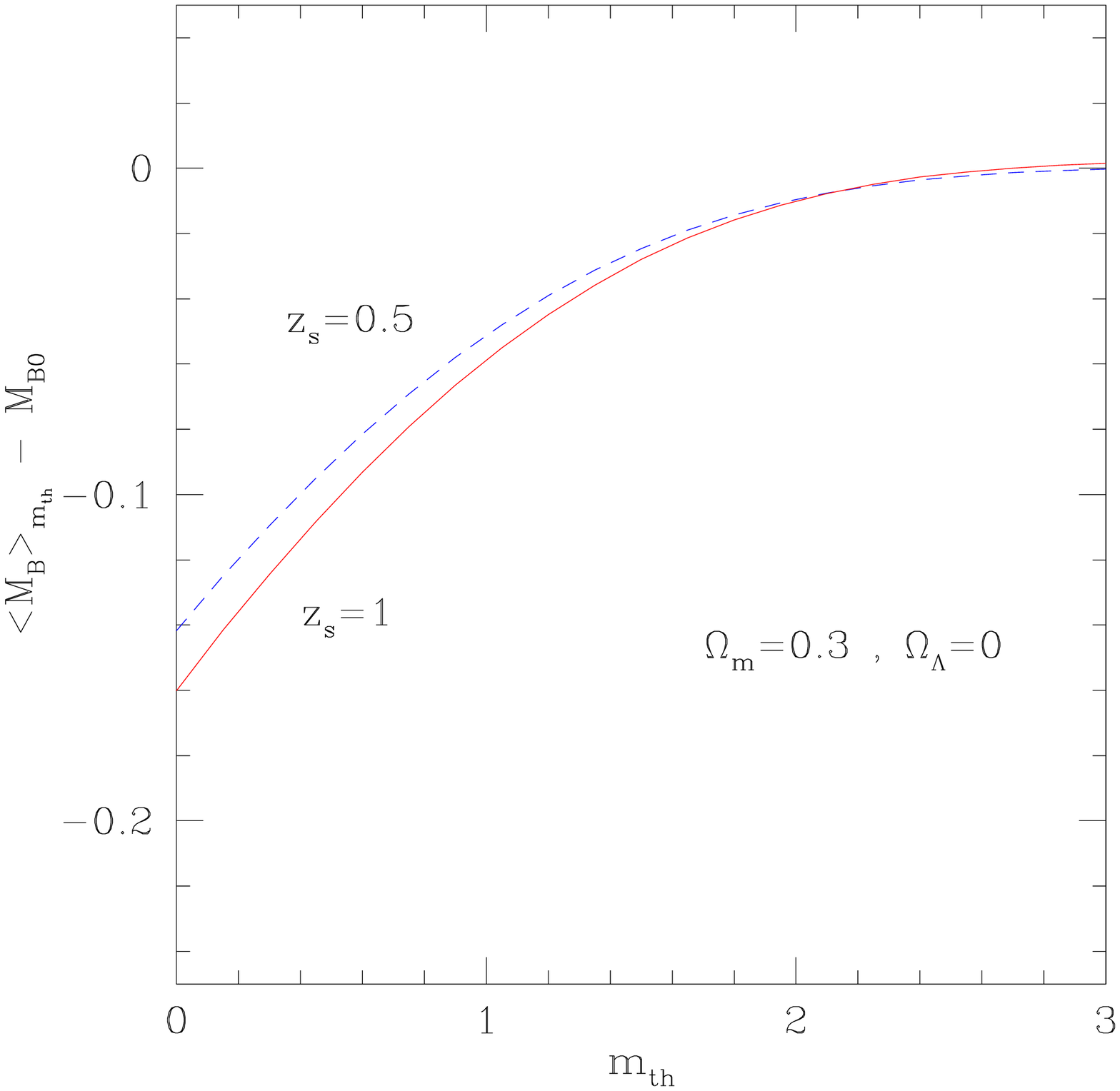} }
{\epsfxsize=8 cm \epsfysize=5.4 cm \epsfbox{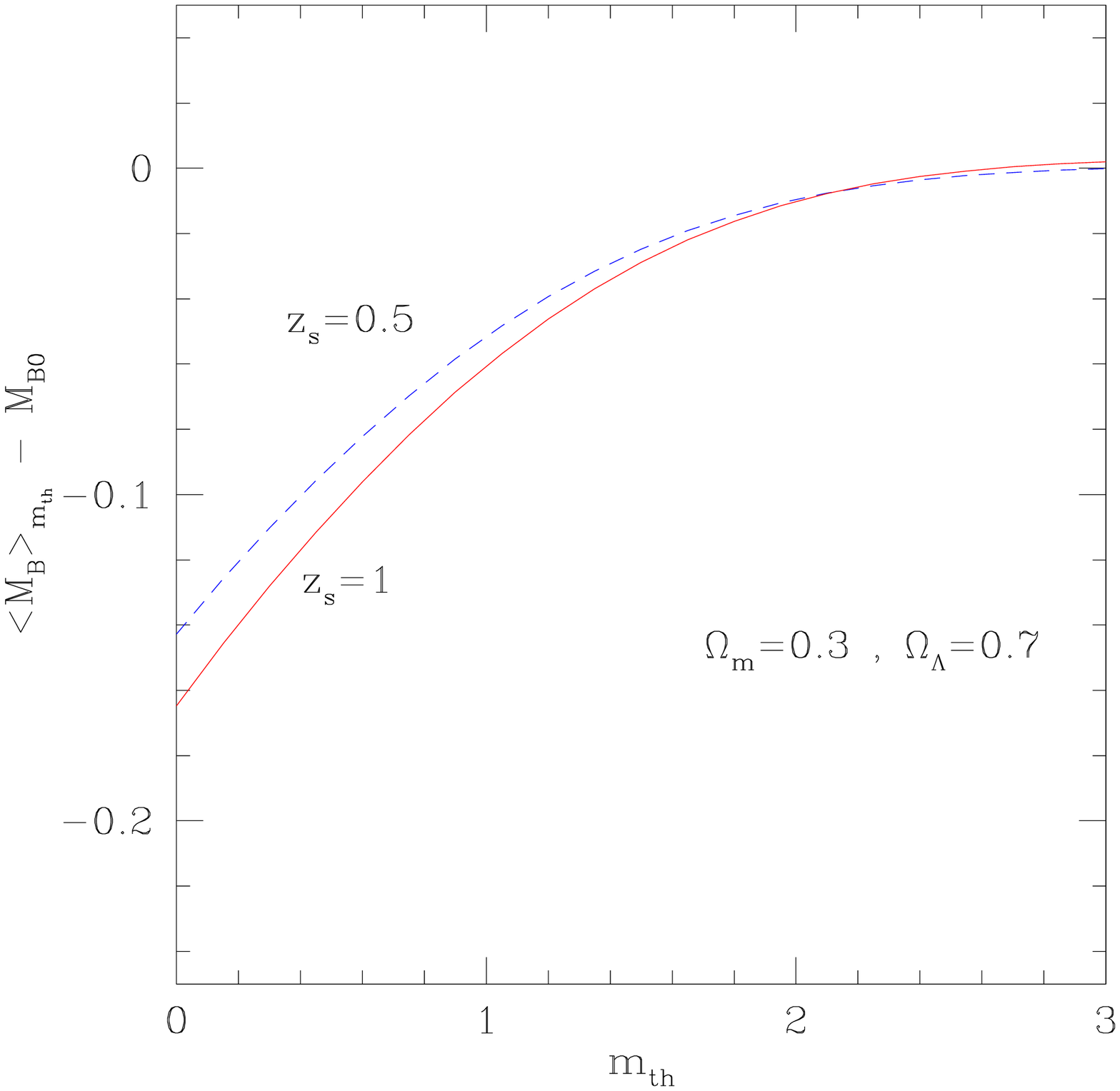} }

\caption{The mean magnitude $\lag M_B\rag_{M_{Bth}}$ of a survey limited by the apparent magnitude $M_{Bth}$ for $z_s=0.5$ (dashed lines) and $z_s=1$ (solid lines). We note $m_{th} = (M_{Bth}-M_{B0})/\sigma_B$ the deviation of the threshold $M_{Bth}$ from the intrinsic mean $M_{B0}$ in units of the dispersion $\sigma_B$.}

\label{figMagbias}

\end{figure}

We present in Fig.\ref{figMagbias} the deviation from the intrinsic mean $M_{B0}$ of the average magnitude measured by a survey with the upper absolute magnitude threshold $M_{Bth}$, given by the ``reduced magnitude'' threshold $m_{th}$ as defined in (\ref{mreduc}). For bright magnitude thresholds the apparent mean is close to $M_{Bth}$ while for faint $M_{Bth}$ the survey is almost complete and $\lag M_B\rag_{M_{Bth}} \simeq M_{B0}$. As explained above, even for $M_{Bth} \rightarrow \infty$ the observed mean is not equal to $M_{B0}$. However, as we can see in (\ref{MBmeanL}) and in Fig.\ref{figMagbias} the deviation is very small for the redshifts of interest $z_s \la 1$ since the variance $\ximu$ of the magnification is small, see Fig.\ref{figXi}. In particular, we note that the curve $\lag M_B\rag (M_{Bth})$ shows a very small dependence on redshift. In practice, in order to measure the cosmological parameters $\Omega_m$ and $\Omega_{\Lambda}$ one observes SNeIa at low ($z_s \sim 0.04$) and high ($z_s \sim 0.8$) redshift. The low $z_s$ data gives the normalization of the curve $m_{Bapp}(z)$, where $m_{Bapp}$ is the apparent magnitude, while the large $z_s$ data constrains the cosmological parameters. Thus, the determination of the cosmological parameters is only sensitive to the difference between the deviations $\lag M_B\rag_{M_{Bth}} - M_{B0}$ at low and large $z_s$. Provided both surveys have a sufficiently faint absolute magnitude threshold: $M_{Bth} > M_{B0} + 2 \sigma_B$, the bias $\Delta M_B$ due to weak lensing effects is:
\beq
\begin{array}{l} {\displaystyle  \Delta M_B(z_1,z_2) = ( \lag M_B\rag_{M_{Bth2}} - M_{B0} ) (z_2) } \\ \\ {\displaystyle \hspace{3cm} - ( \lag M_B\rag_{M_{Bth1}} - M_{B0} ) (z_1) }  \end{array} 
\eeq
where $z_1 < z_2$ are the redshifts of both surveys. Thus, if both surveys are almost complete we have:
\beq
\Delta M_B(z_1,z_2) \la \frac{2.5}{2 \; \mbox{ln} 10} \; \ximu(z_2) 
\eeq
For $z_2<1$ we get $\Delta M_B(z_1,z_2) < 5.4 \; 10^{-3}$, see Fig.\ref{figXi}. Thus this effect is negligible for the determination of cosmological parameters by SNeIa. Of course, if the low and large redshift surveys have bright magnitude thresholds $M_{Bth} \la M_{B0}$ which are different it is not possible to estimate $\Omega_m$ nor $\Omega_{\Lambda}$. This does not seem to be the case in practice (Perlmutter et al.1999).

\begin{figure}

{\epsfxsize=8 cm \epsfysize=5.4 cm \epsfbox{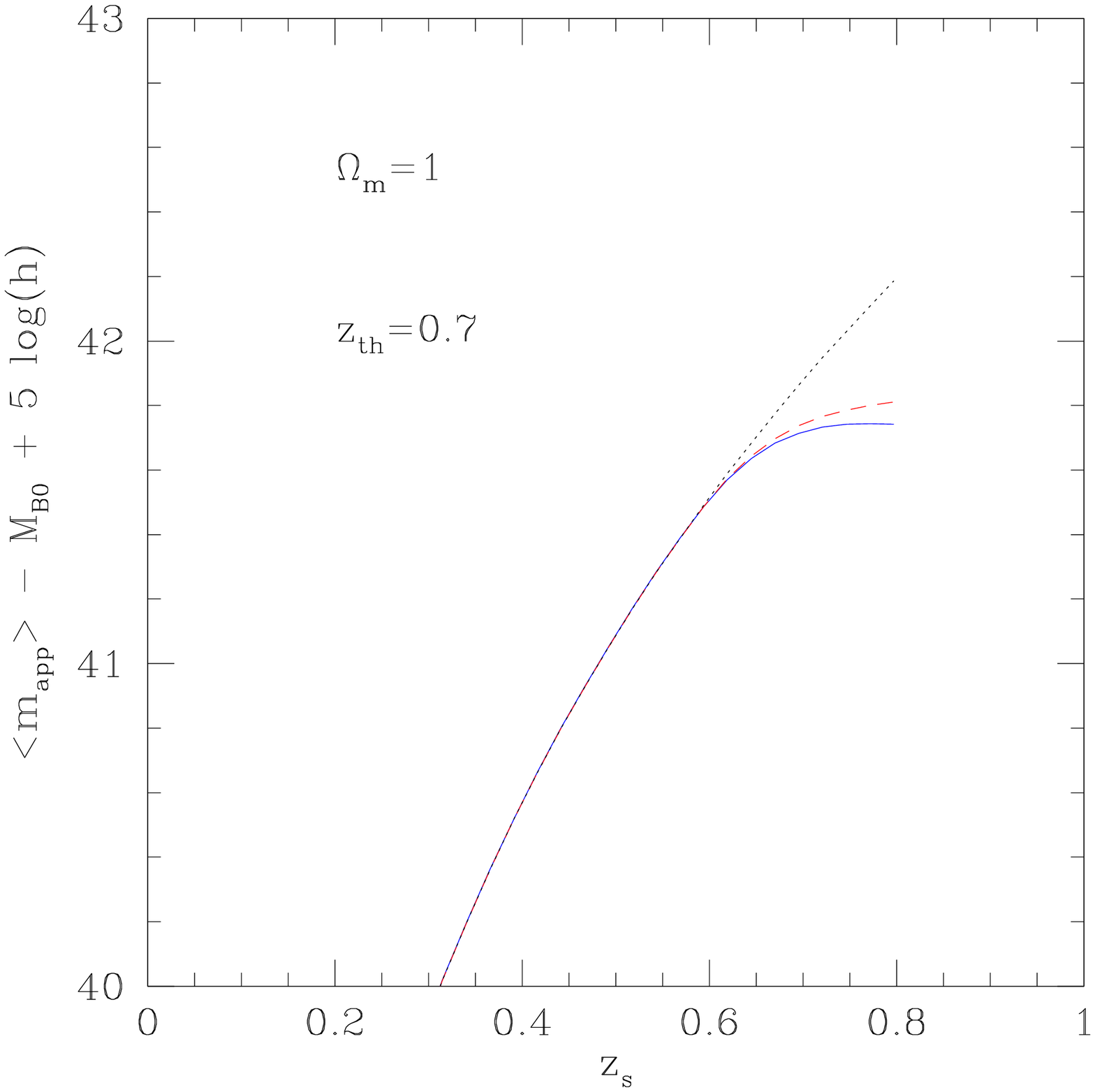} }
{\epsfxsize=8 cm \epsfysize=5.4 cm \epsfbox{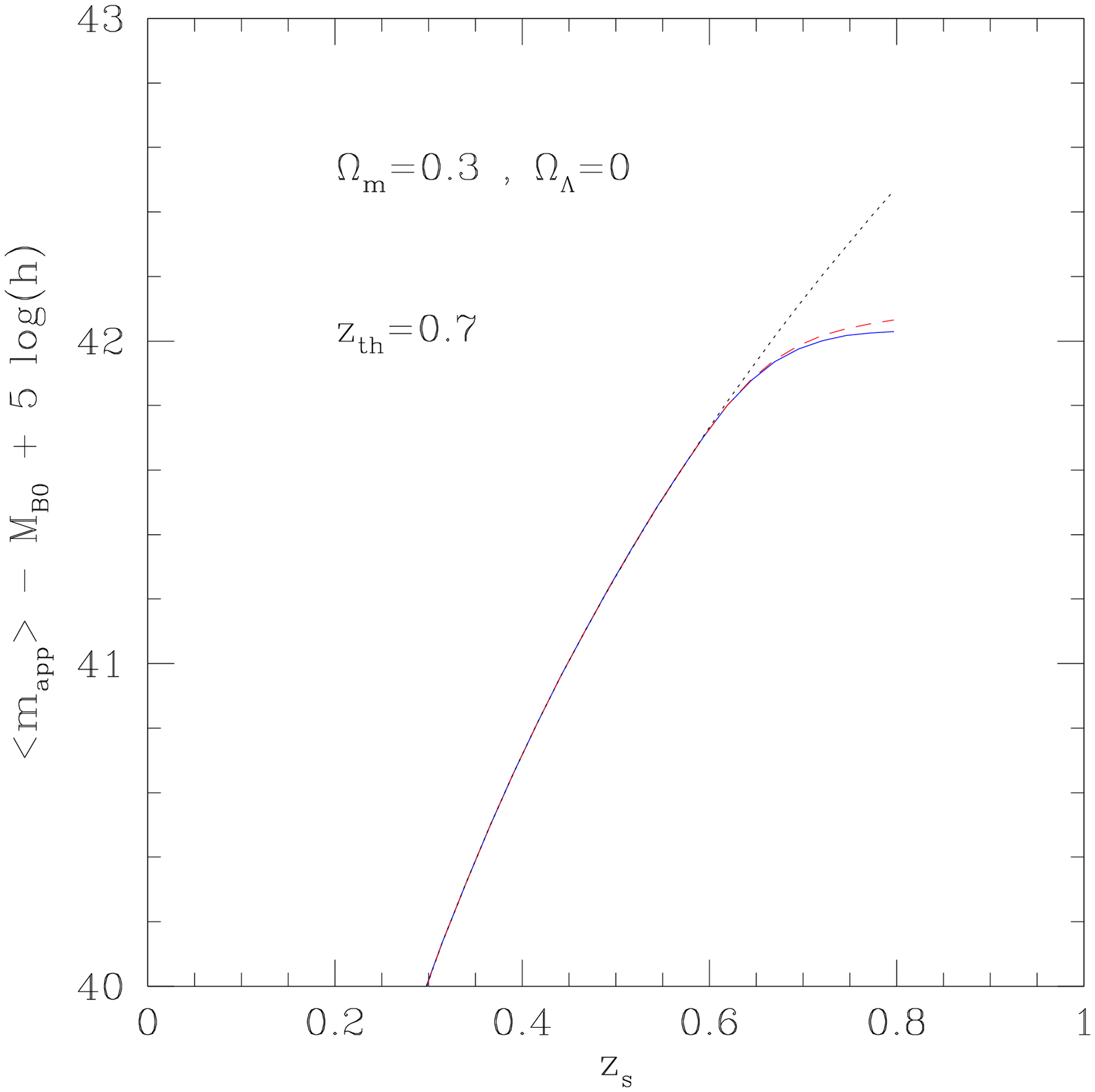} }
{\epsfxsize=8 cm \epsfysize=5.4 cm \epsfbox{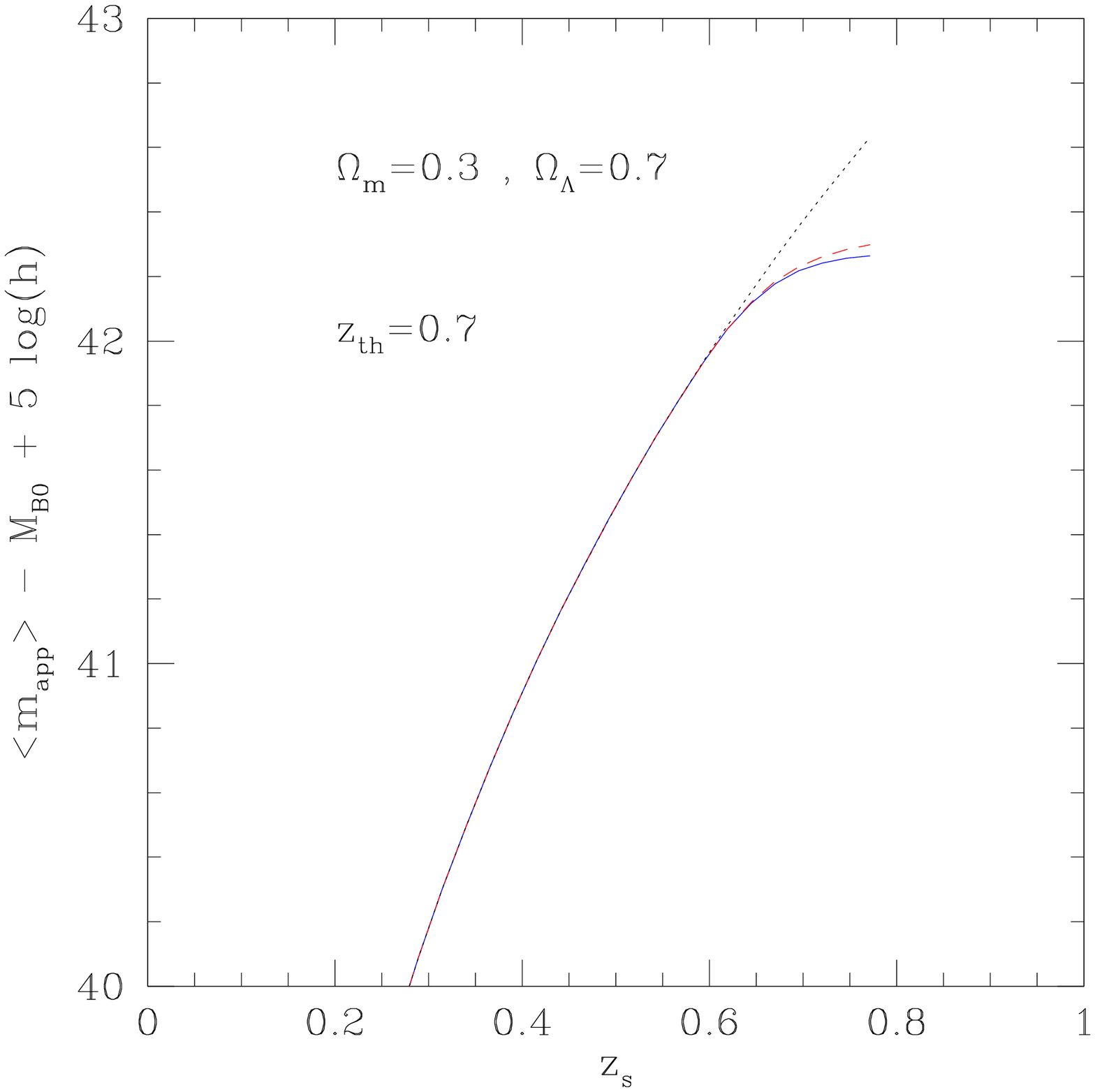} }

\caption{The redshift $\leftrightarrow$ apparent magnitude diagram for three cosmologies. We display the quantity $\lag m_{app}\rag-M_{B0} + 5 \log(h)$ as a function of the redshift $z_s$ of the source. The dotted curves show the apparent magnitude-redshift relation (\ref{mapp3}) obtained for a survey which is not limited by a flux threshold. The solid (resp. dashed) curves show the effect of an apparent magnitude threshold when weak gravitational lensing is (resp. is not) taken into account.}

\label{figMagz}

\end{figure}

In order to derive from observations the cosmological parameters $(\Omega_m,\Omega_{\Lambda})$ one draws a redshift $\leftrightarrow$ apparent magnitude diagram. Moreover, the apparent magnitude $m_{app}$ is ``corrected'' thanks to the light-curve width-luminosity relation (e.g. Perlmutter et al.1999). Of course, because there is some scatter in this latter relation there is still a small lightcurve-width-corrected dispersion $\sigma_B=0.17$ mag. The curve $m_{app}(z)$ depends on the cosmology, through the luminosity distance $d_L$, which allows one to derive $(\Omega_m,\Omega_{\Lambda})$:
\beq
m_{app} = M_B + 5 \log \left( \frac{d_L}{10 \mbox{pc}} \right)
\label{mapp3}
\eeq
where the observed apparent magnitude has been corrected for $K$ and extinction corrections. However, due to the apparent magnitude threshold of SNeIa surveys a Malmquist bias appears at large redshifts where the mean apparent magnitude of supernovae is close to this threshold. Indeed, the curve one obtains from observations is not (\ref{mapp3}) but:
\beq
\lag m_{app}\rag_{M_{Bth}(z)} = \lag M_B \rag_{M_{Bth}(z)} + 5 \log \left( \frac{d_L}{10 \mbox{pc}} \right)
\label{mapp4}
\eeq
where $\lag M_B \rag_{M_{Bth}(z)}$ is the mean magnitude of a survey limited by the absolute magnitude threshold $M_{Bth}(z)$, as in (\ref{meanMag}). The latter is obtained from a given apparent magnitude threshold $m_{appth}$ by:
\beq
m_{appth} = M_{Bth}(z) + 5 \log \left( \frac{d_L}{10 \mbox{pc}} \right)
\eeq
We display in Fig.\ref{figMagz} the curves $\lag m_{app}\rag(z_s)-M_{B0} + 5 \log(h)$ we obtain for three cosmologies (the term $\log(h)$ removes the dependence on the Hubble constant). The dotted curves, which correspond to (\ref{mapp3}) (no magnitude threshold), show the (small) dependence on cosmology of the apparent magnitude-redshift relation: low-density universes lead to larger luminosity distances hence to larger apparent magnitude (fainter object) at fixed $z_s$, this effect is larger for the flat model than for the open case. The solid (resp. dashed) curve shows the effect of an apparent magnitude threshold when weak gravitational lensing is (resp. is not) taken into account. The threshold is chosen so that $M_{Bth}(z)=M_{B0}$ at redshift $z_{th}=0.7$. Thus, we see in the figure that at lower redshifts $z_s \leq 0.6$ the apparent magnitude threshold of the survey plays no role: all curves superpose onto (\ref{mapp3}). At higher redshift $z_s \geq 0.6$, because of the threshold $m_{appth}$, the survey only detects the brightest SNeIa which means that the average $\lag m_{app}\rag_{M_{Bth}(z_s)}$ is biased towards small magnitudes (large luminosities). This leads to a clear deviation of the observed magnitude-redshift relation from (\ref{mapp3}). Thus, the break in the curve clearly marks the redshift beyond which the cosmological parameters cannot be derived from observations with this apparent magnitude threshold. Taking into account the weak lensing effects (solid lines) slightly amplifies this bias towards large luminosities because the random magnification by density fluctuations along the line of sight increases the large apparent luminosity tail of the SNeIa distribution.

\section{Conclusion}

In this article, we have shown how one can obtain the {\it probability distribution of the magnification} of distant sources by weak gravitational lensing, using a {\it realistic description of the density field} which has already been checked against numerical simulations of structure formation within hierarchical scenarios. Thus, this work improves the results obtained by previous analytical studies which used for instance a ``Swiss cheese'' model to describe the universe (Kantowski 1998) or considered a collection of virialized halos amid an empty space. We recover the behaviour observed in numerical simulations, which is not surprising since we consider similar density fields. Thus, the probability distribution of the magnification shows a maximum at a value slightly smaller than the mean $\lag \mu\rag=1$ and it shows an extended large $\mu$ tail. Moreover, the variance $\ximu$ increases at larger redshifts while the deviation from a gaussian gets higher at lower redshifts. The advantage of our approach is that we obtain a direct connection of the weak lensing properties with the characteristics of the underlying non-linear density field. In particular, {\it the non-gaussian behaviour of the magnification is expressed in terms of the non-gaussian properties of the density field}, through its many-body correlation functions.

Then, we have applied our results to the magnification of distant Type Ia supernovae. We have shown that {\it the inaccuracy introduced by weak lensing} in the derivation of the cosmological parameters {\it is not negligible}: $\Delta \Omega_m \ga 0.3$ for two observations at $z_s=0.5$ and $z_s=1$. However, {\it observations can unambiguously discriminate between $\Omega_m=0.3$ and $\Omega_m=1$}. Moreover, in the case of a low-density universe one can clearly discriminate between $\Omega_{\Lambda}=0$ and $\Omega_{\Lambda}=1-\Omega_m$. Besides, the accuracy increases as the number of SNeIa gets larger (there are already 42 available SNeIa, see Perlmutter et al.1999). On the other hand, if it were possible to measure the distortions due to weak lensing one would obtain some valuable information on the properties of the underlying non-linear density field, since we have shown that the probability distribution of the magnification can be directly expressed in terms of the probability distribution of the density contrast at the non-linear scale (typical of present galaxies) where the local slope of the initial linear power-spectrum is $n=-2$. However, this would require a rather high accuracy of the observations, as we have shown that the probability distribution of the magnification is not very extended (the typical deviation is of order $\Delta \mu \sim 0.08$ at $z_s=1$, hence $\Delta$mag $\sim 0.08$). A more detailed discussion of the properties of the {\it p.d.f.} $P(\kappa)$ and $P(M_{\rm ap})$ of the convergence $\kappa$ and the aperture mass $M_{\rm ap}$ is presented in other articles (Valageas 1999b; Valageas 2000; Bernardeau \& Valageas 2000).

\begin{acknowledgements}

The author would like to thank F.Bernardeau for useful discussions.

\end{acknowledgements}

\end{document}